\documentclass{aa}
\usepackage{txfonts}
\usepackage{general}
\usepackage{dcolumn}
\usepackage[version=3]{mhchem}
\graphicspath{{fig/}{radex-mcmc/}}

\begin{document}

\newcommand\tul{\ensuremath{T_{ul}}}
\newcommand\qpart{\ensuremath{{\cal Q}}}
\newcommand\pip{\ensuremath{\Pi^+}}
\newcommand\pim{\ensuremath{\Pi^-}}
\newcommand{\csuro}{{\ensuremath{\rm C/O}}}
\newcommand{\coo}{{\ensuremath{\varpi}}}
\newcommand{\ctot}{{\ensuremath{\rm C_{gas}}}}
\newcommand{\otot}{{\ensuremath{\rm O_{gas}}}}
\newcommand{\stot}{{\ensuremath{\rm S_{gas}}}}
\newcommand{\scas}{{\ensuremath{\rm CAS(S)}}}
\newcommand{\cgas}{{\ensuremath{\rm C_{gas}}}}
\newcommand{\ogas}{{\ensuremath{\rm O_{gas}}}}
\newcommand{\sgas}{{\ensuremath{\rm S_{gas}}}}
\newcommand{\cd}[1]{{\ensuremath{N(\ce{#1})}}}
\newcommand{\nM}{\ensuremath{n_{\rm M}}}
\newcommand{\nE}{\ensuremath{n_{\rm e}}}
\newcommand{\kdr}{\ensuremath{k_{\rm DR}}}
\newcommand{\kr}{\ensuremath{k_{\rm R}}}
\newcommand{\NNS}{\ensuremath{N_{\ce{NS}}}}

\newcommand{\nelec}{\ensuremath{n_{\rm e}}}
\newcommand{\nmol}{\ensuremath{n_{\rm M}}}
\newcommand{\xm}{\ensuremath{\ab{m+}}}
\newcommand{\xa}{\ensuremath{\ab{a+}}}
\newcommand{\xel}{\ensuremath{\ab{e-}}}
\newcommand{\xs}{\ensuremath{\ab{S+}}}

\newcommand{\err}[3]{\ensuremath{#1_{-#2}^{+#3}}}

\newcommand{\new}{}
\renewcommand{\vfig}[2]{\includegraphics[height=#1cm]{#2}}
\renewcommand{\hfig}[2]{\includegraphics[width=#1\hsize]{#2}}
\def\phb#1{\textbf{#1}}

\title{Sulfur gas-phase abundance in dense cores\thanks{To the memory of Charles Malcolm Walmsley (1941-2017).}}
\author{%
  P.~Hily-Blant$^{1,2}$
  \and G. Pineau des For\^ets$^{3,4}$
  \and A.~Faure$^{1}$
  \and F.~Lique$^{2,5}$}

\institute{%
    Univ. Grenoble Alpes, CNRS, IPAG, 38000 Grenoble, France
    \email{pierre.hily-blant@univ-grenoble-alpes.fr}
    \and Institut Universitaire de France
    \and Université Paris-Saclay, CNRS, Institut d’Astrophysique Spatiale, 91405 Orsay, France
    \and Observatoire de Paris, PSL university, Sorbonne Université, CNRS, LERMA, 75014 Paris, France
    \and LOMC-UMR 6294, CNRS-Université du Havre, 25 rue Philippe Lebon, BP F-1123-76 063 Le Havre cedex, France}
\date{}

\abstract{The abundance of volatile sulfur in dense clouds is long-standing problem in studies of the physics and chemistry of star-forming regions. Sulfur is an important species because its low ionization potential may possibly make it an  important charge carrier. The observed sulfur-bearing species in the gas-phase of dense clouds represent only a minor fraction of the cosmic sulfur abundance, which has been interpreted as a signature of sulfur depletion into ices at the surface of dust grains. However, atomic sulfur, which could be the main gas-phase carrier, cannot be observed directly in cold cores. We present measurements of the nitrogen sulfide (NS) radical toward four dense cores performed with the IRAM-30m telescope. Analytical chemical considerations and chemical models over a wide parameter space show that the NS:\ce{N2H+} abundance ratio provides a direct constraint on the abundance of gas-phase atomic sulfur. Toward early-type cores, we find that $n(\rm S)/n_{\rm H}$ is close, or even equal, to the cosmic abundance of sulfur, 14$\times 10^{-6}$, demonstrating that sulfur is not depleted and is atomic, which is in agreement with chemical models. More chemically evolved cores show sulfur depletion by factors up to 100 in their densest parts. In L1544, atomic sulfur depletion is shown to increase with increasing density. Future observations are needed to discover the solid-phase carrier of sulfur. The initial steps of the collapse of pre-stellar cores in the high sulfur abundance regime also need to be explored from their chemical and dynamical perspectives.}

\keywords{Astrochemistry; ISM: abundances; Individual objects: L1521E, L1521B, L1498, L1544}
\maketitle
\section{Introduction}

%\phb{recent paper: \cite{laas2019}}

\begin{figure*}[t]
        \centering
        \includegraphics[width=\hsize]{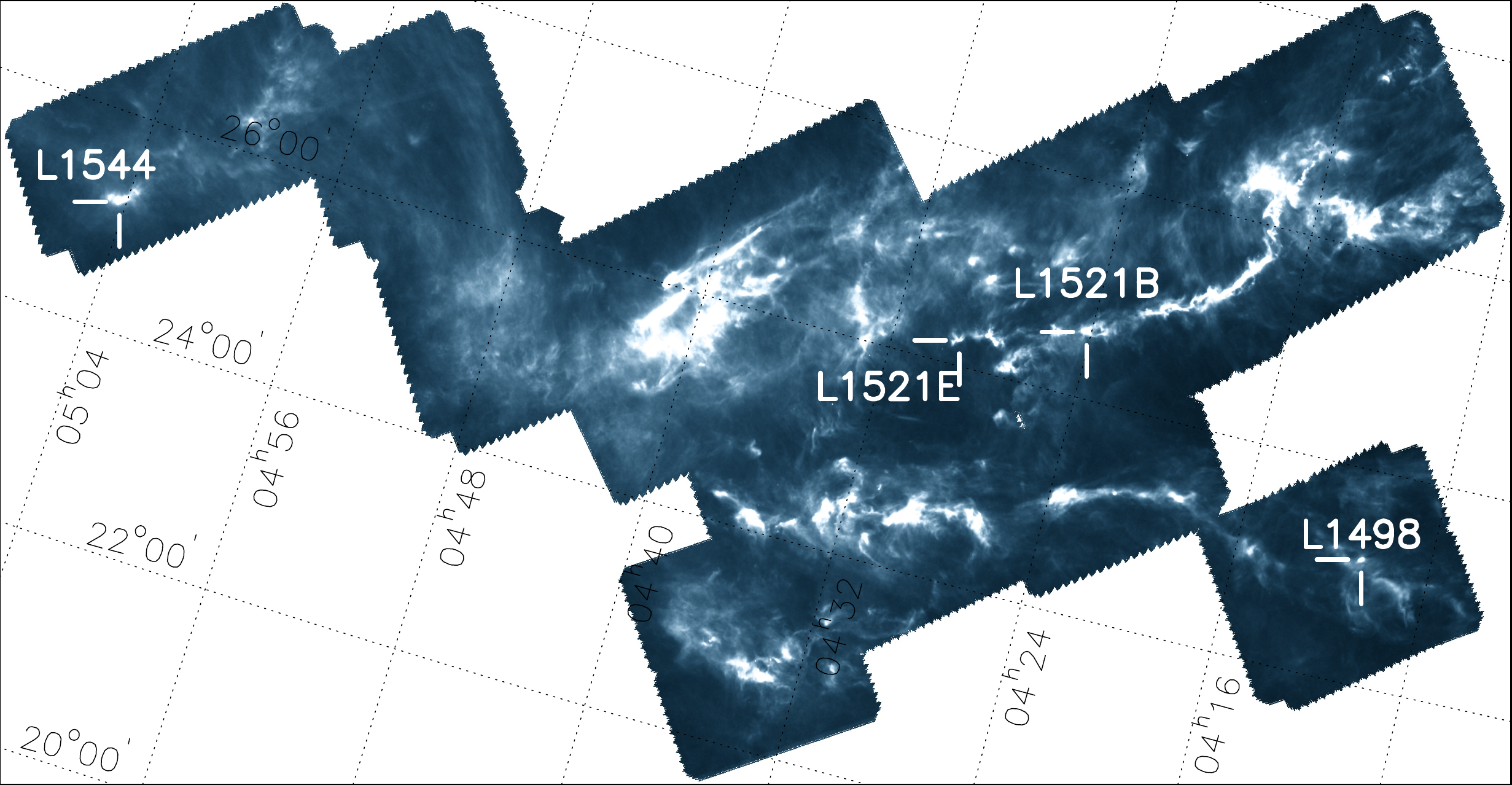}
        \caption{Large-scale environment of the four cores studied in this work. The color scale is the dust emission at 500\micr\ measured with the Herschel/SPIRE instrument \citep{andre2010}. At the distance of the Taurus molecular cloud (140 pc), 1\degr\ corresponds to 2.4 pc.}
        \label{fig:large}
\end{figure*}

The partitioning of interstellar elements between the solid, gas, and ice phases is the result of the complex interplay of refractory grain formation, gas-phase reactive collisions, and adsorption and desorption processes, which {all} depend in a non-linear fashion on the environmental conditions (density, temperature, radiation field, etc.). Beyond the important successes of astrochemistry, the current understanding of these processes is still incomplete and large uncertainties on the abundances of important elements remain \citep{legal2014}.

In the interstellar medium (ISM), some metals are consumed to build up the solid, refractory material of interstellar grains. The propensity of a given chemical element toward condensation into solids has been constrained primarily through absorption-line studies in the ultraviolet \citep{jenkins2009}, showing that it depends on the element. Atoms not locked into grain cores remain in the gas phase and constitute the volatile phase of the ISM (gas and ices). All gas-phase species are expected to freeze-out at sufficiently high density \citep{walmsley2004, friesen2014},  not with the same degree of efficiency, however \citep{tafalla2004a, hilyblant2008cn}.

The total abundance (solid, ice, and gas) of any element is equal to its cosmic abundance, or cosmic abundance standard (CAS), also called elemental abundance. The values for CAS are obtained through laboratory measurements in chondrites \citep{lodders2003} and photospheric spectroscopy of the Sun \citep{asplund2009} or of young, B-type stars located within a $\sim 1$~kpc from the Sun \citep{nieva2012}. It may be emphasized that photospheric determinations are a difficult task. The composition of the refractory phase is constrained indirectly by the wavelength-dependent reddening of background starlight \citep[e.g.,][]{draine2007}. The composition of ices, which become an important reservoir at gas densities \nh\ typically above a few 10\ccc\ \citep[][ Fig.~3]{whittet2010}, is mostly derived from infrared absorption spectra and has been so far limited to bright, star-forming regions \citep{boogert2015}. At the lower densities of the diffuse ISM where ices are not present and chemical elements are essentially atomic, elemental gas-phase abundances can be measured directly through visible and UV spectroscopy \citep{sofia2001}. In contrast, elemental abundances in high-density regions, such as dense cores, star-forming regions, or protoplanetary disks, have never been obtained directly because the main carriers in the gas and ice of the major metals are usually not observable. Thus, the partitioning between dust, gas-phase, and ice can only be inferred by combining observations of trace species and chemical models. The resulting uncertainties are usually larger than the statistical ones, although it is usually not possible to properly quantify by what factor. Part of the uncertainties of the models predictions come from the volatile abundance, that is CAS minus refractory, of the {main elements}: carbon, oxygen, nitrogen, and sulfur. These uncertainties then propagate in a complex way throughout the entire chemical network. Obtaining direct constraints on the volatile abundance of these elements in the dense ISM is thus of paramount importance to obtain quantitative and accurate predictions from astrochemical models.

The volatile abundances (gas plus ice) of carbon and oxygen are uncertain by typically a factor of two \citep{whittet2010, jenkins2014}, leading to large variations of the abundances predicted by astrochemical models \citep{bergin1997a, pratap1997}. These uncertainties can be traced back to measurements of the refractory or gas-phase reservoirs. In particular, it was shown that the C/O ratio, namely, the relative amounts of gas-phase carbon and oxygen, can have a strong impact on the abundances of gas-phase N and \ce{N2}, presumably the main carriers of nitrogen in dense clouds \citep{hilyblant2010n, legal2014}.

In contrast with carbon and oxygen, sulfur shows no tendency toward depletion into the solid phase \citep{jenkins2009}. {Thus, it is expected that sulfur is entirely volatile that is,} sulfur-bearing species are either in the gas or in the ice. The cosmic abundance of sulfur is well determined from spectroscopy of solar and B-type stars, with measurements converging toward $14.1_{-1.5}^{+1.7}$~parts per million (ppm) \citep{daflon2009}. In this paper, we thus adopt 14~ppm as a reference value for the volatile abundance of sulfur. {This value is in harmony with the abundance of gas-phase atomic sulfur in the Orion Molecular Cloud \citep{goicoechea2021}.}

In the dense ISM, only trace sulfur-bearing species are observed, which sum up to $\sim$1\% of the cosmic abundance, and chemical models are unable to consistently reproduce the observed abundances \citep{tieftrunk1994, woods2015}. Solutions to the problem generally involve species in ices, such as organo-sulfur molecules or sulfur chains or rings \citep{jimenez2011, jimenez2014, martin2016, laas2019}. However, {in ices}, it is only OCS that has been detected so far, toward massive YSOs, with an abundance of at most 0.06~ppm, while only upper limits have been obtained for \ce{H2S}, which is not consistent with it being the main carrier of sulfur \citep{smith1991, boogert2015}. Specifically, gas-grain models predict that S-bearing species in ices represent a minor reservoir \citep{woods2015}. Recent chemical models using an updated gas-phase and gas-grain chemical network suggest that sulfur is, indeed, only weakly depleted (by a factor of three) from the gas-phase {in dense clouds, but leaving open the question of the main sulfur reservoirs} \citep{vidal2017}. This is also consistent with the predictions from PDR models, which best reproduce the observations toward the Horsehead nebula when assuming that sulfur is not depleted and in the form of gas-phase S atoms \citep{goicoechea2006}. Further supporting evidence  comes from the finding that gas-phase atomic sulfur is an important, if not dominant, reservoir of sulfur in the dense ISM, derived from the detection of the fine-structure line emission of S\textsc{i} and leading to a lower limit on the abundance of S atoms of 5-10\% with respect to the cosmic abundance \citep{anderson2013}. However, constraints on the abundance of atomic sulfur in pre-stellar cores are scarce and highly indirect.

In this paper, we present observations of the NS molecule toward a sample of low-mass, starless, {and} pre-stellar cores that are combined with published \ce{N2H+} column densities. Based on astrochemical model calculations over a wide parameter range, we show that the \ce{NS}:\ce{N2H+} abundance ratio is a direct tracer of the abundance of atomic sulfur in the gas-phase, a result that can be understood through a simple analysis of the chemistry of NS.

\begin{table*}[t]
        \centering
        \caption{\label{tab:obs}Source sample properties.}
        \begin{tabular}{lccccc}
                \toprule
                Source & Type & RA,DEC & $v_{\rm LSR}$&$n(\hh)$&References\\
                (1)&(2)&(3)&(4)&(5)&(6)\\
                \midrule
                L1521E
                & Starless& 04:29:15.70, 26:14:05.0 & 6.9 & 2.7& (a), (e) \\
                L1521B$^\dagger$
                & Starless& 04:24:12.70, 26:36:53.0 & 6.7 & 1.9& (b), (e) \\
                L1498& Starless & 04:10:51.50, 25:09:58.0 & 7.9 & 1.0 & (c) \\
                L1544& Pre-stellar & 05:04:16.90, 25:10:47.7 & 7.25 & 200&(d)\\
                \bottomrule
        \end{tabular}
        \tabnotes\ $\dagger$ The reference position corresponds to the carbon-chain peak from Hirota et al 2004, but the dust peak is at Ra,Dec=04:24:34, 26:36:43 (J2000.0). \textit{Columns}: (2) See Section \ref{sec:sample}. (3) Reference position (EQ J2000.0) in the maps of Fig.~\ref{fig:maps}; (4) Systemic LSR velocity (in \kms); (5) Central \hh\ densities in units of \dix{5}\ccc. (6) References: (a) \cite{tafalla2004b}; (b) \cite{hirota2004}; (c) \cite{magalhaes2018a}; (d) \cite{bizzocchi2013}; (e) this work.
\end{table*}

\def\wa{3.8cm}
\begin{figure}[t]
        \centering
        \includegraphics[height=\wa]{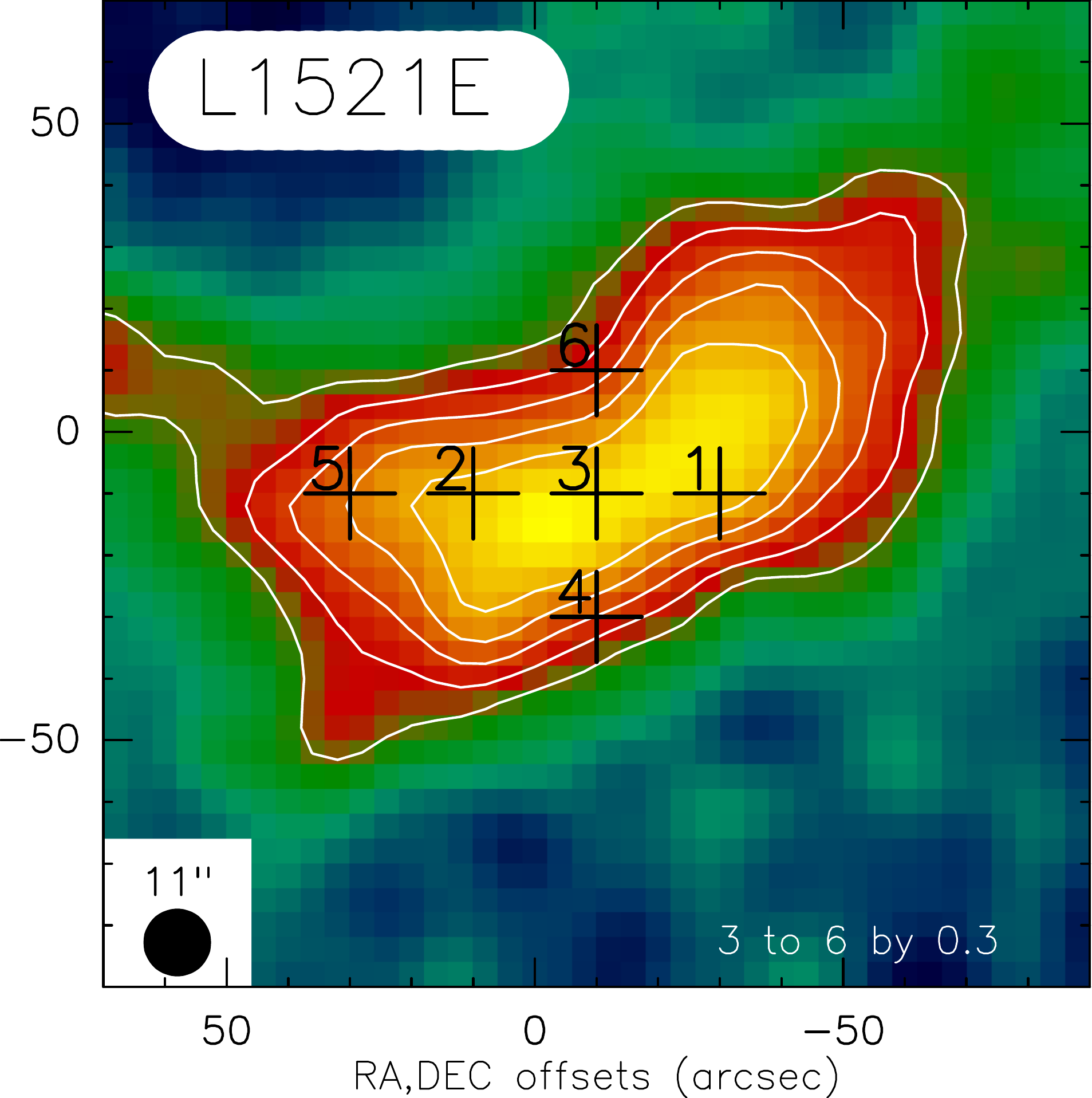}\hfill%
        \includegraphics[height=\wa]{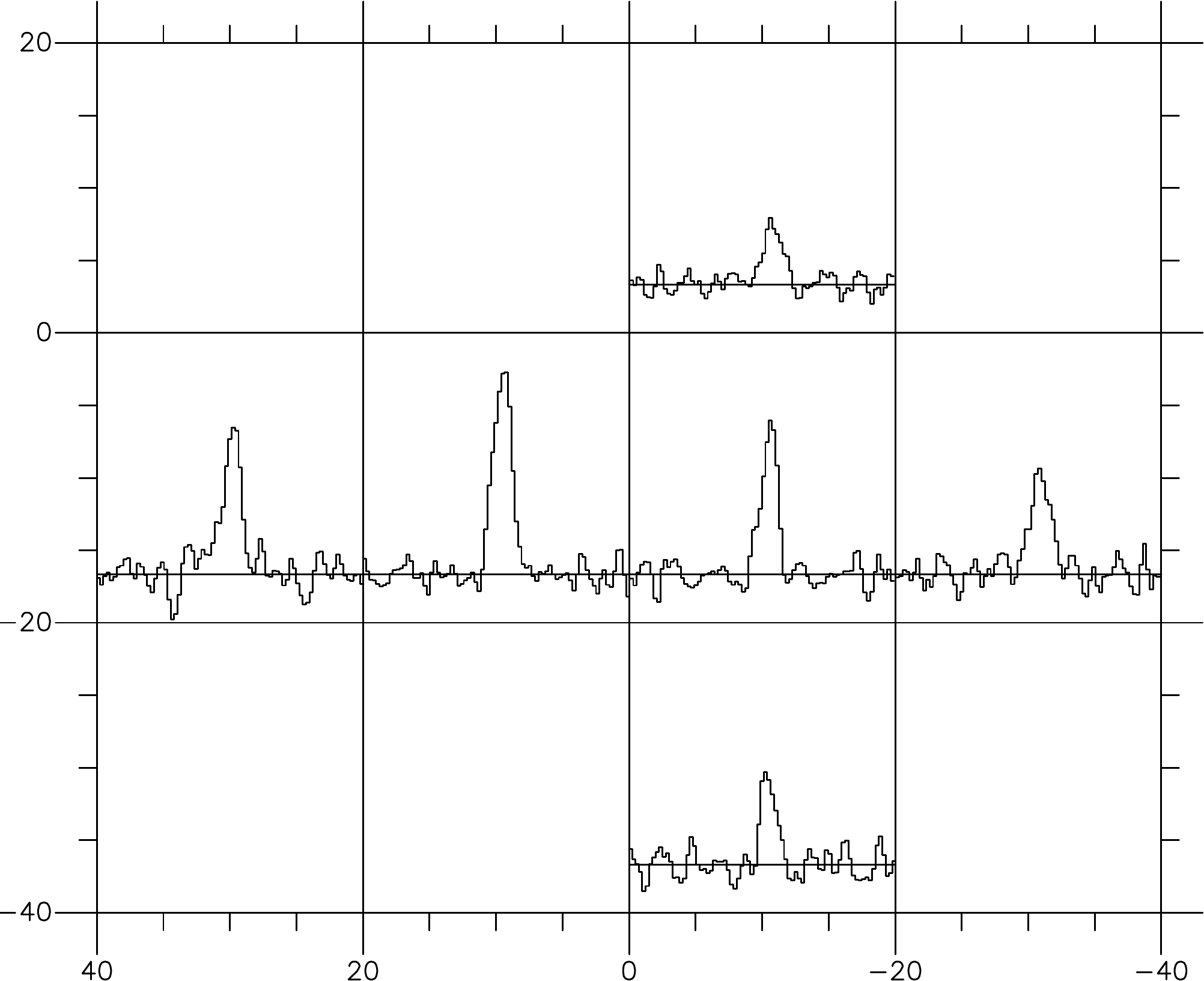}\bigskip\\
        \includegraphics[height=\wa]{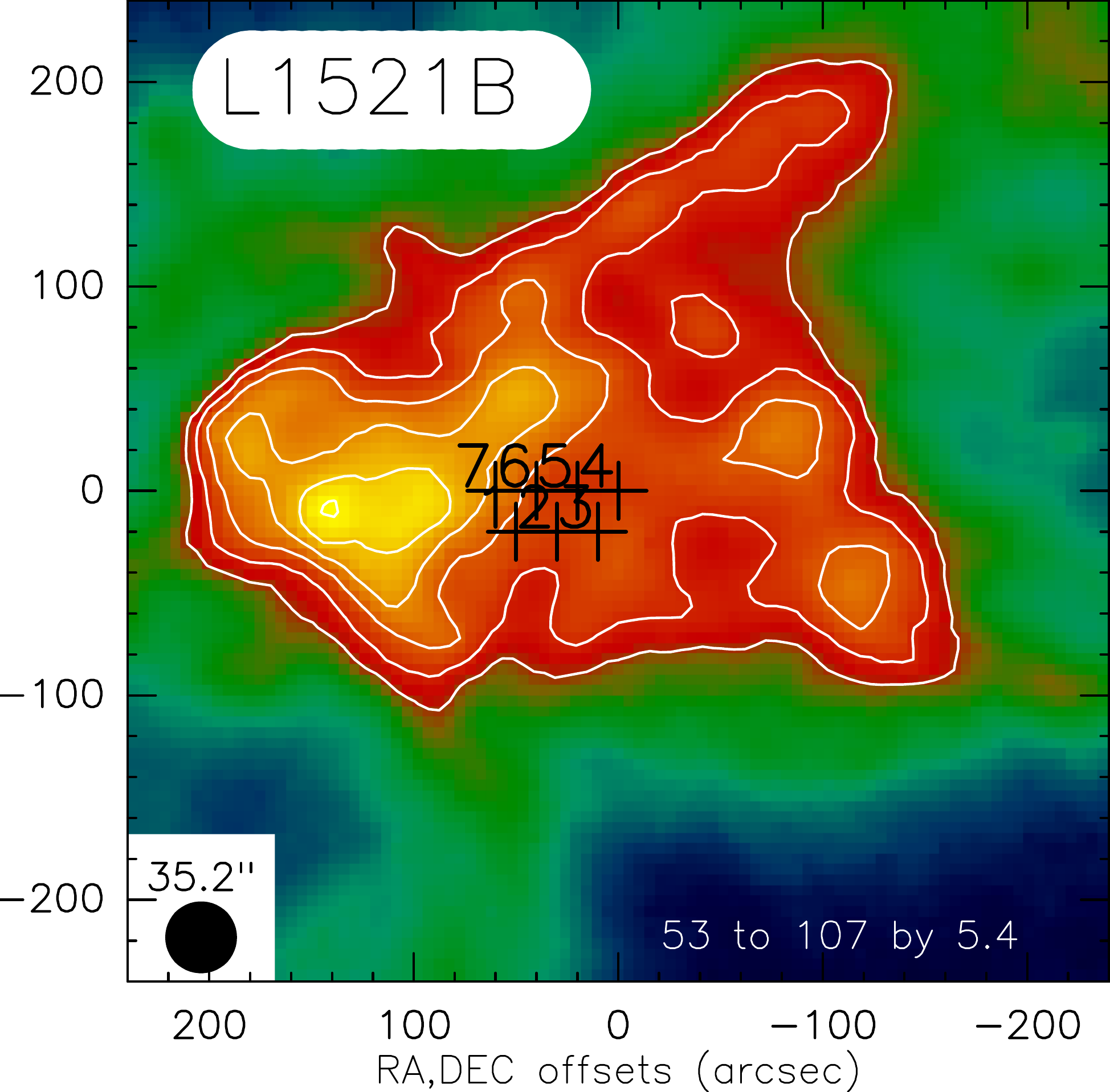}\hfill%
        \includegraphics[height=\wa]{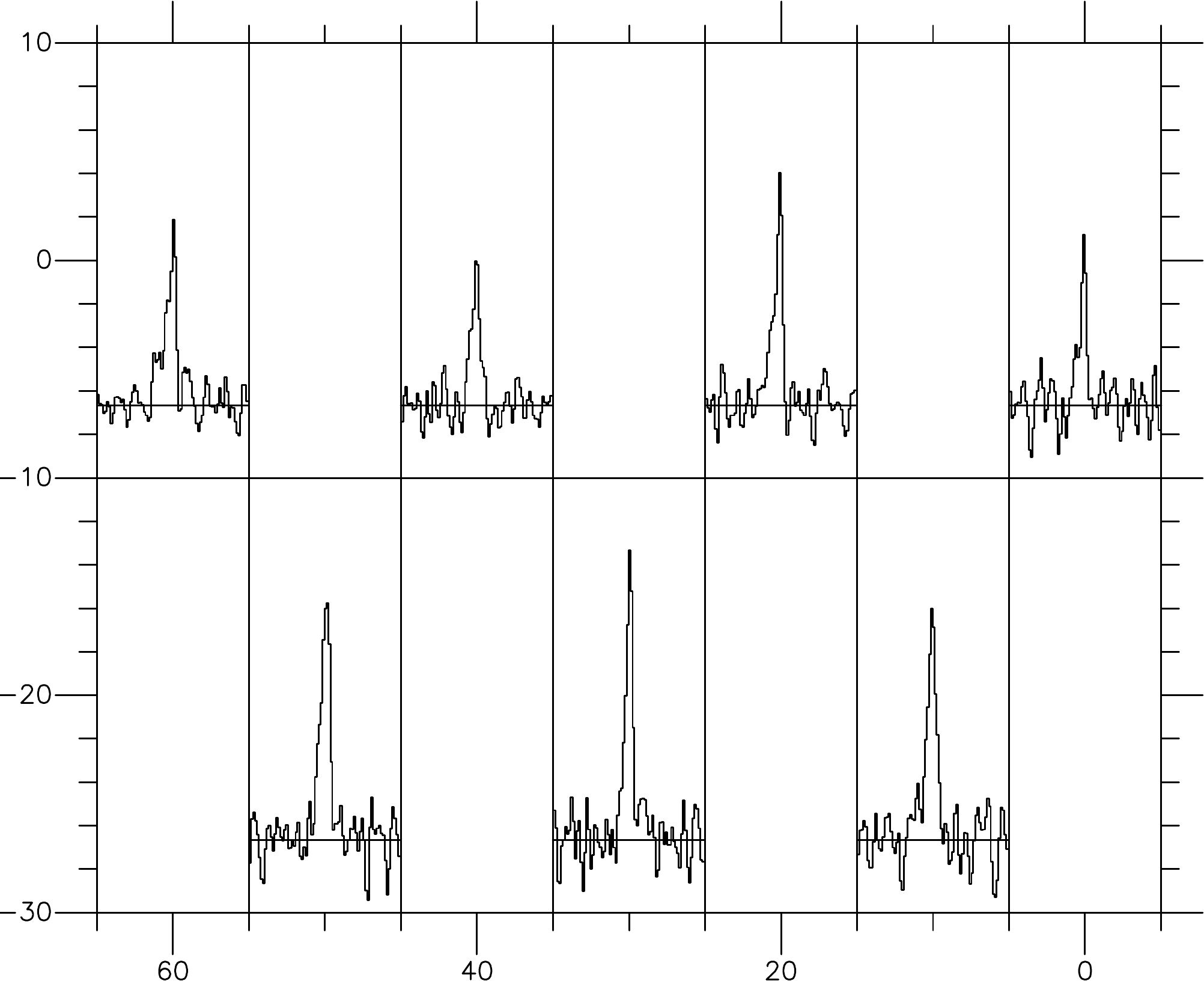}\bigskip\\
        \includegraphics[height=\wa]{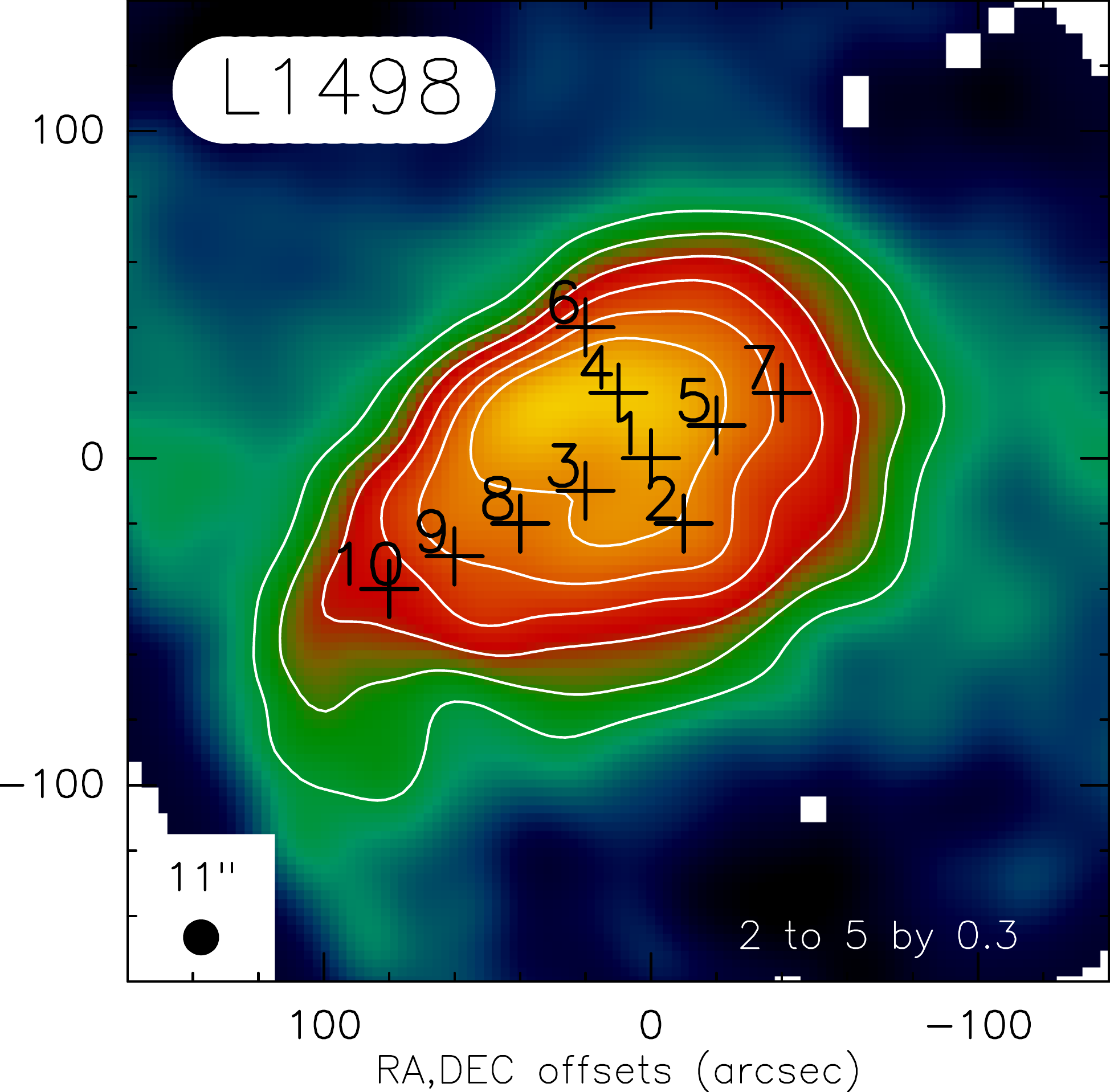}\hfill%
        \includegraphics[height=\wa]{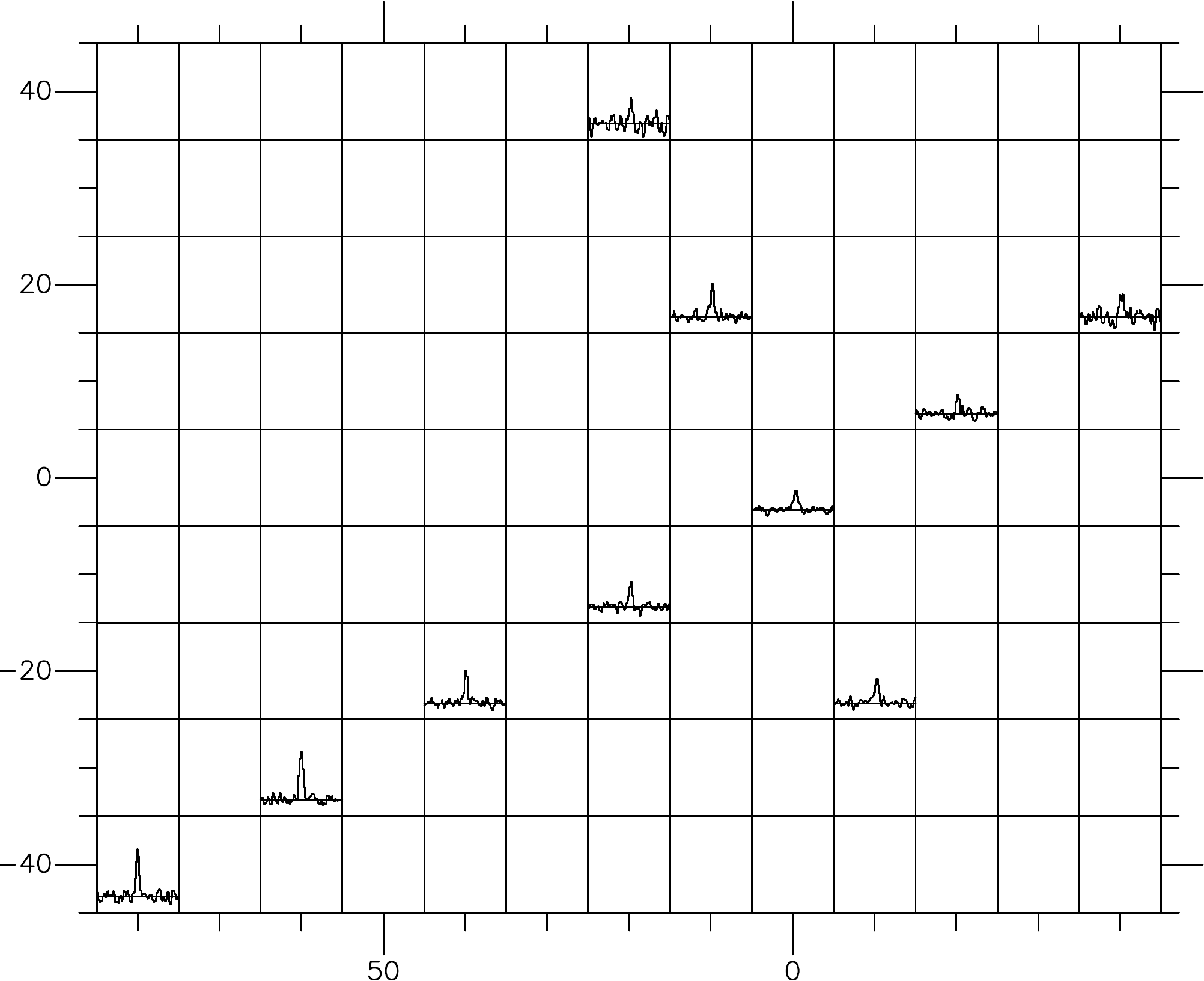}\bigskip\\
        \includegraphics[height=\wa]{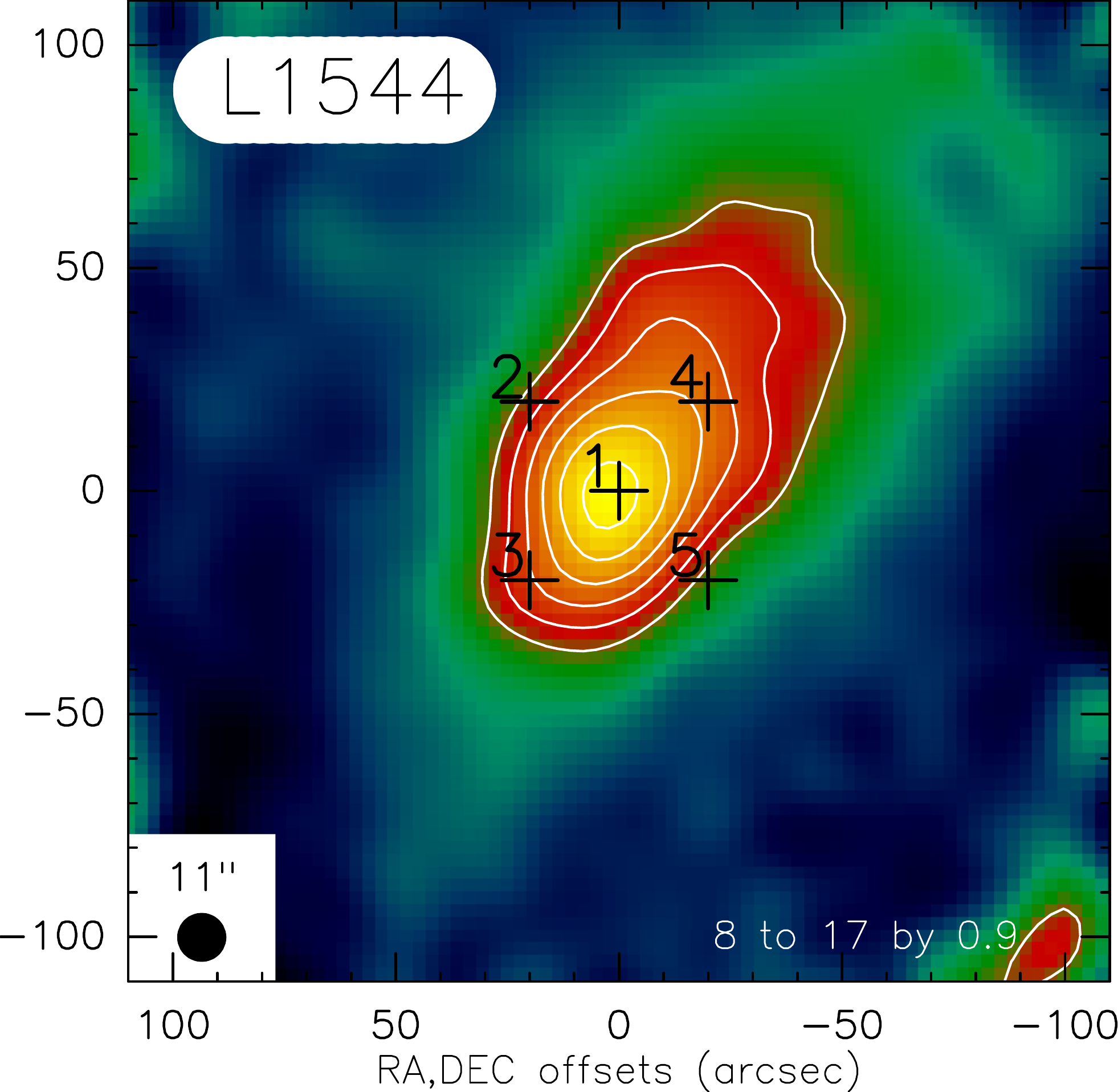}\hfill%
        \includegraphics[height=\wa]{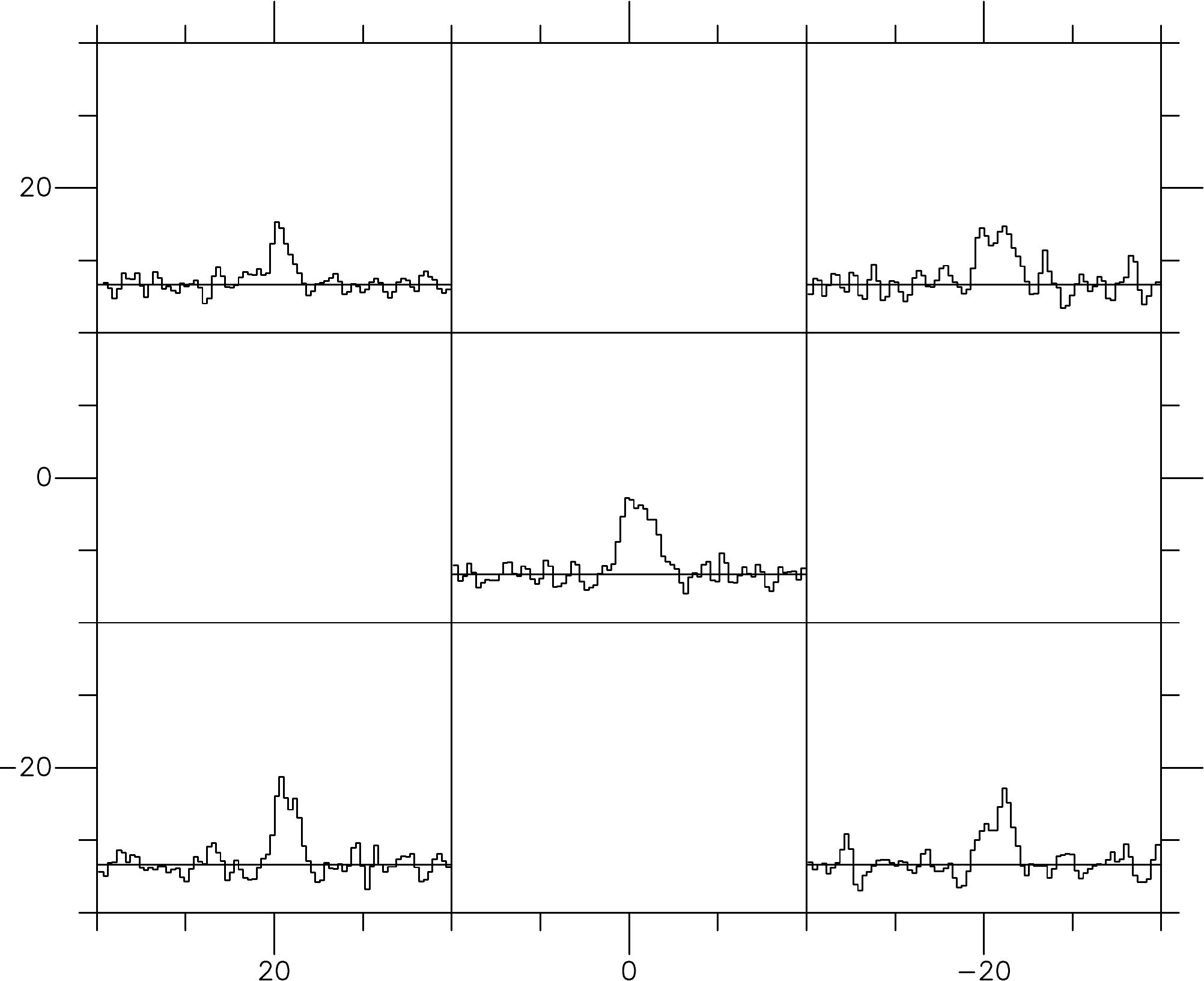}\bigskip\\

        \caption{Maps of the dust emission and positions of the NS spectra for each source in the sample.\textit{Left}: Dust emission maps at $\lambda=$1.3mm (IRAM/30m) toward L1521E, L1498, and L1544 \citep{tafalla2002, tafalla2004a}, and at $\lambda=$0.5mm (Herschel/SPIRE) toward L1521B. Contour levels {(in MJy/sr)} and HPBW are indicated in each panel. The reference positions (0",0 offsets) are given in \rtab{obs}. \textit{Right}: Map of the NS emission line (only the main hyperfine component of the $\Pi^+$ manifold at 115 GHz is shown; see also Fig.~\ref{fig:ns115}). The intensity scale (\tant, -0.1 to 0.5 K) is the same for all sources.}
        \label{fig:maps}
\end{figure}

\section{Observations}
\label{sec:obs}

\subsection{Source sample}
\label{sec:sample}

Our source sample is made of low-mass dense cores located in the Taurus molecular cloud (see Fig.~\ref{fig:large}) and includes both starless cores  (which do not show in-fall signatures) and one pre-stellar core (L1544), and none of the observed cores harbor a protostar. The reference position and some basic properties of each core are summarized in Table \ref{tab:obs}.

L1521E is a young, starless core located in the B218 filament in the Taurus molecular cloud \citep{tafalla2004b} that is characterized by a very low depletion level, exemplified by its high CO:\nnhp\ column density ratio. L1521B is another starless core located in the B216 filament in the Taurus molecular cloud and is classified as a chemically young core on the basis of its large abundance of carbon-chain molecules (e.g., CCS) over late-type species such as ammonia \citep{hirota2004}. Furthermore, the B216 region is believed to be the most recently condensed filament in the L1495 filamentary complex \citep{seo2015}. Next is L1498, a well-studied starless core situated in the periphery of the Taurus molecular cloud. L1498 is believed to be in a more advanced stage of evolution than the young L1521E and L1521B, on the basis of clear CO-depletion patterns and high abundance of \ce{N2}-daughter molecules \citep{tafalla2004a, magalhaes2018a}. Finally, L1544, the most evolved of the cores in our sample, is a pre-stellar object also located at the outskirts of the Taurus molecular cloud. Contrary to the other cores of our sample, it shows signposts of infall in the very inner regions, as traced by inverse P-Cygni profile of water lines \citep{caselli2012}. It is also the core with the highest central density.

In view of computing abundances, we have estimated the total molecular hydrogen column density, \Nhh, in 26"-beam or 35"-beam, from published radial density profiles or from multi-wavelength dust emission maps analyzed in a standard fashion assuming a black-body radiation combined with a frequency-dependent dust opacity. The minimization was performed using a Monte-Carlo Markov-Chains (MCMC) approach, as described in {detail in Appendix~\ref{app:NH2}.}

\subsection{IRAM/30m observations}

The observed millimeter transitions of the NS radical (mass number 46) involve the rotational levels of the $^2\Pi_{1/2}$ manifold of the fundamental electronic and vibration state and {have a structure} similar to the NO radical \citep{gerin1992}. The lowest rotational level of the $^2\Pi_{3/2}$ manifold lies 322~K above the $J=1/2$ lowest level of the $^2\Pi_{1/2}$ manifold, and is not significantly populated at the low temperatures of dark clouds. The unpaired electron leads to $\Lambda$-type doubling of each rotation level leading to two rotational levels of opposite parity. Each $\Lambda$-level further splits into a set of hyperfine levels due to the coupling of the nuclear spin of the nitrogen atom ($I=1$) with the orbital angular momentum $J$. The hyperfine levels are described by the quantum number $F$ where $\mathbf{F = I+J}$. Radiative transitions are subject to selection rules: $\Delta J = \pm1$, $\Delta F =0, \pm1$, and a change of parity. This results into a set of hyperfine transitions as shown in Fig.\,\ref{fig:spectro}.

The observations have been performed in August 2016 (project 008-16) with the IRAM-30m telescope. The EMIR receivers were connected to the VESPA backends facility to measure simultaneously the $+\rarr -$ and $- \rarr +$ hyperfine transitions between the $J=5/2$ and $J=3/2$ rotational levels (see Table~\ref{tab:spectro} and Fig.~\ref{fig:spectro}) with 20~kHz spectral resolution, or 0.05~\kms. Weather conditions were good and opacity at these frequencies is dominated by atmospheric oxygen. The median receiver temperature, system temperature, and zenith opacity were respectively 60~K, 294~K, and 0.4. In some cases, two $J=7/2 \rightarrow 5/2$ transitions could be observed. Assuming no beam dilution and negligible contribution from the error-beams, the antenna temperature was converted into main-beam temperature as $\tmb=\tant\times\feff/\beff$ with\feff\ and \beff\ the forward and beam efficiencies. Values of \feff\ and \beff\ are 0.94 and 0.78 at 115~GHz, respectively, and 0.93 and 0.71 at 161~GHz. {The half-power beam widths at the frequencies relevant to the present study---including the \nnhp\ lines from the literature---are 26", 21", and 15", at 93, 115, and 161~GHz, respectively.}

\def\wa{0.23\hsize}
\begin{figure*}
  \centering
  \includegraphics[width=\wa]{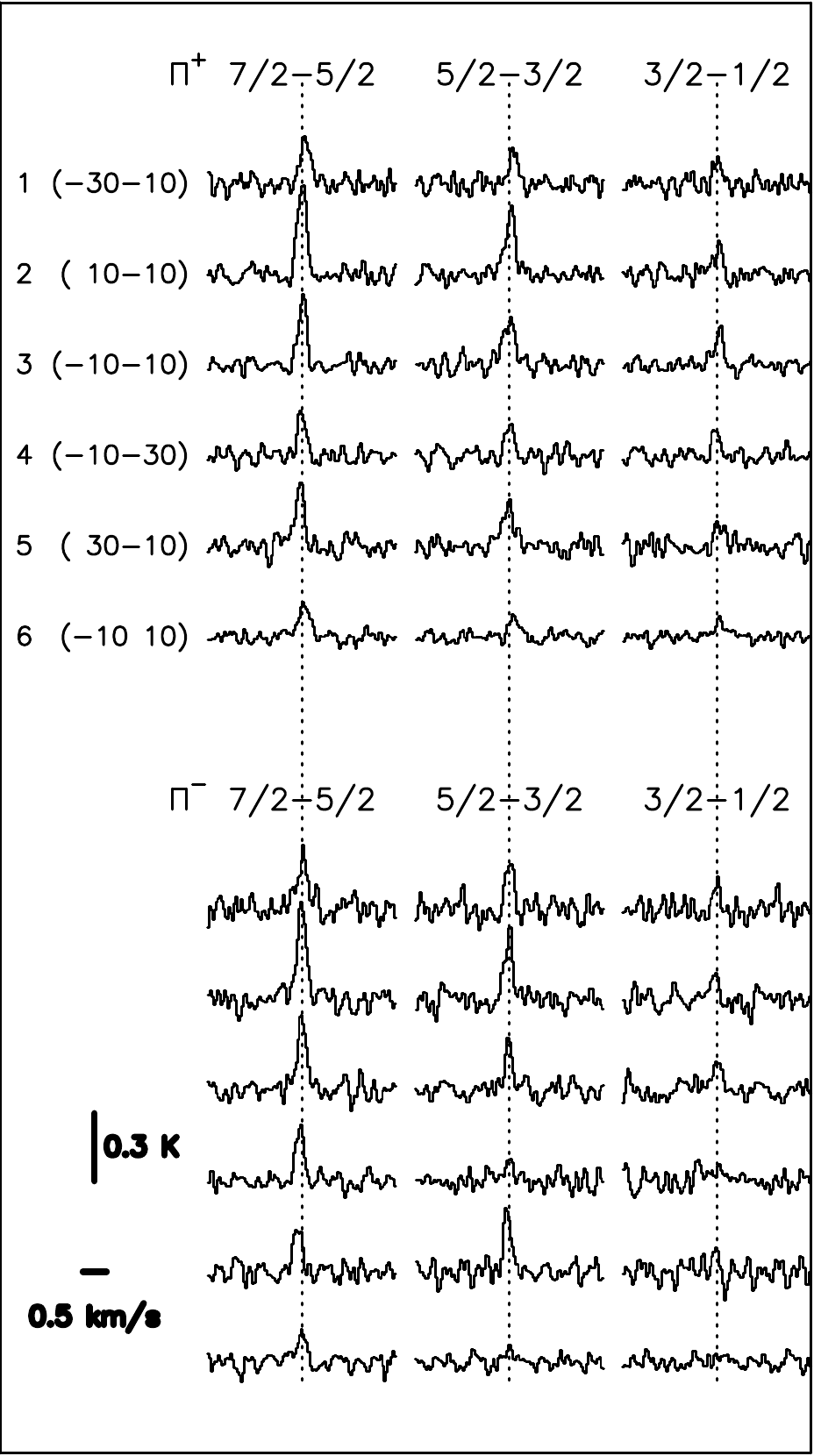}\hfill%
  \includegraphics[width=\wa]{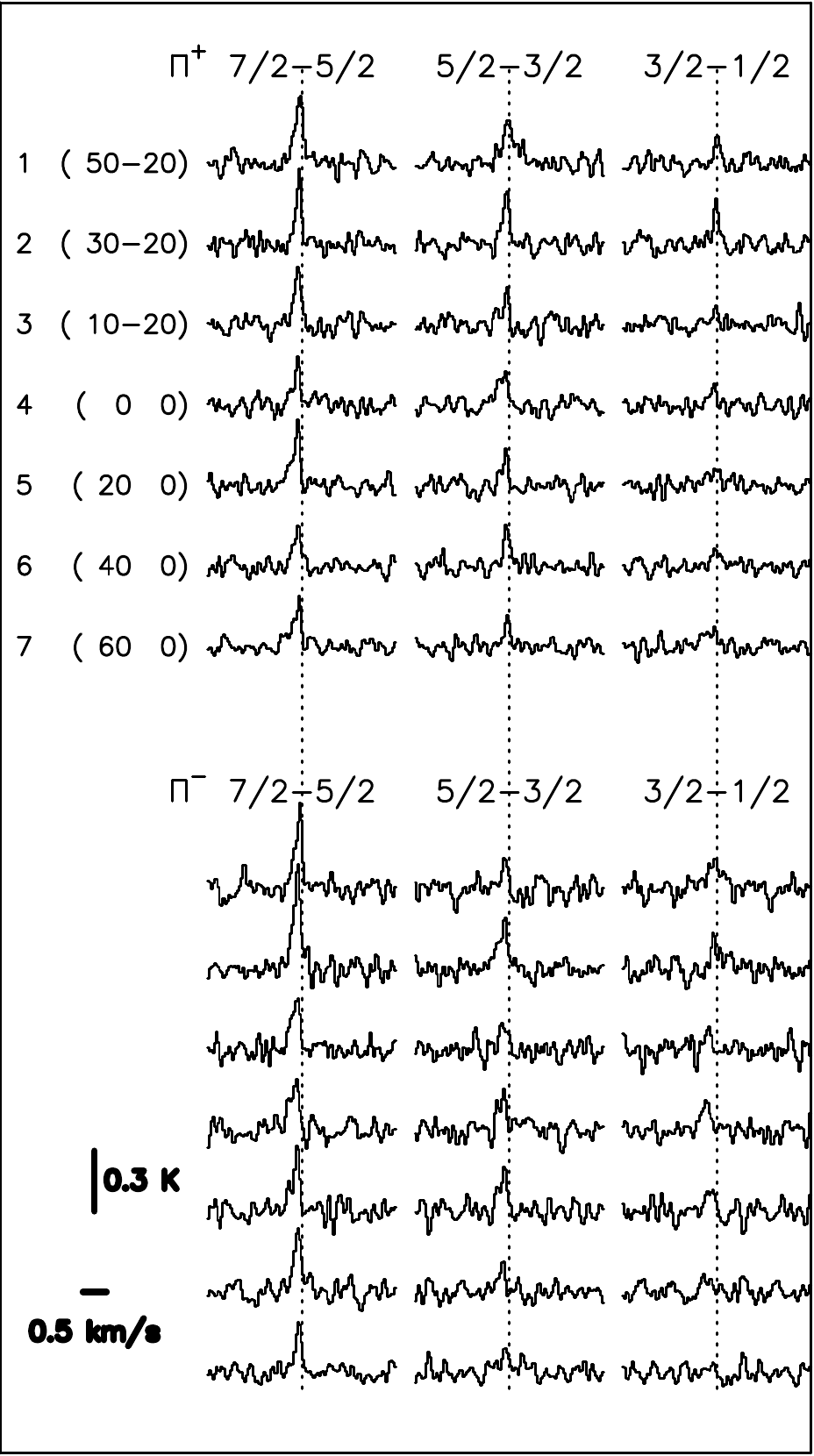}\hfill%
  \includegraphics[width=\wa]{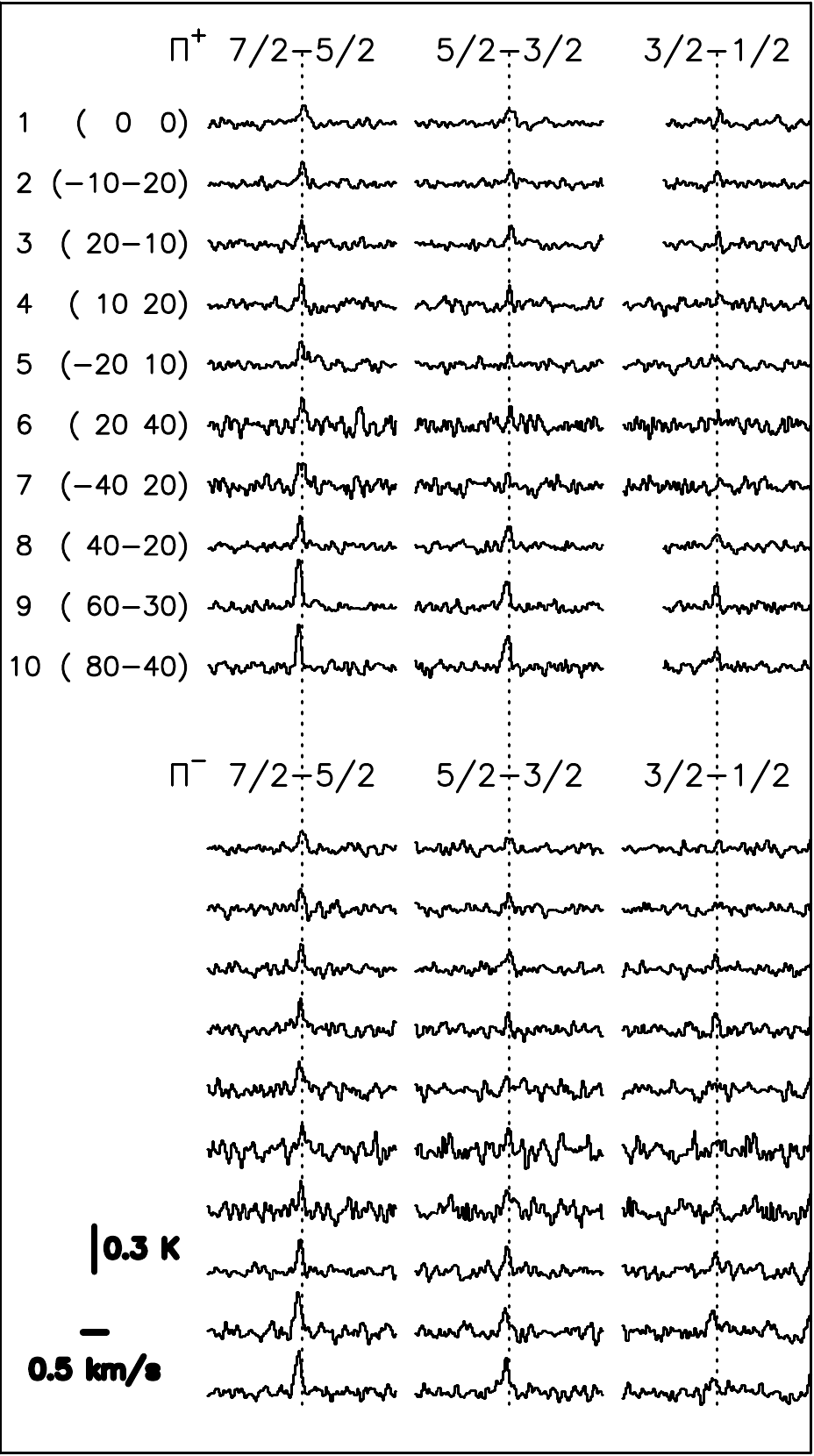}\hfill%
  \includegraphics[width=\wa]{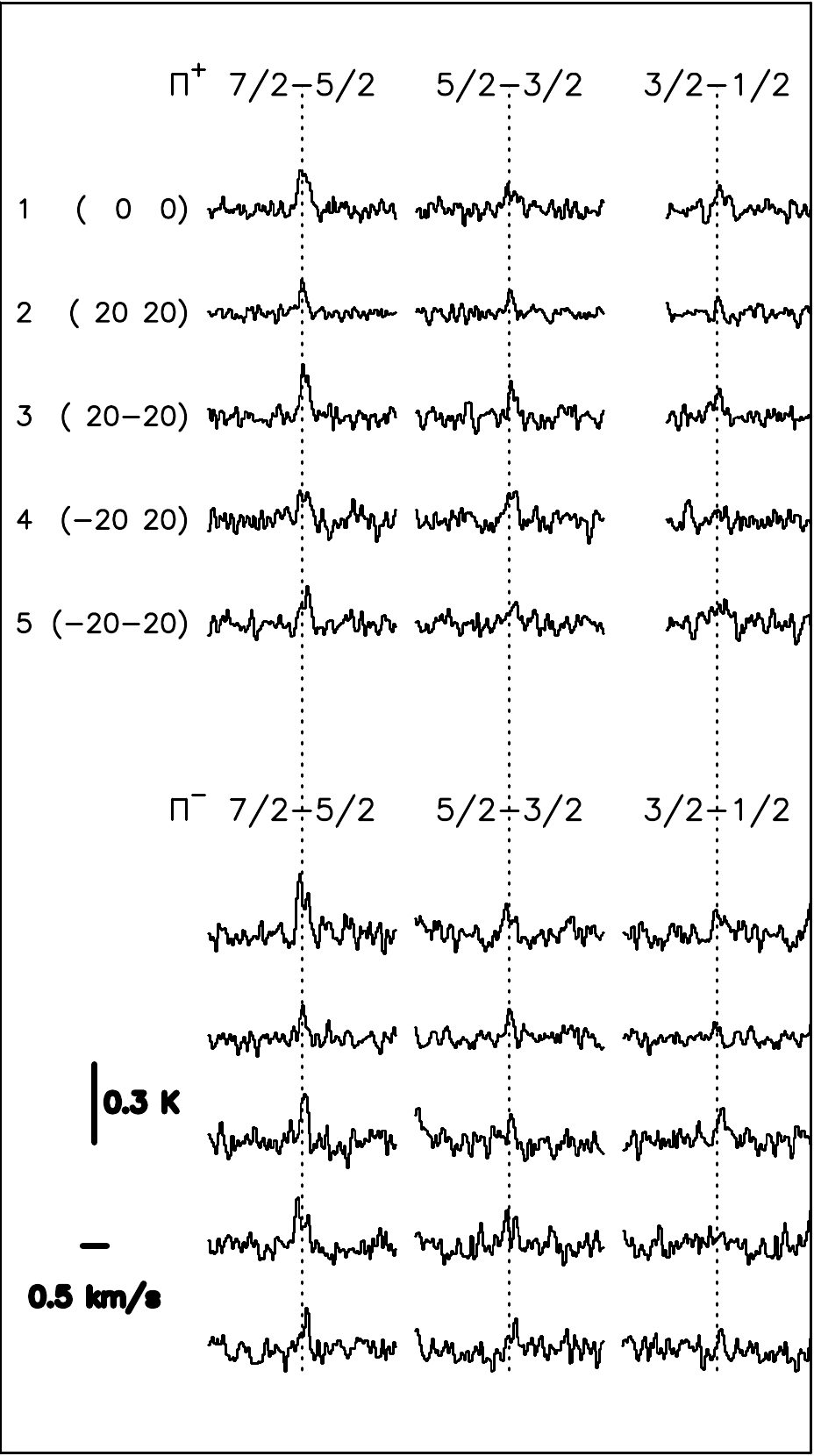}
  \caption{NS($J=5/2-3/2$) spectra, in the \tant\ temperature scale (in K), toward L1521E, L1521B, L1498, and L1544 (from left to right). In each panel, the hf set of lines (identified by their quantum number $F$) is split into the \pip\ (top) and \pim\ (bottom) (see \rtab{spectro}). For the sake of clarity, hf transitions are shown individually. The position number (see \rfig{maps}) and the offsets (in arcsec), with respect to the reference position given in Table~\ref{tab:obs}, are indicated on the left of each spectrum. The vertical dotted lines indicate the adopted systemic velocity. The horizontal and vertical scales are indicated in bold face.}
  \label{fig:ns115}
\end{figure*}

\section{Line properties}
\label{sec:lines}

\new The hyperfine (hf) structure of the two sets of lines corresponding to each parity of the $^2\Pi_{1/2}$ manifold of NS is detected in all sources, and toward most positions (see Figs.~\ref{fig:maps} and \ref{fig:ns115}). The properties of each hyperfine component obtained through Gaussian fitting are summarized in Tables~\ref{tab:l1521e_gfit} to \ref{tab:l1544_gfit}. At a few positions toward L1521E and L1498 (see Fig.~\ref{fig:gfit_l1498}), transitions within the $J=7/2-5/2$ manifold at 161~GHz were observed and detected. Toward L1521B, the observed positions, based on the reference position of \cite{hirota2004}, are shifted by more than 80" from the dust emission peak.

The line intensity varies among the sources, although by small factors, while the variation within a given source are up to a factor two. There is no apparent systematic dependence of the intensity with the dust emission, but in L1498, the line intensity increases when moving away from the central region.

The line properties have been obtained from Gaussian fits, the results of which are summarized in Tables~\ref{tab:l1521e_gfit}--\ref{tab:l1544_gfit}. The line width of the strongest hyperfine transition within $J=5/2-3/2$ toward L1521E and L1521B are typically 0.2-0.3\kms. Toward L1498 and L1544, the lines are narrower, with a full width $\approx 0.15-0.25$\kms, which would correspond to a gas temperature above 20~K, well above the values in these sources (see references in Table~\ref{tab:obs}). Thus, the 115~GHz lines are emitted in regions where non-thermal (turbulent) broadening is actually dominant.

It is instructive to compare the relative line intensities, normalized to that of the brightest component, to the theoretical line ratios expected for thermalized level populations in the optically thin case (see Fig.~\ref{fig:gfit_tpeak}). It is found that the intensity of the weakest line is compatible, in many cases, with its theoretical relative strength of 0.375. The same is true for the second brightest hyperfine line (with a relative intensity 0.63) although some departures are observed. Overall, this suggests that the NS lines are not optically thick and that the level populations are close to be thermalized at a single excitation temperature.

\section{Methodology}

\subsection{Column density of \ce{N2H+}}
\label{sec:n2hp}

When the total opacity of a \ce{N2H+} transition is high, measuring the column density must take into account deviations from the single excitation temperature assumption because of hyperfine trapping effects \citep{tafalla2004a, daniel2007}. To do so requires both a source structure (density, temperature, velocity, line broadening) and an abundance profile \citep[e.g.,][]{magalhaes2018a}.

For L1498, we have used the value of \cite{daniel2007}, $N(\ce{N2H+})=7.3(3)\tdix{12}\cc$, obtained by taking into account above effects. Indeed, this column density is a factor 1.5 larger than the one derived by \cite{tafalla2004a} who neglected hyperfine trapping. The use of updated, and more accurate, collision rate coefficients also explain the different column density. The \nnhp\ column density obtained by \cite{daniel2007} corresponds to offsets (+10",0") relative to the dust peak, or (+20",+20") in our reference frame. Nevertheless, from the constant \nnhp\ abundance derived by \cite{tafalla2004a} and the\hh\ column density profile derived from the density profile from \cite{magalhaes2018a}, we infer the variation of \cd{N2H+} over 20" to be 10\%.

In L1521E, we derive the 26"-beam averaged column density of \nnhp\ toward offsets (-10",-10") using the constant abundance, \nnhp:\hh$=2\tdix{-11}$ (or \ab{N2H+}=\dix{-11}) from \cite{tafalla2004b}, in combination with our \hh\ column density, $\Nhh=2.4\tdix{22}\cc$ (see Appendix~\ref{app:NH2}), and thus $\cd{N2H+}=4.8\tdix{11}\cc$. The abundance in L1521E being 8.5 times lower than in L1498 using the same method as in \cite{tafalla2004a}, the lines are probably not optically thick and we thus did not apply the correction factor of 1.5 {found by \cite{daniel2007} in L1498}. Our estimate of \cd{N2H+} is $\approx 50$ times lower than the peak column density estimated by \cite{nagy2019} from their local thermal equilibrium (LTE) study. However, for the sake of consistency with the \hh\ column density and because the analysis from Tafalla et al. uses a 1D non-LTE radiative transfer, we use their results.
For L1521B, we use the 1$\sigma$ upper limit obtained by \cite{hirota2004} of 1.9\tdix{12}\cc.

Finally, toward L1544, the 26"-beam-averaged column density is $N(\ce{N2H+})=4.1\tdix{13}\cc$ \citep{bizzocchi2013}. In this source, we also estimated the \ce{N2H+} column density at the offsets where NS was observed, by assuming a constant\nnhp:\hh\ (abundance) of 6.5\tdix{-10}, as in \cite{bizzocchi2013}, combined with their source model (from their Fig. 2), which corresponds very closely to a \hh\ density profile of the form $n_{\hh}(r) = n_0 / [1+(r/r_0)^2]$, with $n_0=2\tdix{7}\ccc$ and $r_0=316$~au. The \hh\ column density corresponding to each offset can be computed analytically, from which the column density \ce{N2H+} follows.

\subsection{Column density of NS}
\label{sec:ns}

Two complementary, methods were applied to derive the total column density of NS. One is the usual HFS method in the CLASS software \citep[e.g.,][]{hilyblant2013a}, which assumes a single excitation temperature within a hyperfine manifold. The second uses the escape probability approximation to relax the single excitation temperature assumption. However, as shall be discussed below, degenerate sets of solutions are obtained in some cases and the results from the HFS method are biased toward the high optical depth solutions. Therefore, only the results from the non-LTE approach are discussed and used in what follows.

Non-equilibrium level populations were computed with the public code RADEX \citep{vandertak2007} using new, dedicated NS-\hh\ collision rate coefficients (see Appendix \ref{app:coll}) and spectroscopic properties from the CDMS. In this procedure, the FWHM $\Delta v$ of the lines, obtained from Gaussian fitting and taken to be equal for all hf lines, and the background temperature are held fixed, along with the \hh\ density, the NS column density, $N$\footnote{Actually, this procedure constrains $N/\Delta v$.}, and the kinetic temperature are the free parameters.
        
The three-dimensional parameter space exploration may be performed with a forward-fitting approach, through a regular grid at each point of which the reduced $\chi^2$ is determined. Alternative methods include MCMC constrained by a prior probability on the meaningful parameter space. In this approach, we  obtain the probability of the parameters' given measured values. Both approaches lead to complementary views, but the MCMC is more optimized in that it naturally refines the parameter space exploration close to the various local minimum locations, while the forward-fitting requires progressively refined regular grids. In this study, we used the publicly-available implementation of MCMC of \cite{emcee}.
        
It is widely known that line inversion is often degenerate in cold cores especially when only one rotational transition is available. In the present study, most sources were observed only through the $J=5/2-3/2$ rotational manifold and a degenerate set of solutions, from high-\nhh/low-$N$ to low-\nhh/high-$N$ was effectively found for several lines of sight (see Fig.~\ref{fig:l1521b_cross}). In contrast, when detected, the $J=7/2-5/2$ usually breaks this degeneracy. Furthermore, the density is known to be larger than typically \dix{4}\ccc\ in the sources of our sample, which also helps in alleviating the degeneracy. Eventually, the NS column density could be constrained in each source.

\section{Results}
\label{sec:results}
{The MCMC minimization procedure is described in Appendix~\ref{app:minimize} and the results} are shown in Figs.~\ref{fig:l1521e_cross}-\ref{fig:l1544_cross}. The resulting column densities and ratios are given in Table~\ref{tab:results}.

\begin{table*}
        \centering
        \caption{Column density of \hh NS and \nnhp\ in our source sample and their corresponding abundances and NS:\nnhp\ ratios.}
        \label{tab:results}
        \begin{tabular}{lcrr cccccc}
                \toprule
                Source &\# & \mc{2}{Offsets}
                & $\Nhh^\dagger$
                & $\cd{NS}^\dagger$
                & $\cd{N2H+}^\dagger$
                & \cd{NS}/\cd{N2H+}
                & \cd{NS}/\Nh
                & [\nnhp]$^\ddagger$\\
%               & \stot$^\S$\\

                &&\arcsec&\arcsec&\cc&\cc &\cc& &\tdix{-11} &\tdix{-11}\\
                \midrule
                L1521E
                &1&-30 & -10 & 22.37(8) & $>$12.8     &          &                     & $>13.5$\\
                &2&+10 & -10 & 22.23(9) & $>$12.9   &          &                     & $>23.4$\\
                &3&-10 & -10 & 22.33(7) & 12.8(2)$^e$ & 11.67(7) & $14.8^{+9.3}_{-5.7}$& $14.8^{+9.3}_{-5.7}$ &1.0$^a$ \\
                &4&-10 & -30 & 22.33(8) & $>$12.6     &          &                     & $>9.33$\\
                \\
                L1521B
                &1& 50 & -20 & 22.24(6) & 12.25(10) &             &         &$5.1^{+1.6}_{-1.2}$\\
                &2& 30 & -20 & 22.23(7) & 12.35(10) &             &         &$6.6^{+2.1}_{-1.6}$\\
                &3& 10 & -20 & 22.16(7) & 12.15(10) &             &         &$4.9^{+1.6}_{-1.2}$\\
                &4&  0 &   0 & 22.10(7) & 12.3(1)--13.2(1)       &$<12.3^b$&$>1$ &$7.9^{+2.6}_{-2.0}$ & $< 7.9^b$\\
                &5& 20 &   0 & 22.20(7) & 12.3(1)      &             &         &$6.3^{+2.1}_{-1.6}$ \\
                &6& 40 &   0 & 22.24(7) & 12.3(1)      &             &         &$5.7^{+1.9}_{-1.4}$ \\
                &7& 60 &   0 & 22.23(6) & 12.1(1)      &             &         &$3.7^{+1.9}_{-1.4}$ \\
                \\
                L1498
                &1&  0 &   0 & 22.26(17)&12.3(2)$^e$ &          &         &$5.48^{  +4.55}_{  -2.49}$\\
                &2&-10 & -20 & 22.20(20)&12.2(1)$^e$ &          &         &$5.00^{  +3.37}_{  -2.01}$\\
                &3&+20 & -10 & 22.26(17)&12.15(20)$^{e,f}$ &          &         &$3.88^{+3.22}_{ -1.76}$\\
                &4&+10 & +20 & 22.25(16)&12.7(2)     & 12.86(2)$^c$ & $0.7^{+0.4}_{-0.3}$  &$14.1^{+11.32}_{-6.28}$ & 20.0$^c$ \\
                &5&-20 & +10 & 22.17(20)&12.0(1)     &          &         &$3.38^{  +2.28}_{  -1.36}$\\
                &6&+20 & -40 & 22.28(17)&12.1(1)     &          &         &$3.30^{  +1.90}_{  -1.21}$\\
                &7&-40 & +20 & 22.14(20)&12.0(1)     &          &         &$3.62^{  +2.44}_{  -1.46}$\\
%               &+40 & -20 & 22.29(17)&13.4(5)$^e$ &          &         &$64.4^{+152.9}_{ -45.3}$\\
                &9&+60 & -30 & 22.24(18)&12.6(2)$^{e,f}$ &          &         &$11.5^{  +9.83}_{  -5.29}$\\
                &10&+80 & -40 & 22.14(25)&12.4(1)$^e$ &          &         &$9.10^{  +7.81}_{  -4.20}$\\
   %   0.0     0.0 &$5.48^{  +4.55}_{  -2.49}$
   % -10.0   -20.0 &$5.00^{  +3.37}_{  -2.01}$
   %  20.0   -10.0 &$54.8^{+130.14}_{ -38.57}$
   %  10.0    20.0 &$14.1^{ +11.32}_{  -6.28}$
   % -20.0    10.0 &$3.38^{  +2.28}_{  -1.36}$
   %  20.0   -40.0 &$3.30^{  +1.90}_{  -1.21}$
   % -40.0    20.0 &$3.62^{  +2.44}_{  -1.46}$
   %  40.0   -20.0 &$64.4^{+152.90}_{ -45.32}$
   %  60.0   -30.0 &$11.5^{  +9.83}_{  -5.29}$
   %  80.0   -40.0 &$9.10^{  +7.81}_{  -4.20}$
                \\
                L1544
                &1&  0 &  0  & 22.82(1)$^d$ &12.0(1) & 13.6(1) & 0.023(6) & $0.76^{+0.20}_{-0.16}$ & 33$^d$ \\
                &2& 20 & 20  & 22.64(2)$^d$ &11.9(1) & 13.5(1) & 0.028(6) & $0.91^{+0.24}_{-0.19}$ & 33$^d$ \\
                &3& 20 &-20  & 22.68(1)$^d$ &12.1(1) & 13.5(1) & 0.040(9) & $1.32^{+0.34}_{-0.27}$ & 33$^d$ \\
                \bottomrule
        \end{tabular}
        \tabnotes\ \new Numbers in brackets are 1$\sigma$ uncertainties in units of the last digit. Column (2) gives the spectrum number as shown in Figs.~\ref{fig:maps} and \ref{fig:ns115}. $^\dagger$ Column density (log10) measured in a 35", 21" and 26"-beams for \hh, NS and \nnhp\ respectively (see Appendix~\ref{app:NH2} and Table~\ref{tab:sed}). $^\ddagger$ Abundance relative to H nuclei in units of \dix{-11}. $^a$ Constant \nnhp\ abundance taken from \cite{tafalla2004b}. $^b$ Based on the upper limit on \cd{N2H+} taken from \cite{hirota2004}. $^c$ \nnhp\ abundance and column density from \cite{daniel2007} (see Section~\ref{sec:n2hp}). $^d$ Based on the constant \nnhp/\hh\ abundance of 6.5\tdix{-10} from \cite{bizzocchi2013}. $^e$ Column density of NS determined using both $J=5/2-3/2$ and $J=7/2-5/2$ manifolds. $^f$ Temperature held fixed at 10~K.
\end{table*}

\subsection{L1521E}
\label{sec:l1521e}

The physical properties and column density of NS were constrained toward four positions for which the signal-to-noise ratio (S/N) was sufficient. The results of the minimization process are shown in Fig.~\ref{fig:l1521e_cross}. At offsets (-10",-10"), the $J=7/2-5/2$ line was observed enabling the \hh\ density and NS column density to be determined. The most likely density is $\nhh=4.75^{+0.35}_{-0.40}$ dex\footnote{We use the notation dex for decimal exponent such that a value $10^x$ is equivalent to $x$ dex.} (or $5.6_{-3.4}^{+7.0}\times 10^4\ccc$) to which corresponds\footnote{Uncertainties, at the 1$\sigma$ level, are indicated in brackets, in units of the last digit. The minimization was done on the log10 of the density and column density.} $\cd{NS}=12.8(2)$dex (or \err{6.3}{2.3}{3.7}\tdix{12}\cc).

Toward the other offsets, only the $J=5/2-3/2$ manifold was observed and an ensemble of density and column density are found that can reproduce the line properties, while the kinetic temperature is unconstrained. Nevertheless, the density at these positions is most likely lower than at the peak position at (-10",-10"). At offsets (-30",-10"), although a continuum of solutions cannot be excluded, the priors distributions for the density and column density favor two solutions, at low ($\approx \dix{4}\ccc$) and high ($>$ few \dix{5}\ccc) density. Thus, focusing on the low-density one gives a column density of 13.3(4) dex. A conservative approach consists in adopting 4.75 dex as a lower limit on the \hh\ density, leading to lower limits on the NS column density of 12.8, 12.9, and 12.6 dex, toward offsets (-30",-10"), (+10",-10"), and (-10",-30") respectively. Hence, a trend may be apparent of an outward increase of the NS column density.

\subsection{L1521B}
\label{sec:l1521b}

In the case of L1521B, where only $J=5/2-3/2$ was observed, a continuous set of density and column density is found at all positions (see Fig.~\ref{fig:l1521b_cross}). Nevertheless, at offsets (60",0"), (40",0"), and (50",-20"), which are also the closest to the dust emission peak (see Fig.~\ref{fig:maps}), solutions with \hh\ density above \dix{5}\ccc\ are favored, in agreement with the previous study of \cite{tafalla2004a} who found $\nhh=1.9\tdix{5}$\ccc. In this density regime, the \cd{NS} column density is well constrained. The results are summarized, at each position, in Table~\ref{tab:results}, showing that the column density varies little among the various positions, with \cd{NS}=12.3(2) dex.

Toward the other positions, high \hh\ density (above $\sim\,\dix{5}\ccc$) solutions are also more likely with associated values of \cd{NS} again very close to 12.3 dex. However, it is not possible to firmly exclude lower densities ($\nhh\sim\dix{4}\ccc$), for which \cd{NS}$\approx 13.2$ dex.

\subsection{L1498}
\label{sec:l1498}

In contrast to L1521B, the \hh\ density and NS column density in L1498 are well constrained toward several positions in L1498, in particular, at the four locations where the $J=7/2-5/2$ $\Pi^-$ and $\Pi^+$ manifolds are well detected (see Fig.~\ref{fig:gfit_l1498}). Toward the dust peak position at offsets $(+10",+20")$, we obtain \cd{NS}=12.7(2) dex, with a weak dependence on the temperature. We find significant spatial variations of the NS column density, from 13.3(5) at $(+20",-10")$ and $(+40",-20")$ to 12.0(1) at $(-20",+10")$ and $(-40",+20")$. However, no systematic radial variation is apparent. At other locations, the solutions are degenerate.

\subsection{L1544}
\label{sec:l1544}

L1544 is the densest of the four cores considered in this study, with a central density of $\nhh= 2\tdix{7}$\ccc\ \citep{bizzocchi2013}. The NS lines from L1544 are weaker than in the other three sources and although NS was detected at the five observed positions, the derivation of the column density was performed toward only three, $(0",0")$, $(20",20")$, and $(20",-20")$. Toward $(0",0")$, a two-component Gaussian fit was performed and the minimization was done using the strongest of the two. Toward the other two positions, the line intensity was obtained with a single-component Gaussian fit. The results, shown in Fig.~\ref{fig:l1544_cross}, confirm that the density is higher than \dix{6}\ccc, and most likely closer to 3\tdix{7}\ccc. The corresponding column density are low, $\cd{NS}\approx12.0(1)$ dex.

\subsection{Summary}

The main observational result is that the NS:\nnhp\ column density ratio varies by more than two orders of magnitude, from $\approx$15 in L1521E to 0.02 in L1544. This variation is due to a combination of a decrease of the NS abundance and an increase of that of \nnhp. {Although typical uncertainties of 50-100\% may apply on the \hh\ column density (see Appendix C.3), not only these are much smaller than the source-to-source variations reported here, but the consistent method used in determining the \hh\ column density mitigates the impact of such uncertainties when comparing the NS:\nnhp\ ratio in our sample}. There are trends (at 1$\sigma$ level) of spatial variations of the NS:\nnhp\ ratio within L1544, and of the abundance of NS in L1521B and L1498. The abundance of NS vary between $\approx \dix{-11}$ and 1.5\tdix{-10}. These values are typically a factor of 2 or more lower than to those obtained in TMC-1 and L134N \citep{mcgonagle1994}. This may be due to the LTE assumption which, as already noted, is biased toward larger column densities.

\section{Abundance of gas-phase sulfur}
\label{sec:models}

\begin{table}[t]
        \caption{Initial gas-phase conditions in our grids of chemical models.}
        \label{tab:init}
        \centering
        \begin{tabular}{cc}
                \toprule
                Element & Abundance (relative to H nuclei) \\
                \midrule
                H               & 1.0 \\
                He              & 0.1 \\
                N               & 6.4(-5) \\
                \ce{C+} & 8.3(-5) \\
                \ce{Fe+}& 1.5(-9) \\
                O               & 6.92(-5), 8.30(-5), 1.04(-4), 1.38(-4), 2.08(-4)$^\dagger$ \\
                \ce{S+} & 0.140(-6), 0.443(-6), 1.40(-6), 4.43(-6), 14.0(-6) \\
                \bottomrule
        \end{tabular}
        \tabnote Numbers are written in the form $a(b) = a\tdix{b}$. $^\dagger$ The corresponding values of C/O are 1.2, 1.0, 0.8, 0.6, 0.4.
\end{table}

\def\wa{0.33\hsize}
\begin{figure*}[t]
        \centering
        \includegraphics[width=\wa]{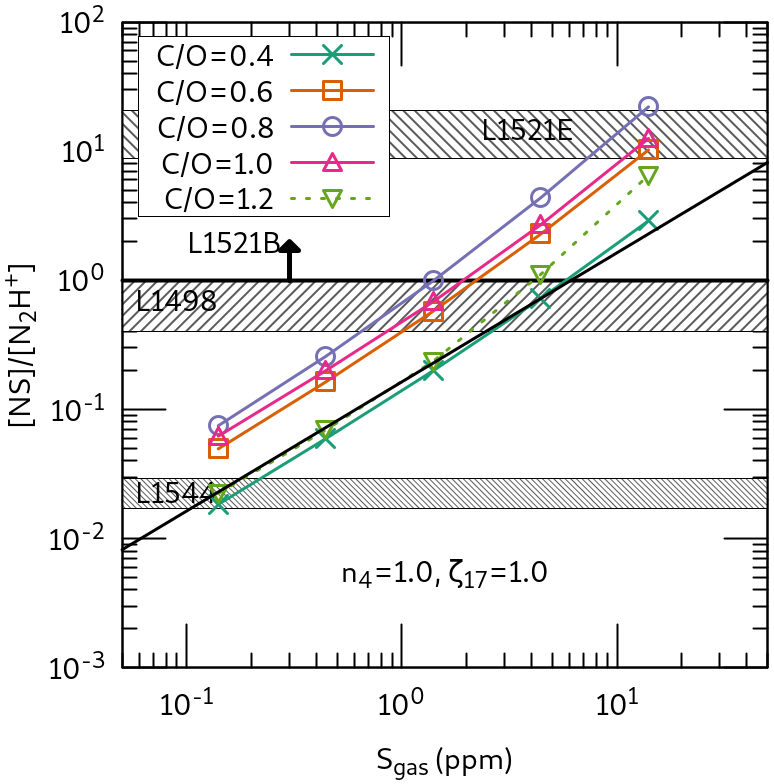}\hfill%
        \includegraphics[width=\wa]{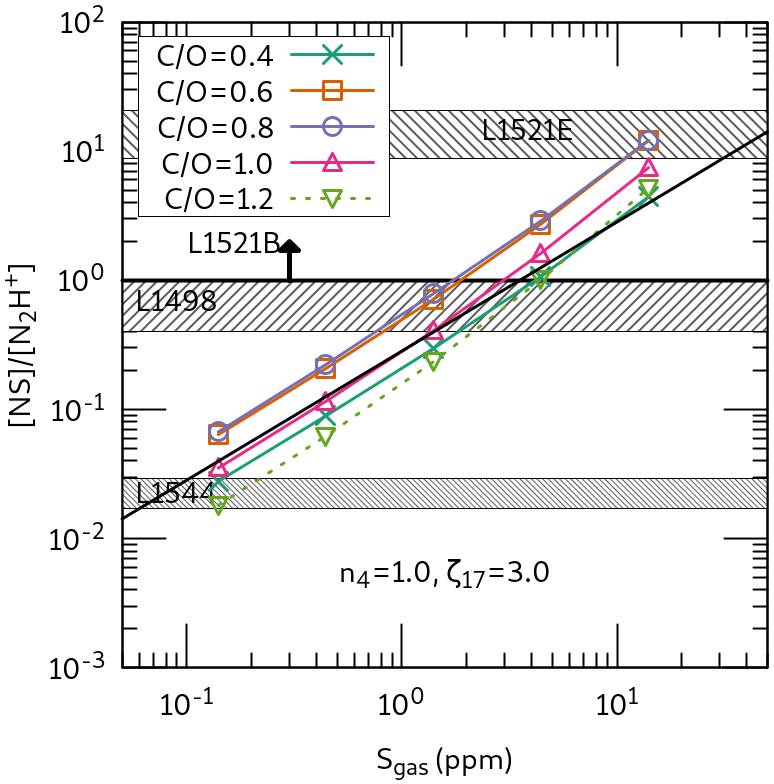}\hfill%
        \includegraphics[width=\wa]{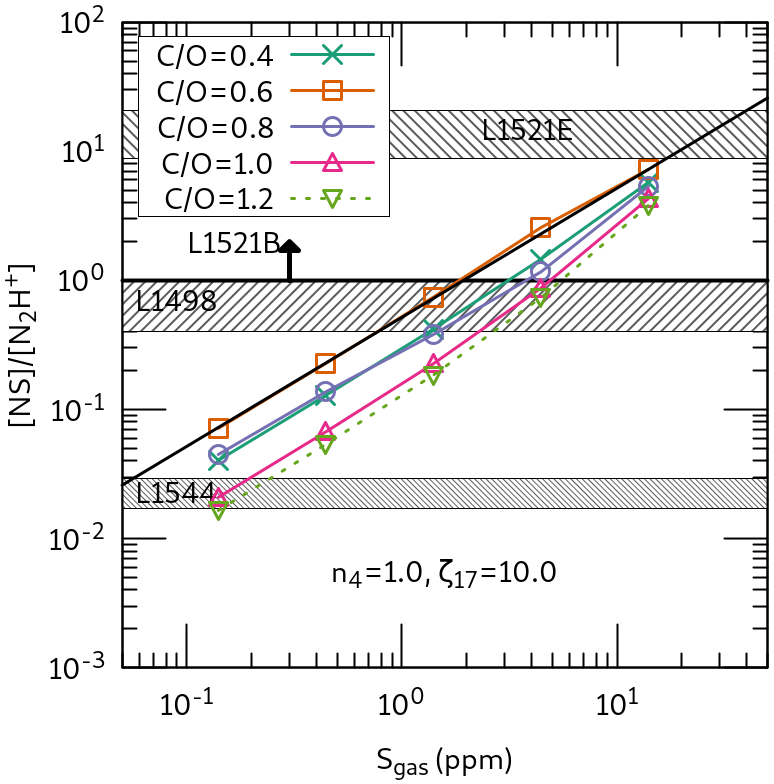}\smallskip\\
        \includegraphics[width=\wa]{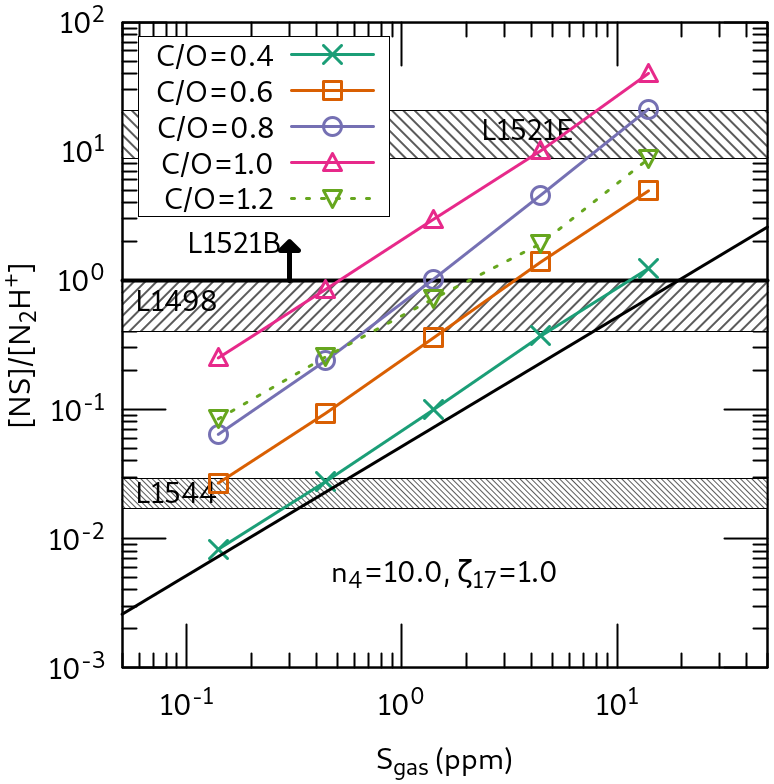}\hfill%
        \includegraphics[width=\wa]{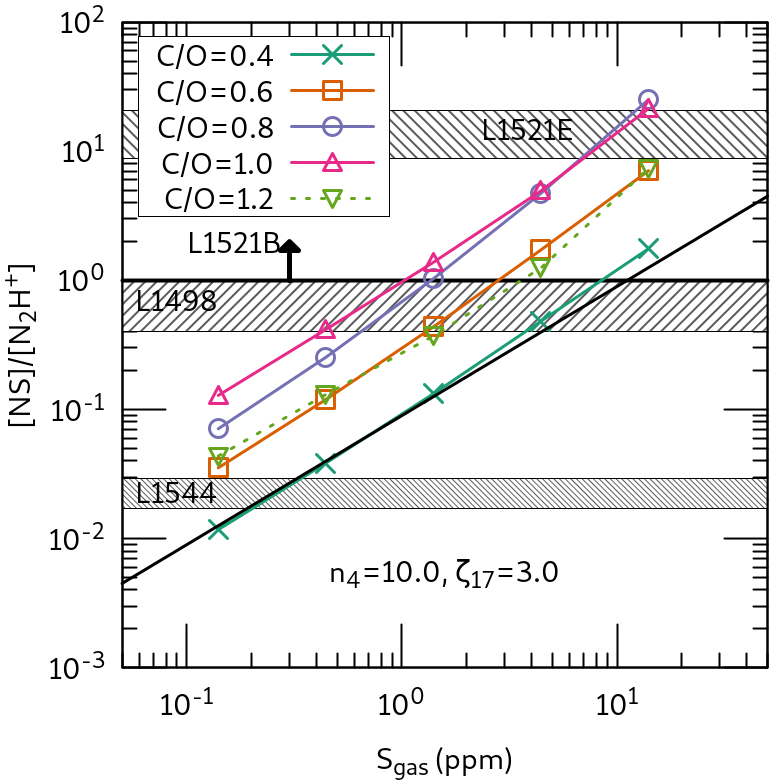}\hfill%
        \includegraphics[width=\wa]{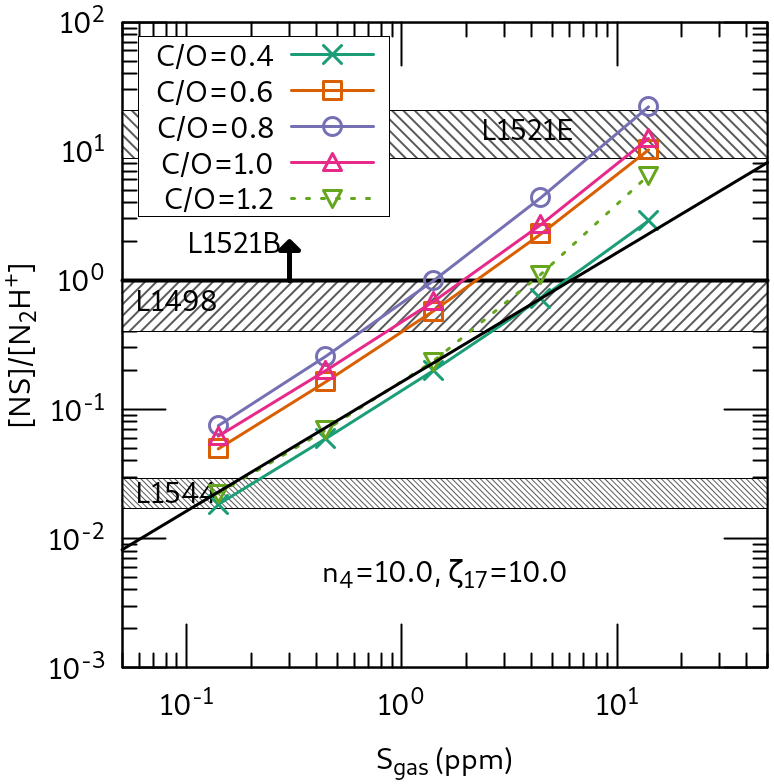}
        \caption{\new  \ab{NS}:\ab{N2H+} ratio as a function of \stot\ at steady-state from chemical models with with $n_4=1$ (top) and 10 (bottom) and $\zeta_{17}=1$, 3, and 10 (left to right). In each panel, five values of the C/O ratio (0.4 to 1.2) are shown. The observed NS:\ce{N2H+} column density ratios (with $\pm1\sigma$ uncertainties) toward L1521E, L1498, and L1544 (central position), are shown as hashed areas, and the lower limit in L1521B is indicated with an arrow. The black, straight, line shows our analytical prediction given in \req{s}.}
        \label{fig:models}
\end{figure*}

\subsection{Chemical models}

We now turn to chemical model calculations to obtain constraints on the abundance of gas-phase sulfur, \stot-- namely, the total abundance of gas-phase sulfur-bearing species --- from the measured NS:\nnhp\ column density ratio. We consider here only steady-state calculations, allowing us to explore a parameter space in which the C/O ratio, {the density}, the cosmic-ray ionization of molecular hydrogen $\zeta_{\hh}$, and the volatile sulfur abundance \stot, are varied. The abundance of gas-phase oxygen \otot\ is varied while that of carbon, \ctot, is kept constant and equal to 83~ppm, in such a manner that the C/O ratio (equal to \ctot/\otot) covers the range from 0.4 to 1.2 by steps of 0.2. Three representative cosmic-ray ionization rates are considered to account for the lack of accurate determinations in the dense gas: $\zeta_{17} = \zeta_{\hh}/(\dix{-17}\pers)=1$, 3, and 10. Finally, the value of \stot\ covers two orders of magnitude, namely 0.14, 0.443, 1.4, 4.43, and 14~ppm. In what follows (except otherwise stated), the gas:grain processes (depletion, hydrogen saturation, and desorption) were not included, although varying the value of \otot\ may be seen as a simple mean of simulating such processes \citep{legal2014}. The UGAN network \citep{hilyblant2018a} has been updated and upgraded, using the KIDA database \citep{wakelam2015} for those reactions involved in the chemistry of NS. The updated and new reactions are listed in \rtab{chemistry}.

In the present calculations, the pre-stellar core is modeled as an {isothermal gaseous sphere with} purely atomic {initial} conditions ({see Table~\ref{tab:init}}) and the abundances are computed until a steady-state is reached. As detailed above, a grid of models has three free parameters (C/O, $\zeta_{17}$, \stot), {while the gas temperature and density are held fixed with \tkin=10~K, and $n_4=\nh/\dix{4}\ccc$={1, 10, and 100.}

\subsection{Results}

\begin{table}
        \centering
        \caption{\label{tab:stot}\new Values of \stot\ (in ppm) in each source based on Fig.~\ref{fig:models}.}
        \begin{tabular}{l ccccc}
                \toprule
                C/O     & 0.4 & 0.6 & 0.8 & 1.0 & 1.2 \\
                \midrule
                \mc{6}{$\zeta_{17}=3$, $n_4=1$}\smallskip\\
                L1521E & $\gg 14$ & 14 & 14 & $\ge14$ & $>14$ \\
                L1521B & $>4.4$ & $>1.4$ & $>1.4$ & $>3$ & $>4.4$ \\
                L1498  & 3 & 1.4 & 1.4 & 2 & 3 \\
                L1544-1$^\dagger$  & 0.14 & $<0.14$ & $<0.14$ & $<0.14$ & 0.14 \\
                L1544-3$^\dagger$ & 1 & 0.3 & 0.3 & 0.3 & 1\\
                \midrule
                \mc{6}{$\zeta_{17}=3$, $n_4=10$}\smallskip\\
                L1521E & $\gg14$ & $\approx14$ & $\approx14$ & $\approx14$ & $\approx14$ \\
                L1521B & $>8$ & $>3$ & $>1.4$ & $>1$ & $>3$ \\
                L1498  & 6 & 2 & 1 & 0.8 & 2 \\
                L1544-1  & 0.3 & $<0.14$ & $<0.14$ & $<0.14$ & $<0.14$ \\
                L1544-3 & 0.6 & 0.14 & $<0.14$ & $<0.14$ & 0.14 \\
                \midrule
                \mc{6}{$\zeta_{17}=3$, $n_4=100$}\smallskip\\
                L1544-1  & 0.6 & 0.2 & $<0.14$ & $<0.14$ & $<0.14$ \\
                \bottomrule
        \end{tabular}
        \tabnotes  Predictions from models with $n_4=1$ and 10 and $\zeta_{17}=3$ are compared to all sources, while models with $n_4=100$ are compared to L1544 only. The quoted values are roughly estimated adopting a linear (C/O-dependent) correlation between NS:\ce{N2H+} and \stot\ from the models. $^\dagger$ L1544-1 and L1544-3 correspond to positions labeled 1 and 3 in Table \ref{tab:results}.
\end{table}

Figure \ref{fig:models} shows the NS:\ce{N2H+} abundance ratio {as a function of \stot} {for two grids of models, with $n_4=1$ (top) and $n_4=10$ (bottom).} A close-to linear {power-law} is evident, which is found to depend only weakly on C/O and $\zeta_{17}$, {while the intercept does depend on these environmental parameters. The origin of the correlation can be understood based on simple chemical considerations (see Section~\ref{sec:chemistry}).}

The primary result of this study is that the tight NS:\ce{N2H+}--\stot\ correlation shows that the sources in our sample have distinct gas-phase sulfur abundances: L1521E has the highest value of \stot, L1544 the lowest, L1498 is intermediate, and the lower limit in L1521B indicates a high value of \stot.

To explore the results in more details, let us first consider the model with $n_4=1$ and $\zeta_{17}=3$ (top and middle panels). For each value of \stot, the predicted NS:\ce{N2H+} ratio depends only weakly on C/O and $\zeta_{17}$, and  only marginally modifies the trend of a decreasing \stot\ from L1521E to L1544. Rough estimates of \stot\ are summarized in Table~\ref{tab:stot}.

Toward L1521E, our observations and models indicate that volatile sulfur abundance is compatible with the cosmic abundance of 14~ppm, provided that C/O = 0.6-0.8, while C/O = 0.4 would require an unrealistic value of \stot\ ($>$ 30 ppm). Toward L1521B, there is a lower limit on \stot, $>1.4$ ppm for C/O = 0.6-0.8, and $>3$ ppm for lower or higher C/O ratios. In L1498, the observed NS:\ce{N2H+} ratio constrains \stot\ to $1.4-3$ ppm over the entire range of C/O. L1544 is the most sulfur-depleted source in this model, with $\stot<0.14$ ppm {toward the innermost position} over the entire $n_4=1$ grid, two orders of magnitude lower than in L1521E.

While models with $n_4=1$ are representative of the outer parts of the cores, examining models with a higher density ($n_4=10$ and 100) enable a study of regions closer to the inner plateau. Compared to models with $n_4=1$, the predicted NS:\nnhp\ ratios for $n_4=10$ are more scattered across the C/O range. Nevertheless, the above trends remain valid. Thus, in L1521E, C/O = 0.4 requires a very large \stot\ and is thus excluded by these models, while C/O = 0.6 to 1.2 are all consistent with \stot =14 ppm. In L1521B, the lower limit on the NS:\nnhp\ ratio gives $\stot>8$ ppm for C/O = 0.4, while for C/O $= 0.6-0.8$, a lower limit of 1.4 ppm on \stot\ is found. In L1498, a range of $\stot = 0.8-6$ ppm is derived for C/O = 0.4 to 1.2, while $\stot = 1-2$ ppm for C/O between 0.6 and 0.8. In L1544, low values of \stot, as in models with $n_4=1$, are derived with $\stot<0.6$ ppm. L1544 is thus the most sulfur-depleted core, followed by L1498, while L1521E and L1521B are both compatible with a non-depleted sulfur gas phase.

The spatial distribution of NS and \ce{N2H+} in L1544 is also of interest. As can be seen from Table~\ref{tab:results}, the abundance of NS and the NS:\ce{N2H+} ratio increase outward by a factor of two over 30\arcsec\ or 4200~au. We note that at these offsets, where the density in L1544 is \nhh = \dix{5}\ccc, similar to that in L1498, \stot\ becomes $\approx0.14$~ppm for C/O = 0.6 (and $n_4=10$), still significantly lower than in L1498 (2 ppm). The spatial variation of the NS:\ce{N2H+} abundance ratio in L1544 provides clear indication that depletion of sulfur increases with density. In L1521B, the abundance of NS also decreases when moving toward the peak position of the core, but the lack of \ce{N2H+} information does not allow us to constrain the NS:\ce{N2H+} ratio and thereby the spatial variation of \stot\ in this source.

\section{Chemical considerations}
\label{sec:chemistry}

\subsection{Chemistry of NS}

The reason why the NS:\ce{N2H+} abundance ratio provides a direct constraint on \stot\ may be understood by analyzing the chemistry of NS under typical dark cloud physical conditions. The main chemical routes involved in the formation and destruction of NS, along with its chemical link with atomic sulfur, are summarized in \rfig{network}. Nitrogen sulfide is formed primarily by the neutral-neutral reactions:
\begin{eqnarray}
  \ce{S + NH -> NS + H}\label{eq:form_S},\\
  \ce{N + SH -> NS + H}\label{eq:form_N},
\end{eqnarray}
both with temperature-independent rates of \dix{-10}\cccs. The dissociative recombination of \ce{HCNS+} also contributes to the formation of NS, but to a lesser extent. In parallel, NS is destroyed via reactions with atomic oxygen and carbon:
\begin{eqnarray}
\ce{NS + O -> NO + S}, &&\quad k=3\tdix{-11,}\cccs\label{eq:destO}\\
\ce{NS + C -> CN +S}, &&\quad k=2\tdix{-10}\cccs\label{eq:destC}.
\end{eqnarray}
We note that rates for reactions (1)-(4) have been estimated and evaluated by the KIDA committee, but measurements at a low temperature are required.

The relative weight of reactions (1) and (2) depends primarily on \stot: reaction (1) is always the main formation route (more than 50\%) while the importance of reaction (2) increases with \stot, becoming typically half that of reaction (1) for \stot=14 ppm.

As already emphasized by \cite{mcgonagle1994} and \cite{cernicharo2018}, the abundance of NS is sensitive to the carbon to oxygen gas-phase abundance ratio. Indeed, the relative importance of reactions (3) and (4) depend on the relative abundance of atomic oxygen and carbon in the gas phase: the oxygen route will dominate if \dens{O}:\dens{C} is larger than the ratio of the two rates, $>2\tdix{-10}/3\tdix{-11}=6.7$.  Reaction (3) is found to dominate destruction ($>55$\%) when C/O $\le 0.6$; at higher C/O ratios, reaction (4) becomes the main destruction route ($\approx 35\%$) but other routes, involving collisions with \ce{H+} or atomic nitrogen, N, also become non-negligible.

\begin{figure}[t]
        \begin{center}
                \includegraphics[width=\hsize]{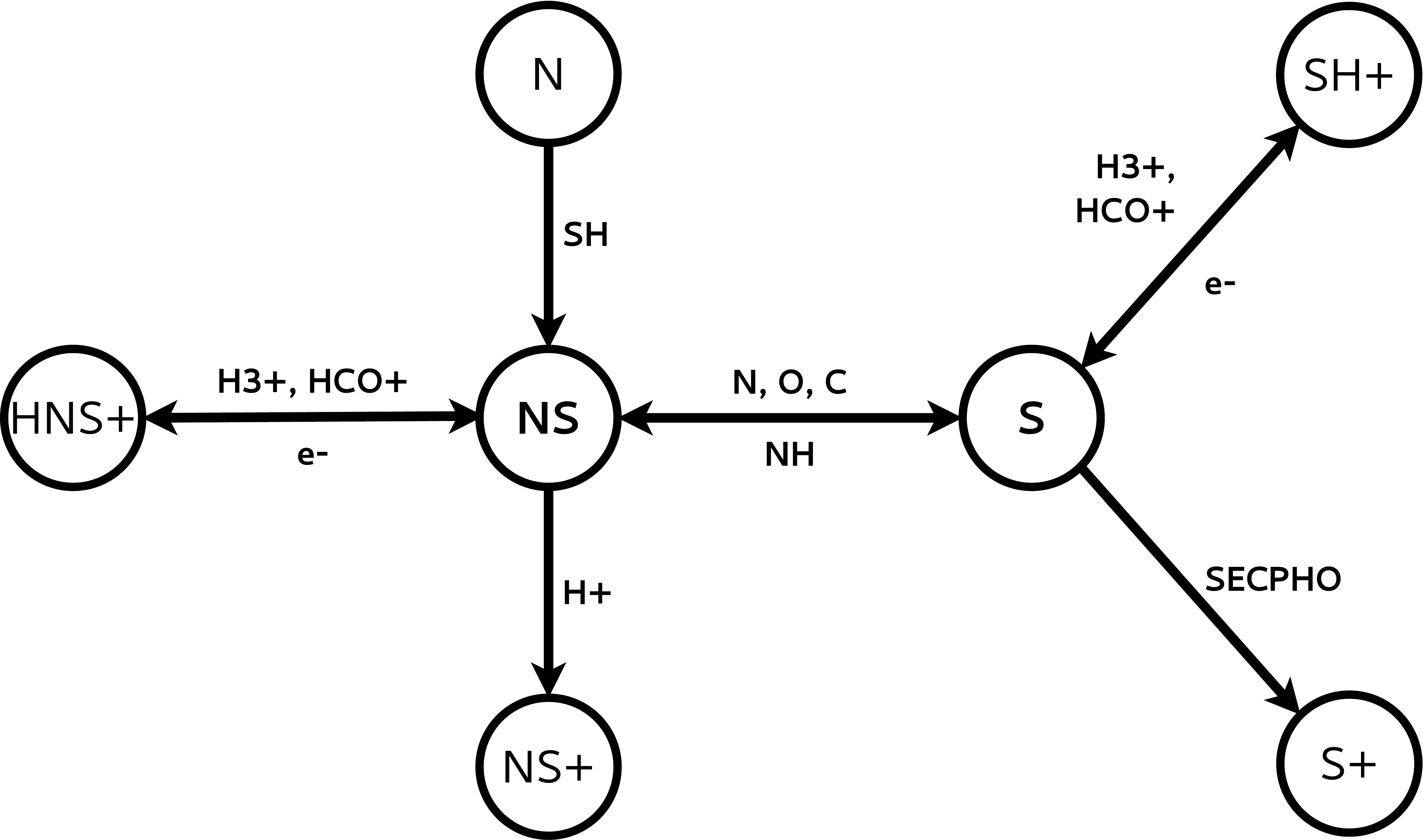}
                \caption{Main chemical routes involved in the formation and destruction of NS under dark cloud conditions and its chemical link with atomic sulfur. We note that \texttt{SECPHO} stands for secondary photons.}
                \label{fig:network}
        \end{center}
\end{figure}

\subsection{Steady-state abundance at low C/O}
From the previous remarks, it transpires that a single analytical expression predicting the abundance of NS over the entire parameter space explored in this study is highly unlikely. Nevertheless, focusing on the low C/O-regime ($\le 0.6$), a simple expression can be obtained. Values of C/O ratios lower than 1 were found in conditions typical of pre-stellar cores but C/O ratio larger than 1 cannot be ruled out \citep{pratap1997, bergin1997a, legal2014}.

From reactions (1)-(4), the steady-state abundance of NS is given by
\begin{equation}
  \ab{NS} = \dix{-10}\frac{\ab{S}\ab{NH}+\ab{N}\ab{SH}}
  {3\tdix{-11}\ab{O} + 2\tdix{-10}\ab{C}}
,\end{equation}
where we note \ab{X}=\dens{X}/\nh\ the abundance relative to hydrogen nuclei. The abundance of atomic sulfur may then be written as:
\begin{equation}
  \ab{S} = (0.3 + 2\ab{C}/\ab{O}) \frac{\ab{NS}\ab{O}}{\ab{NH}}
                        -\frac{\ab{N}\ab{SH}}{\ab{NH}}.
  \label{eq:s_full}
\end{equation}
Over most of the parameter space, reaction (\ref{eq:form_S}) is actually the primary formation route --- although its importance reduces to $\approx 60\%$ when \stot=14 ppm --- such that the right-hand side in Eq.~\ref{eq:s_full} can reasonably be reduced to its first term. Regarding the destruction, ours limiting to $\csuro\le0.6$ ensures that reaction (3) is the main route such that Eq.~\ref{eq:s_full} can be simplified into:
\begin{equation}
        \ab{S} \approx 0.3\ab{NS}\ab{O}/\ab{NH}.
        \label{eq:s_approx}
\end{equation}
The steady-state abundance of NH may be obtained considering that the formation is primarily through the dissociative recombination of \ce{N2H+} \citep{hilyblant2010nh, vigren2012} and its destruction proceeds by the neutral-neutral reaction with atomic oxygen \ce{NH + O -> NO + H}, such that:
\begin{equation}
  \ab{NH} = \frac{\kdr \ab{N2H+}\ab{e-}}{6.6\tdix{-11}\ab{O}}
,\end{equation}
where $\kdr=2.1\tdix{-8}(T/300)^{-0.74}$\cccs\ is the dissociative recombination rate toward NH. It thus follows that in the low-C/O regime:
\begin{equation}
  \ab{S} \approx \frac{0.3\times6.6\tdix{-11}}{\kdr}\,
  \frac{\ab{O}^2}{\ab{e-}}\,
  \frac{\ab{NS}}{\ab{N2H+}}
  \label{eq:sss}
.\end{equation}
At 10~K, $\kdr=2.6\tdix{-7}\cccs$, and the (dimensionless) numerical coefficient is  7.6\tdix{-5}. Since atomic sulfur is, in all our models, the main sulfur carrier, this expression shows how the NS:\ce{N2H+} abundance ratio can be used to measure \stot. We now discuss the $\ab{O}^2/\ab{e-}$ term.

\subsection{Electronic abundance}
\label{sec:xel}

The electronic density, \nelec,\ results from the competition between ionization and recombination processes and is thus driven by the cosmic-ray ionization rate and the total density. The abundance of electrons also depends on the total grain area per unit volume although to a lesser extent \citep{walmsley2004, flower2007}. The formation rate of electrons is due to the ionization of \ce{H2} and He by cosmic rays, with rates $\zeta_{\hh}$ and 0.5$\zeta_{\hh}$, respectively, while the destruction proceeds primarily through dissociative recombination with molecular ions with rate $k_R$. This leads to (assuming \nh=2\dens{H2} and 10\% of He):
\begin{equation}
        \label{eq:xe0}
        \nelec/\nh = \ab{e-} = 0.55 \frac{\zeta_{\hh}}{\kr \nh}\frac{1}{\xm}
,\end{equation}
where we have noted \xm\ the fractional abundance of positively charged molecular ions. Provided that molecular ions dominate over atomic ions, charge neutrality translates into $\xm = \xel$, and the usual $(\zeta_{\hh}/\nh)^{0.5}$ dependence is recovered \citep[e.g.,][]{walmsley2004}. Numerically, with $\kr \approx 2\tdix{-6}$\cccs\ at 10~K, we obtain
\begin{equation}
  \label{eq:xe1}
  \ab{e-} \approx 1.7\tdix{-8}\,(\zeta_{17}/n_4)^{1/2},
\end{equation}
with $n_4$ the proton density in units of \dix{4}\ccc\ and $\zeta_{17}$ {the \hh\ cosmic-ray ionization rate in units of \dix{-17}\pers}. These values compare well with the results of our numerical models (see Tables~\ref{tab:abundances_n4} and \ref{tab:abundances_n5}) as long as \stot\ and the C/O ratio remain lower than a few ppm and $\approx 0.6,$ respectively.

However, the high-\stot\ regime deserves special attention because \ce{S+} becomes the main charge carrier so that the approximation $\ab{e-} = \ab{m+}$ is no longer valid. In such instances, we obtain the following:
\begin{equation}
\label{eq:xe2}
\ab{e-}\propto \zeta_{17}/n_4,
\end{equation}
instead of Eq.~\ref{eq:xe1}. Moreover, the chemical state of the gas-phase -- and more specifically its ionization state -- may become increasingly sensitive to the amount of molecular oxygen and charged species \citep{pineau1992, lebourlot1995, boger2006}.

\subsection{Abundance of atomic oxygen}
The abundance of atomic oxygen in the gas phase depends on the partitioning of oxygen among its main carriers in solid, ice, and gas forms, and there are indications that this partitioning is not well understood \citep{whittet2010, jenkins2014}. In cores exposed to the ambient interstellar UV field, the abundance of oxygen in the gas-phase is essentially driven by the competition between adsorption and photo-desorption of water ice \citep{hollenbach2009}. On observational grounds, the abundance of oxygen-bearing species in ices in low-mass cores are reviewed in \cite{boogert2015}. The abundance of water, CO, and \ce{CO2} in ices are 39, 12, and 13~ppm, respectively, which, together with other trace species, account for $\approx80$~ppm. On the other hand, 240~ppm of oxygen is captured into refractory material \citep{jenkins2009}, so that the amount of oxygen locked in solid (ice and refractory) is $\approx320$~ppm \cite[see also][]{hilyblant2018a}.

The abundance of elemental oxygen in the local ISM is not precisely known, but it is most likely in the range 442 to 575~ppm \citep{sofia2001, nieva2012}, {from which  the corresponding} total gas-phase oxygen abundance is $\otot = 442-320 = 120$~ppm and $\otot = 575-320 = 255$~ppm, respectively.

In the gas-phase of dense cores, the main carriers are CO and atomic oxygen, with molecular oxygen being a minor species \citep{goldsmith2011}. In low-C/O environments, approximately 83~ppm of the oxygen (according to our assumed abundance of gas-phase carbon) will be locked into CO; namely, what remains in the gas phase, presumably in the form of atomic oxygen, ranges from $120-83=37$ to $255-83 = 172$~ppm, or $\ab{O}=3.7\tdix{-5}$ to 1.7\tdix{-4}. As noticed by \cite{whittet2010}, and highlighted by \cite{jenkins2014}, the higher limit would imply that a large fraction of oxygen is locked into an unidentified form. In taking a conservative approach, we use the lower limit in what follows, namely, $\ab{O}=3.7\tdix{-5}$. It is worth stressing that this reasoning applies only if C/O$<1$. In addition, we expect that the abundance of gas-phase oxygen will decrease as the C/O ratio increases.

\subsection{The abundance of atomic sulfur for C/O$<1$}
Inserting Eq.~\ref{eq:xe1} into Eq.~\ref{eq:sss} together with $\ab{O}=37$~ppm, and assuming a kinetic temperature of 10~K, the abundance of atomic sulfur in the gas phase is expressed as:
\begin{equation}
%    \ab{S} = 5.2\tdix{-6} \frac{\ab{NS}}{\ab{N2H+}}
    \ab{S} = 6.1\tdix{-6} \frac{\ab{NS}}{\ab{N2H+}} 
    \left(\frac{\zeta_{17}}{n_4}\right)^{-0.5}.
    \label{eq:s}
\end{equation}
In all our models, atomic sulfur is the main carrier of sulfur (see Tables~\ref{tab:abundances_n4} and \ref{tab:abundances_n5}) such that one may adopt $\ab{S}=\stot$. Therefore, the previous expression may be compared to the results of our model calculations (black line in Fig.~\ref{fig:models}).

Despite the various assumptions applied to derive \req{s}, the overall agreement between the analytical and numerical predictions may appear surprisingly good. This is especially the case at low C/O, such as 0.4, for both $n_4=1$ and 10, and regardless of $\zeta_{17}$. The scatter of the model predictions is less than 1 dex over most of the parameter space; however, the models with $n_4=10$ and $\zeta_{17}=1$ has the largest dispersion of predicted ratios, and the gradient of \stot\ among the sources in the sample is less clear unless a consistent value of C/O is used in the four sources.
Incidentally, since the NS:\nnhp\ ratio is directly related to \ab{S}, the high value of this ratio in L1521E is a direct proof that atomic sulfur is the main carrier of sulfur.

\section{Discussion}
\label{sec:discussion}

\def\wa{0.33\hsize}
\begin{figure*}
        \centering
        \includegraphics[width=\wa]{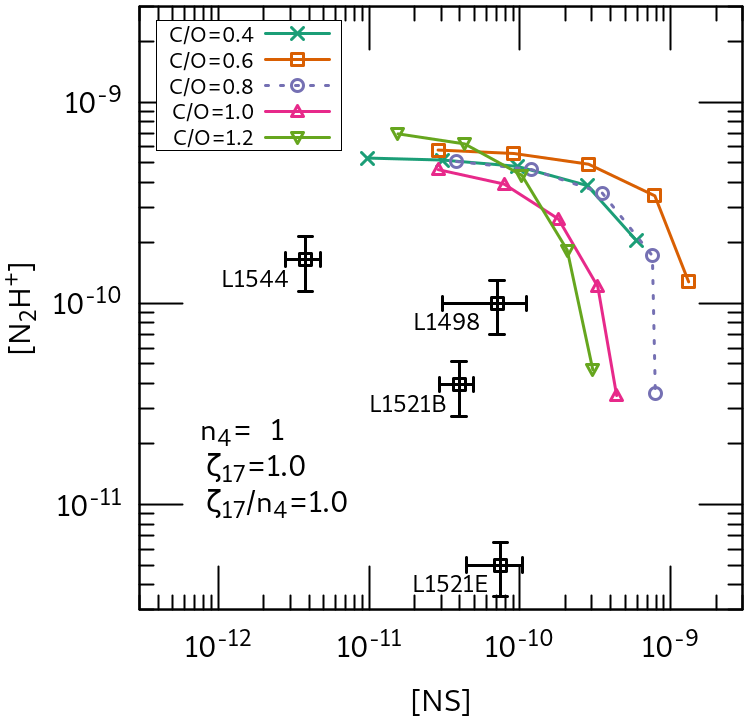}\hfill%
        \includegraphics[width=\wa]{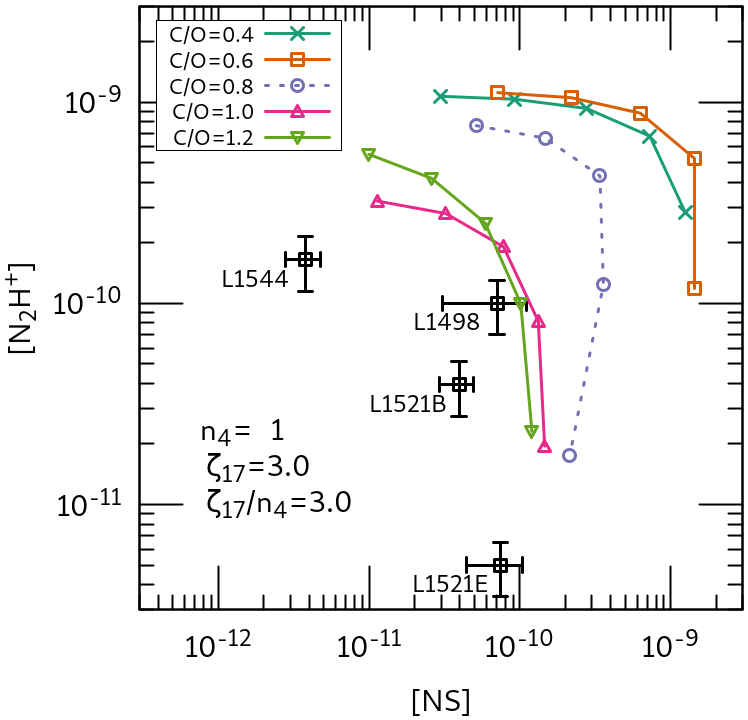}\hfill%
        \includegraphics[width=\wa]{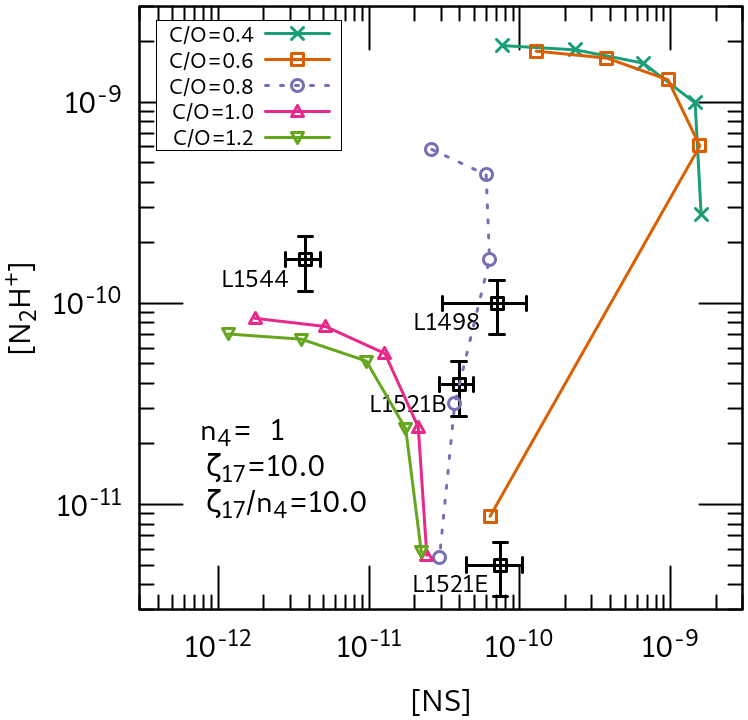}\smallskip \\
        \includegraphics[width=\wa]{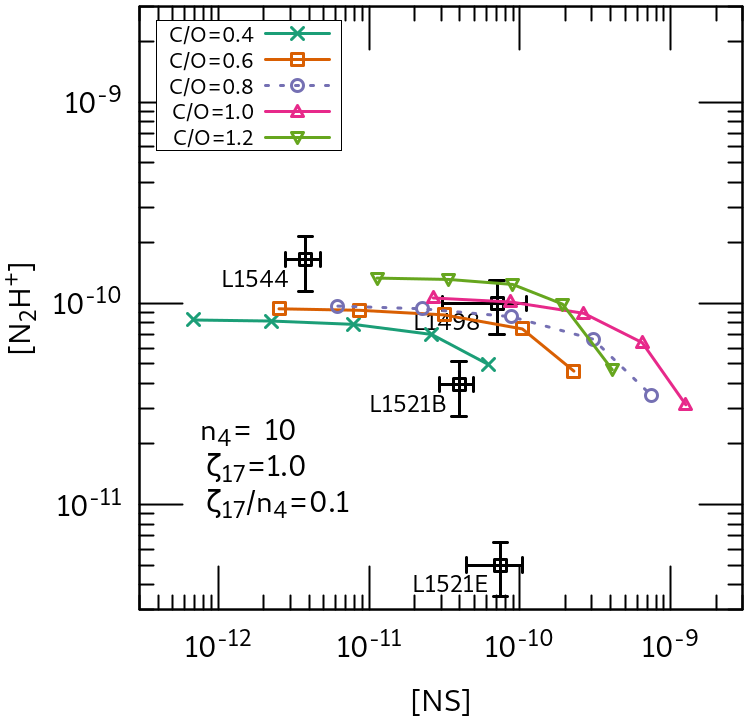}\hfill%
        \includegraphics[width=\wa]{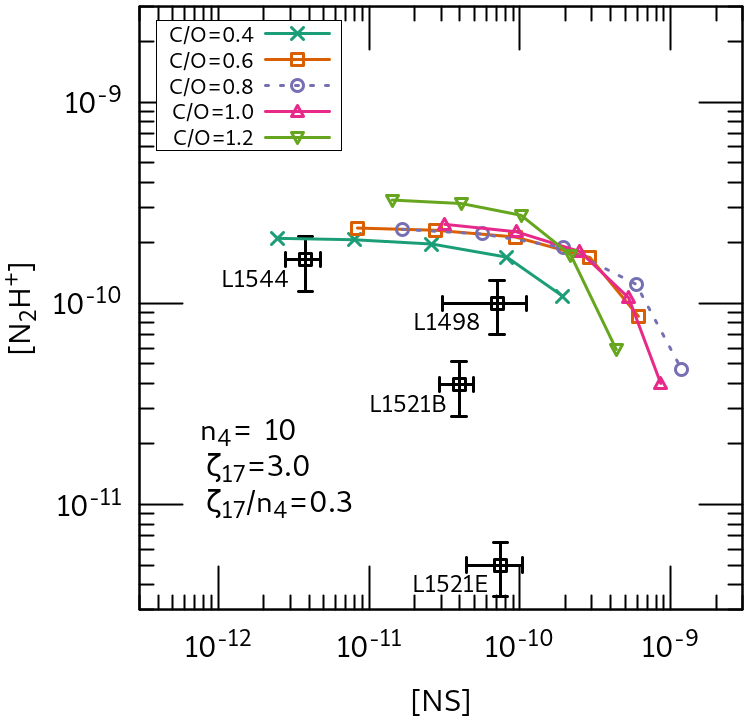}\hfill%
        \includegraphics[width=\wa]{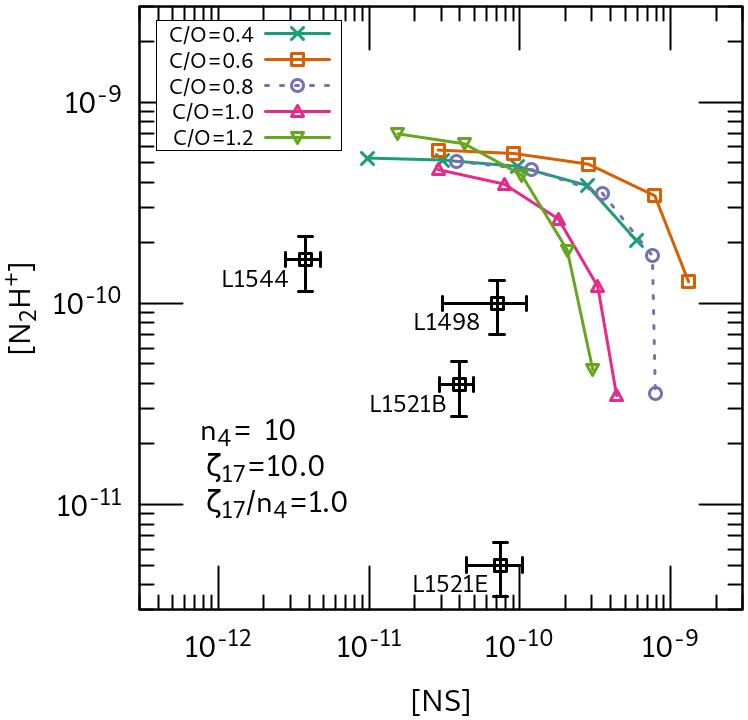}
        \caption{Steady-state abundance of NS and \nnhp\ from our grids of models (top: $n_4=1$; bottom: $n_4=10$) are compared to the observational estimates listed in Table ~\ref{tab:results}. The value of $n_4$, $\zeta_{17}$, and the ratio $\zeta_{17}/n_4$ are indicated in each panel. Five values of \stot\ (see Table \ref{tab:init}) are associated with each C/O ratio.}
        \label{fig:ab}
\end{figure*}

The observable NS:\ce{N2H+} abundance ratio has been shown to provide a direct constraint to the total gas-phase abundance of sulfur, \stot, and, hence of the abundance of gas-phase atomic sulfur \ab{S}. Furthermore, the case of L1521E provides a clear confirmation that atomic sulfur is the main carrier of sulfur, which is in agreement with our and published models \citep{boger2006}.

The differences of \stot\ in our source sample may be interpreted in terms of evolutionary stage. In early-type--chemistry cores, such as L1521E and L1521B, our observations indicate that sulfur is essentially not depleted, with 14~ppm in the gas-phase in the form of atomic sulfur. As cores evolve chemically, the amount of atomic sulfur decreases, which may be interpreted as gas-phase depletion in ices. Thus, in the chemically more evolved core L1498, the gas-phase abundance of sulfur is significantly lower than in L1521E (see Table \ref{tab:stot}), while L1544 has the highest level of sulfur depletion.

Our results are consistent with the non-depletion of sulfur in refractory material as derived from observations of ionized sulfur in the diffuse medium \citep{jenkins2009, jenkins2014}. Moreover, these results are in agreement with the high abundance of atomic sulfur, $\ab{S}=3.5$~ppm, needed to reproduce the observations of CS toward the Horsehead nebula \citep{goicoechea2006}. We note that this abundance is very close to that in L1498 in our model with $n_4=10$. Indeed, the density in their photo-chemical models ($\nhh=0.5-1\tdix{5}\ccc$) is also comparable to that toward L1498. The present study is also in agreement with the finding by \cite{nagy2019} of a higher abundance of sulfur species in L1521E compared to L1544. However, our method provides a direct quantitative estimate of \stot. Our observations also provide observational support to recent chemical models {showing that} the abundances of S-bearing species toward the TMC-1(CP) dense cloud can be reasonably reproduced assuming a $\stot=5-14$~ppm \citep{vidal2017}, in agreement with our measurements toward L1521B, and L1521E. In the translucent part of TMC-1, observations of CS, SO, and isotopologues, and \ce{HCS+}, indicate that sulfur is significantly depleted, with $\stot = 0.8$~ppm \citep{fuente2019}, which is interpreted as a signature of strong sulfur depletion during the atomic-to-molecular transition: this behavior is thus different to what was observed in the Horsehead nebula, although the ratio of the radiation field to the \hh\ density is similar ($G/\nhh=6-12\tdix{-4}$ in the Horsehead and $2-10\tdix{-4}$ in TMC-1). It has also been proposed that sulfur depletion increases with density as S/H~$\sim \nh^{-0.6}$ \citep{rodriguezbaras2021}, based on the behavior of sulfur-bearing species. The NS:\nnhp\ method could be used to confirm this is true also for \stot.

Spatial variations of \stot\ among a given source could only be established in L1544. In the outer parts of L1544, a core for which signatures of collapse were found in the innermost regions from water line observations \citep{caselli2012}, we find lower levels of sulfur depletion than in its central region. The increase in NO synthesis, located at 4200 au, where complex organic molecules (COMs) have been detected \citep{vastel2014, jimenezserra2016, jimenezserra2021}, may indicate that the production of NS could be enhanced by the same mechanism as COMs \citep{vasyunin2017}. Alternatively, this could be a demonstration of the progressive depletion of sulfur toward the inner regions. Indeed, in L1521B, there is a trend of decrease of the NS abundance toward the dust peak. However, since the current spectra were measured by taking the \nnhp\ observations of \cite{hirota2004} as a reference, it is necessary to obtain new NS and \nnhp\ observations covering the dust peak.

The cores included in this study being located within the same large-scale environment, the differences in \stot\ among the various sources are not due to variations in the elemental abundances or in the cosmic-ray ionization rate. Furthermore, the C/O ratio may also be comparable in the four cores since this ratio is dictated by the values of \ctot\ and \otot, which are primarily induced by the composition of the refractory material. However, the fact that the C/O ratio is effectively comparable {in all sources} should be established from the observations.

From Table \ref{tab:stot}, a C/O ratio of 0.4, to which would correspond an unrealistic amount of sulfur in L1521E, may be excluded. Assuming that all the sources have a similar carbon and oxygen volatile budgets, our results suggest C/O ratios of 0.6 to 0.8. For these values, the sulfur reservoir decreases by more than two orders of magnitude in L1521E, L1498, and L1544, respectively, reflecting their evolutionary stage. The young L1521B core is also compatible with an non-depleted sulfur reservoir although the non-detection of \nnhp\ prevents any conclusion. A value of C/O~$\approx 0.8$ was also derived, but in a different environment, toward the envelope of the low-mass protostar \citep{legal2014}. Yet, in the TMC-1 ridge, a lower ratio (0.4-0.5) was proposed \citep{pratap1997}, while \cite{fuente2019} suggest higher values up to unity. The present results are unlikely to be used to measure the C/O ratio; a more promising method may be through the CN:NO abundance ratio \citep{legal2014}. In any instance, independent observational constraints on the C/O ratio are needed in combination with the new NS:\nnhp\ method, to measure \stot.

One important assumption in our model is the abundance of atomic oxygen in \req{sss}, for which we adopted the lower limit of 37~ppm from the studies of \cite{jenkins2009} and \cite{whittet2010}. Nevertheless, our grid of models, while exploring a range of C/O ratios by varying the total oxygen abundance \otot, which also influences \ab{O}, show that our analytical derivation remains overall a good approximation. We have also tested the influence of oxygen depletion by performing calculations with gas-grain chemistry producing saturated species which are returned into the gas-phase by photo-desorption \citep[see ][]{hilyblant2018a}. The impact of oxygen depletion was, however, not as direct as anticipated from our chemical analysis for C/O~$<1$ because carbon-destruction routes then become more efficient, leading to essentially unchanged NS:\ce{N2H+} ratios.

In Eq.~\ref{eq:s_approx}, the abundance of sulfur was expressed in terms of the NS:NH ratio, and the abundance of NH was derived assuming pure gas-phase formation. It has been proposed that a fraction of the intermediate radicals NH and \ce{NH2} could be released in the gas-phase during the hydrogenation of nitrogen in ices \citep{hickson2015}. However, laboratory experiments failed to detect these radicals indicating that hydrogenation proceeds faster than desorption \citep{hidaka2011}. Furthermore, gas-phase processes alone can reproduce the observed amounts of nitrogen hydrides, as well as their ortho:para ratios, and their deuterated forms in low-mass star-forming environments \citep{legal2014, hilyblant2018a}.

Although this work is devoted to the NS:\nnhp\ ratio, as ratios are less sensitive to uncertainties in the derivation of the \hh\ column density, we compare in Fig.~\ref{fig:ab} the estimated abundances of NS and \nnhp\ listed in Table \ref{tab:results} to those predicted by our models. Our model predictions cover the entire range of observed values of both NS and \nnhp, while previous attempts failed to reproduce the NS abundance \citep{millar1990, hasegawa1993, lee1996b}. This is, at least in part, due to our NS abundances being smaller by factors of two or more, compared to the earlier values in low-mass dense cores \citep{mcgonagle1994}. {This figure shows that lowering the density in our model both increases and broadens the range of the predicted \nnhp\ abundances, while  only slightly increasing the NS abundance. On the other hand, it is evident that models with the $n_4=1$, $\zeta_{17}=1$ and with $n_4=10$, $\zeta_{17}=10$ are similar, suggesting that $\zeta_{17}/n_4$ is a good control parameter. It is then apparent that the spread in predicted abundances increases with $\zeta_{17}/n_4$. Furthermore, from the previous discussion, C/O ratios between 0.6 and 0.8 may be favored, suggesting that a model with $(n_4,\zeta_{17})=(1,10)$ and C/O $=0.8$ provides a reasonable agreement for three of the four sources, while over-predicting \nnhp\ and NS in L1544. Nevertheless, from the models with $\zeta_{17}/n_4=0.1$ and 0.3, it appears that the NS:\nnhp\ ratio L1544 can be explained with a C/O ratio within 0.6 and 0.8. More generally, it seems possible to find, for a given C/O similar in all sources, a set of models with different $\zeta_{17}/n_4$ ratios accounting for the NS:\nnhp\ ratio in our sample.} Going further in reproducing the abundances is, however, beyond the present study, as this would {also} require us to follow the dynamical collapse \citep[e.g.,][]{hilyblant2018a}, and, more importantly, to simultaneously reproduce  more nitrogen- and sulfur-bearing species.

New observations of NS and \ce{N2H+} should be performed toward the dust peak of L1521B in order to obtain a measure of the ratio and confirm the absence of depletion in this young object. It would also be interesting to obtain maps of the \ce{N2H+} column density toward these cores, similarly to what was done for L1544, in order to confirm the depletion of sulfur with density. Detailed {one-dimensional} radiative transfer calculations of NS, similar to what \cite{daniel2007} did for \ce{N2H+}, may be useful in assessing the level of deviation from single excitation temperature within a rotational manifold at the hyperfine {level}.

From a theoretical perspective, chemical models are needed to explore the high sulfur-abundance regime in more detail, especially the electronic fraction. Also, the abundance of nitrogen-bearing species was found to decrease with increasing \stot\ in \cite{legal2014} and the \ce{NH2}:NH and \ce{NH3}:NH abundance ratios, measured in the envelope of the I16293-2422 protostar, put an upper limit of 8~ppm on \stot. According to the present results, this would indicate that the envelope is in an advanced stage of chemical evolution. The dynamics of the collapse may likely be impacted by the present findings as we would expect a strong coupling of the gas with magnetic fields in the high-\stot\ regime.

\section{Conclusion}

We provide a new and direct method to measure the total abundance of sulfur, that is, of atomic sulfur, in the gas-phase of dense starless cores. We also provide a simplified analysis of the chemistry of NS, leading to a simple relation between the NS:\nnhp\ ratio and the abundance of gas-phase atomic sulfur, which is in good agreement with the numerical results. This is strong suggestion that the new method is robust over a broad parameter space. Our results provide a clear demonstration of the progressive depletion of atomic sulfur as cores evolve and, for a given core, with density. Our observations also show that sulfur is entirely in the gas-phase in atomic form in chemically young cores.  Comparisons with spatially resolved observations of sulfur-bearing species in ices at densities below and above $\nhh=10^5$\,\ccc\ with the JWST may reveal the progressive uptake of sulfur in ices that we have evidence of in L1544, and to elucidate the main carriers of sulfur in ices.

\begin{acknowledgements}
We are indebted to the anonymous referee for very careful reading and for questions and comments which led to a significant improvement of the paper. The following public softwares were used: GILDAS, Gnuplot, emcee. We also thank M\'elisse Bonfand who contributed to this work during an internship at IPAG in 2015. PHB and FL acknowledge the \emph{Institut Universitaire de France} (IUF) for financial support. This work was supported by the Programme National “Physique et Chimie du Milieu Interstellaire” (PCMI) of CNRS/INSU with INC/INP co-funded by CEA and CNES.

This research used the facilities of the Canadian Astronomy Data Centre operated by the the National Research Council of Canada with the support of the Canadian Space Agency. The James Clerk Maxwell Telescope is operated by the Joint Astronomy Centre on behalf of the Science and Technology Facilities Council of the United Kingdom, the Netherlands Organization for Scientific Research, and the National Research Council of Canada.

\end{acknowledgements}

\bibliographystyle{aa}
\bibliography{general,chemistry,lique}

\clearpage
\newpage
\appendix

\section{\hh\ Column density}
\label{app:NH2}

\begin{table*}
  \centering
  \footnotesize
  \caption{\label{tab:sed} Specific intensity (in \Mjysr, before background subtraction) toward the dust emission peak of L1521E and toward offsets observed in NS.}
  \begin{tabular}{r cccccc ccc}
    \toprule
    Offsets & $I_{1300}$ & $I_{850}$ & $I_{500}$ & $I_{350}$ & $I_{250}$ & $I_{160}$ & $T$ & \logd\Nhh & $\beta$ \\
    arcsec &&&&&&& K & \cc \\
    \midrule
        \mc{10}{L1521E$^1$}\\
    -15, -5 &   5.3 &  14.1 &  56.2 & 110.4 & 138.9 &  67.6 & 10.9(4) & 22.37(7) & 2.2(1) \\
    -10,-10 &   5.1 &  13.4 &  56.1 & 108.8 & 134.7 &  66.8 & 11.0(4) & 22.33(7) & 2.1(1) \\
     10,-10 &   4.5 &  11.4 &  52.3 & 101.0 & 124.1 &  65.5 & 11.4(4) & 22.23(9) & 2.1(1) \\
    -30,-10 &   5.3 &  14.0 &  54.3 & 107.9 & 139.0 &  67.5 & 10.9(4) & 22.37(8) & 2.2(1) \\
    -10,-30 &   5.1 &  13.4 &  56.1 & 108.8 & 134.7 &  66.8 & 11.1(4) & 22.33(8) & 2.1(1) \\
    $I_{\rm bg}^\dagger$
        &       0.5 &      -0.9 &       3.2 &       6.8 &       9.9 &       3.8\\
        $1\sigma^\ddagger$
        &       2.0 &       3.0 &       1.0 &       2.0 &       3.0 &       5.0\\
    \midrule
   \mc{10}{L1521B}\\
    % Background subtracted results
        % V2 -- with convolution by kernels = sqrt(35**2-hpbw**2)
        % 110,-15 &      56.2 &      96.4 &     108.0 &      51.9
   115,-15 &  & &   55.2 &      96.0 &     108.1 &      51.7& 11.0(2) & 22.24(5) & 1.8(1)\\
    60,  0 &  & &   46.8 &      81.4 &      89.9 &      42.9& 10.6(3) & 22.23(6) & 1.9(1)\\
    40,  0 &  & &   43.7 &      74.0 &      80.9 &      37.5& 10.4(3) & 22.24(7) & 1.9(1)\\
    20,  0 &  & &   42.2 &      71.3 &      77.8 &      36.1& 10.5(3) & 22.20(7) & 1.9(1)\\
     0,  0 &  & &   41.6 &      70.7 &      77.7 &      38.0& 10.9(3) & 22.10(7) & 1.8(1)\\
    50,-20 &  & &   45.0 &      77.3 &      85.6 &      39.8& 10.5(3) & 22.24(6) & 2.0(1)\\
    30,-20 &  & &   42.6 &      72.1 &      78.8 &      36.1& 10.4(3) & 22.23(7) & 2.0(1)\\
    10,-20 &  & &   41.9 &      71.1 &      77.6 &      37.0& 10.6(3) & 22.16(7) & 1.9(1)\\
% V1 -- with maps convolved by 35" kernel
%   110,-15 &&&      56.2 &      94.6 &     106.8 &      51.6
%    115,-15 &&&      55.2 &      94.2 &     107.3 &      51.4 & 11.0(2) & 22.24(5) & 1.8(1)\\
%     60,  0 &&&      46.8 &      81.0 &      89.5 &      42.6 & 10.6(3) & 22.23(6) & 1.9(1)\\
%     40,  0 &&&      43.7 &      74.7 &      81.8 &      37.6 & 10.4(3) & 22.24(7) & 1.9(1)\\
%     20,  0 &&&      42.2 &      71.8 &      78.3 &      36.3 & 10.5(3) & 22.20(7) & 1.9(1)\\
%      0,  0 &&&      41.6 &      70.5 &      77.8 &      38.0 & 10.9(3) & 22.10(7) & 1.8(1)\\
    $I_{\rm bg}^\dagger$
        & & &   3.7 &   6.7 &   8.9 &      8.5 \\
        $1\sigma^\ddagger$
        & & &   0.7 &   1.2 &   1.6 &      1.8 \\
    % (115,-15) &       &       &  57.5 &  95.4 & 106.6 &  47.8 & 10.8(2) & 22.27(5) & 1.8(1)\\
    % (60,0)    &       &       &  46.8 &  81.0 &  89.5 &  42.6 & 10.6(3) & 22.23(6) & 1.9(1)\\
    % (40,0)    &       &       &  43.7 &  74.7 &  81.8 &  37.6 & 10.4(3) & 22.24(7) & 1.9(1)\\
    % (20,0)    &       &       &  42.2 &  71.8 &  78.3 &  36.3 & 10.5(3) & 22.20(7) & 1.9(1)\\
    % ( 0,0)    &       &       &  41.6 &  70.5 &  77.8 &  38.0 & 10.9(3) & 22.10(7) & 1.8(1)\\
    \midrule
    \mc{10}{L1498}\\
     10, 10&     5.1 & &        56.8 &     105.4 &     124.4&&11.1(8)  & 22.26(16) & 1.9(2)\\
      0,  0&     5.0 & &        55.9 &     104.5 &     124.6&&11.1(9)  & 22.26(17) & 1.9(2)\\
    -10,-20&     4.8 & &        54.3 &     102.4 &     123.9&&11.3(10) & 22.20(20) & 1.9(3)\\
     20,-10&     5.0 & &        56.8 &     105.3 &     123.7&&11.0(9)  & 22.26(17) & 1.9(3)\\
     10, 20&     5.1 & &        56.8 &     105.4 &     124.4&&11.1(7)  & 22.25(16) & 1.9(3)\\
    -20, 10&     4.5 & &        52.1 &      99.5 &     122.6&&11.4(10) & 22.17(20) & 1.9(3)\\
     20,-40&     5.0 & &        56.8 &     105.3 &     123.7&&11.0(9)  & 22.28(17) & 1.9(3)\\
    -40, 20&     4.1 & &        49.3 &      95.6 &     120.6&&11.6(10) & 22.14(20) & 1.9(3)\\
     40,-20&     4.9 & &        56.0 &     104.1 &     122.8&&10.9(8)  & 22.29(17) & 2.0(3)\\
     60,-30&     4.2 & &        51.1 &      97.3 &     118.2&&11.0(9)  & 22.24(18) & 2.0(3)\\
     80,-40&     2.9 & &        41.7 &      81.3 &     103.4&&11.1(14) & 22.14(25) & 2.1(3)\\
    $I_{\rm bg}^\dagger$
        &       0.7 &  &     5.5 &      12.3 &      19.4\\
        $1\sigma^\ddagger$
        &       1.2 &  &     1.0 &       2.1 &       3.5\\
    \midrule
        \mc{10}{L1544$^2$}\\
        0,  0 &      12.8 &&     126.1 &     209.9 &     217.9 && 9.6(1) & 22.82(1) & 2 \\
        -20,-20 &       8.1 &&      98.9 &     172.5 &     197.5 && 10.0(1) & 22.63(2) & 2\\
        -20, 20 &      11.3 &&     118.6 &     201.1 &     218.5 && 9.8(1) & 22.75(1) & 2\\
        20,-20 &       9.2 &&     106.8 &     184.4 &     204.4 && 9.9(1) & 22.68(1) & 2 \\
        20, 20 &       8.8 &&     101.2 &     177.5 &     199.4 && 10.0(1) & 22.64(2) & 2 \\
    $I_{\rm bg}^\dagger$
        &       0.0 &&      13.8 &      30.7 &      49.6\\
        $1\sigma^\ddagger$
        &       1.5 &&       1.3 &       3.8 &       5.7\\
    \bottomrule
  \end{tabular}
  \tabnotes Statistical uncertainties are given at the 1$\sigma$ level in brackets in units of the last digit: (22.37(7) means $22.37\pm0.07$). $^\dagger$ $I_{\rm bg}$ is the background intensity measured in the hatched region of Fig.~\ref{fig:NH2_l1521e}. $^\ddagger$ The 1$\sigma$ uncertainty was estimated in the same region. $^1$~In L1521E, the statistics at 1300 and 850\micr\ were derived in a wide region surrounding the core. $^2$ In L1544, the spectral index was held fixed with $\beta=2$.
\end{table*}

We present, in greater detail, the derivation of the \hh\ column density for L1521E, L1521B, and L1498.

\subsection{L1521E}

The \hh\ column density was estimated following two approaches. One is based on the radial density profile from \cite{tafalla2004b}, $n(r)=n_0/[1+(r/r_0)^2]$, which is integrated to give the \hh\ column density at a distance $b$ from the peak, $\Nhh(b)=\pi n_0 r_0 / [1+(b/r_0)^2]$. With $n_0=2.7\tdix{5}\ccc$ and $r_0=30\arcsec$, the 11"-beam averaged \Nhh\ toward the dust peak (at offsets $-15",-5"$) is thus $5.3\tdix{22}\cc$, which compares well with the value of 3.9\tdix{22}\cc\ that can be derived directly from the dust emission, 24~mJy/beam \citep{tafalla2002}. Averaged in the 26"-beam of the \nnhp\ spectra, the column density at offsets $(-10",-10")$ is reduced by only 6\% compared to its peak value of 5.1\tdix{22}\cc\ (17\% for $b=20\arcsec$), or 4.8\tdix{22}\cc.

As a sanity check, we also performed a second calculation based on dust emission maps at several wavelengths, thus providing a measurement of \Nhh\ independent of the radial density profile. Thus the dust emission spectral energy density (SED) was fitted by a modified black-body:
\begin{equation}\label{eq:sed}
I_\nu\,[\Mjysr] = B_\nu(T) [1-e^{-\tau_\nu}] \approx \kappa_\nu \mu \mh \Nhh B_\nu(T),
\end{equation}
with $B_\nu(T)$ the black-body spectrum (in \Mjysr) at a dust temperature $T$, $\mu$ the mean molecular weight, equal to 2.8, \mh\ is the proton mass (taken equal to the atomic mass unit), and \Nhh\ is the \hh\ column density. The mass absorption coefficient is $\kappa_\nu=\kappa_0 (\nu/\nu_0)^\beta$ and we adopted $\kappa_0=0.09 \cmm\per{g}$ at $\lambda=250$~\micr\ \citep[][ and references therein]{magalhaes2018a}. The $T$, \Nhh, and $\beta$ parameter space was explored using the emcee MCMC sampler. 

Observations at wavelengths from 160 to 1300~\micr\ from IRAM, JCMT, and Herschel facilities have been used \citep{tafalla2004b, kirk2005, andre2010, marsh2016}. At 70 \micr, the source is not detected. The maps, convoluted by a 35"-HPBW Gaussian kernel and re-sampled on the 500 \micr\ grid, are shown in the left panel of Fig.~\ref{fig:NH2_l1521e}. The resulting specific intensity are listed in Table~\ref{tab:sed}. Prior to minimization, a background was subtracted to each intensity, which was estimated in a nearby region outside the core (indicated by a hatched histogram in Fig.~\ref{fig:NH2_l1521e}.) At some wavelengths for which the maps are too small, the background was estimated from the histogram of the intensity.

The results of parameter space exploration for the dust peak position at offsets $(-15",-5")$ are shown in the middle panel of Fig.~\ref{fig:NH2_l1521e} and the corresponding SED in the right panel. We note that the peak position is not strictly the same at all wavelengths, with $\approx5"$ differences, much smaller than the 35" beam at 500~\micr. Results toward the offsets observed in NS are summarized in Table~\ref{tab:sed}. The derived dust temperature and spectral index are similar at all positions and typical of starless cores. The \hh\ column density toward the peak position is 22.37(7)dex, or $2.4\pm0.3\tdix{22}$ \cc, a factor of two smaller than the value derived above from the density profile. Given the uncertainties induced by the mass absorption coefficient (with $\beta=2.2$, our value at 1.3~mm is 0.0024~\cmm\per{g}, half that adopted by Tafalla et al 2004 in their study), and the fact that the density profile results from fitting an azimuth average of the map, an agreement within a factor of two can be considered a good one. Our results agree well with those obtained by \cite{makiwa2016}, although our study includes longer wavelength maps, thus providing tight constraints to the spectral index.

\begin{figure*}
        \centering
        \vfig{5}{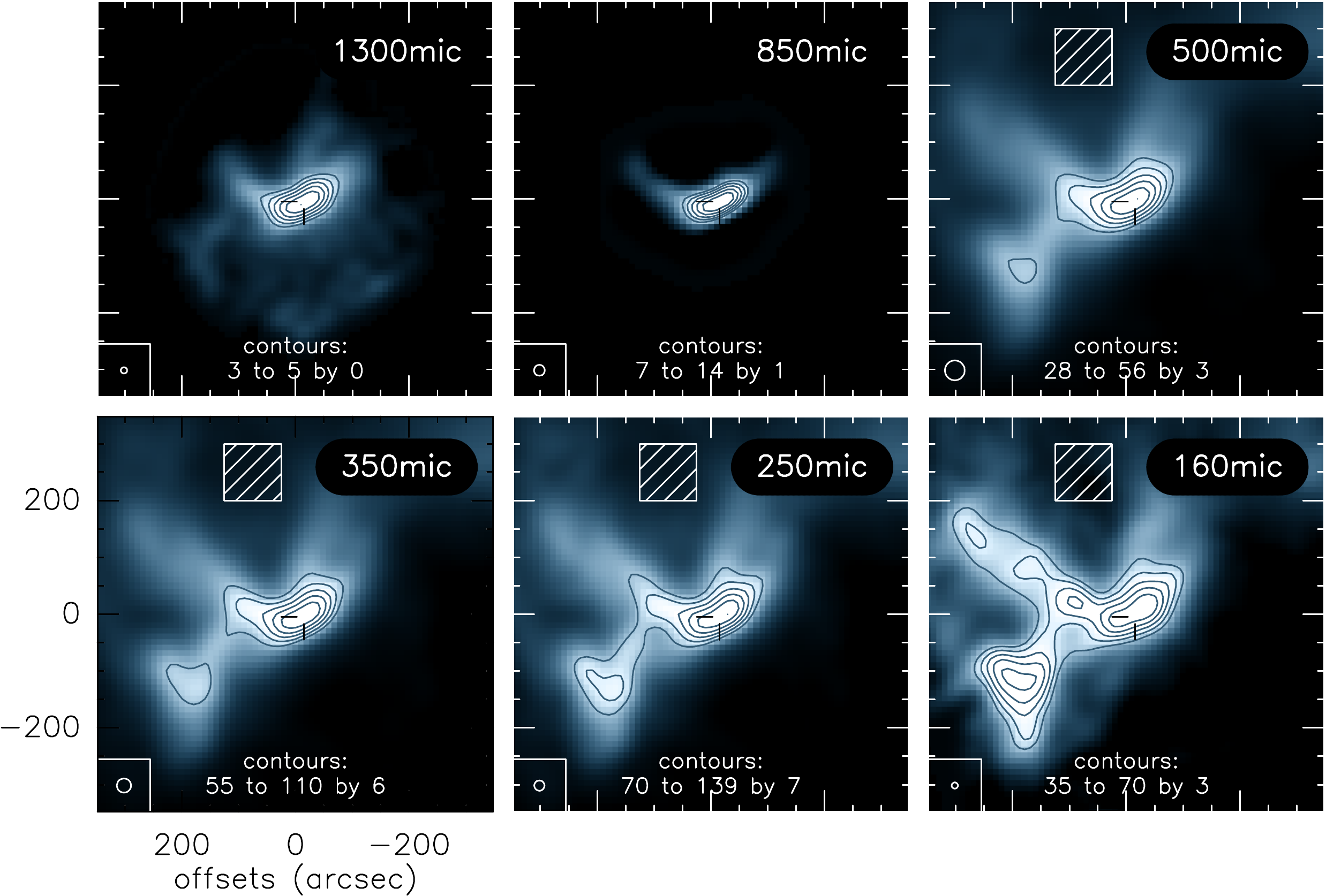}\hfill%
        \vfig{5}{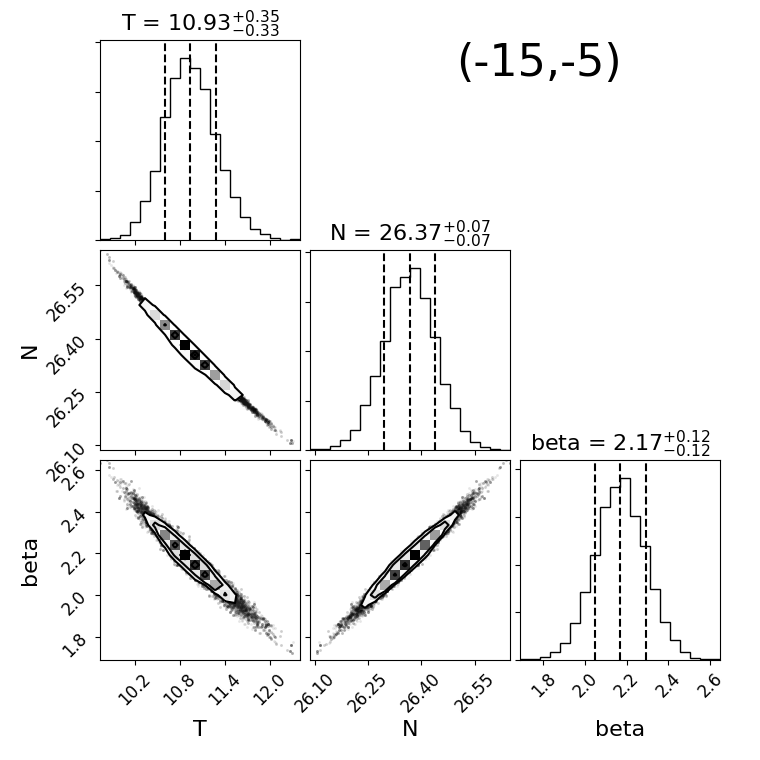}\hfill%
        \vfig{5}{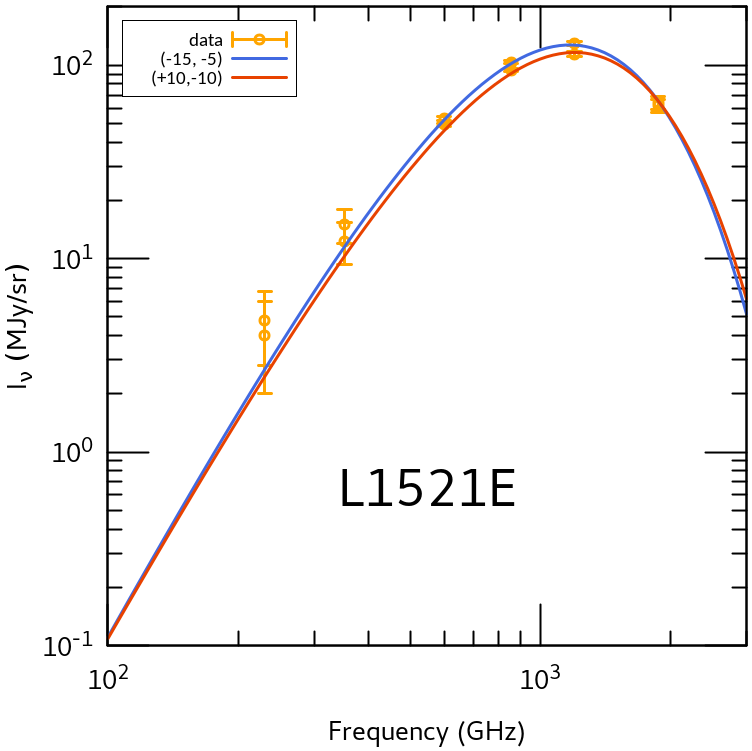}
        \caption{Determination of \Nhh\ in L1521E through SED-fitting by a modified black-body spectrum. \textit{Left} Dust emission, measured in L1521E with the {IRAM-30m/MAMBO, Herschel/PACS, and SPIRE facilities}, used to compute the \hh\ column density (see Section \ref{app:NH2}). The white cross-hairs locate the dust emission peak, at offsets $(-15",-5")$. The contours levels (start, end, increment, in \Mjysr) are indicated in each panel. The hatched area indicates the region used to compute the background level and rms. \textit{Middle} Results toward offsets $(-15",-5")$. Shown is a so-called "corner plot" showing the distribution of each free parameter ($T$, $\logd (\Nhh/\mm)$, and spectral index $\beta$) in the diagonal panels, and the correlation between any pair of them (off-diagonal panels). The most likely parameters based on the $\chi^2$ likelihood are shown by the central dashed, vertical line, in each diagonal panel (the other vertical lines show the 10\% and 86\% quantiles). \textit{Right:} Observed intensities and fitted SED (see Table~\ref{tab:sed}).}
        \label{fig:NH2_l1521e}
\end{figure*}

\subsection{L1521B}
\begin{figure*}
        \centering
        \vfig{6}{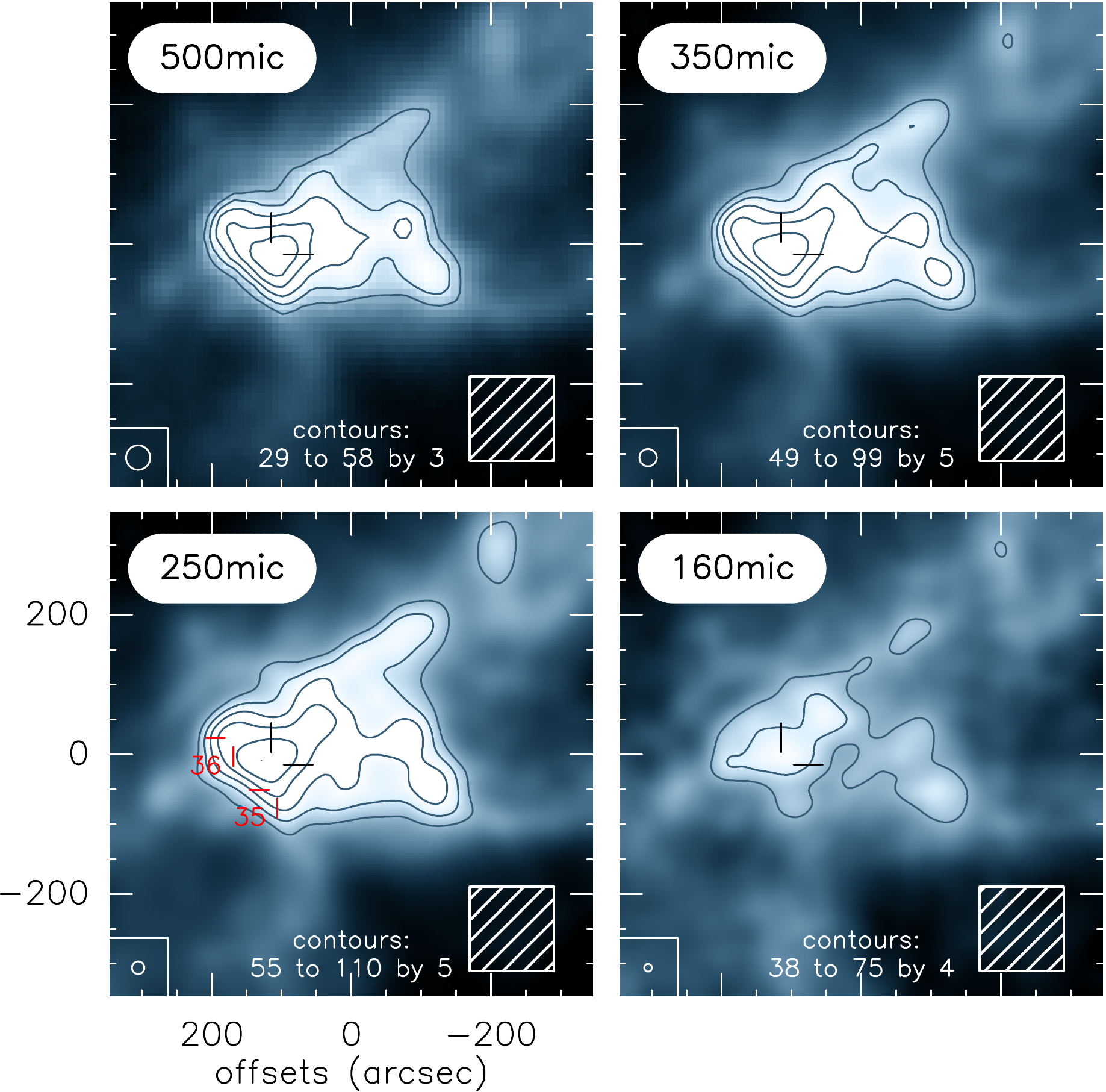}\hfill%
        \vfig{6}{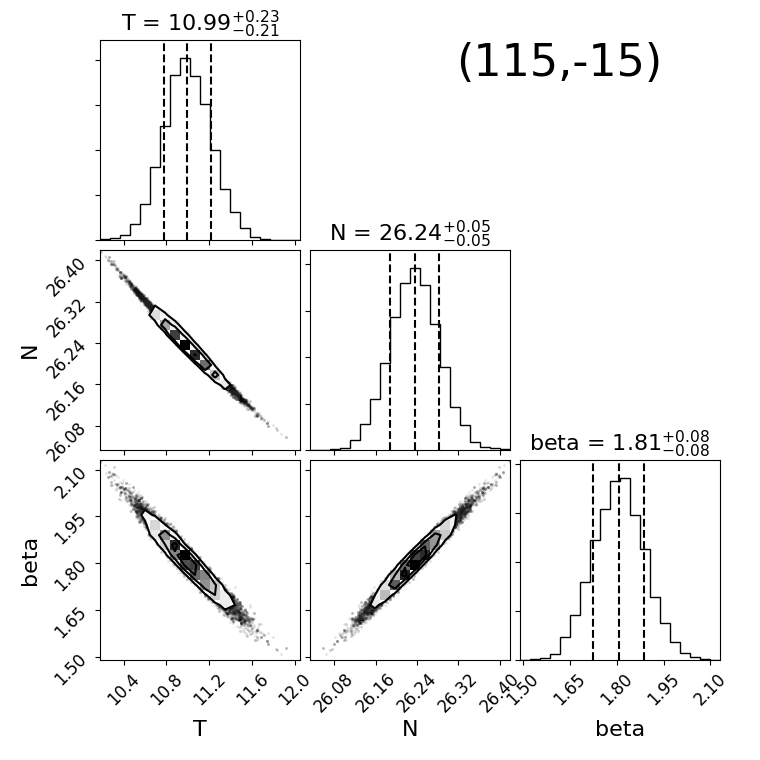}\hfill%
        \vfig{6}{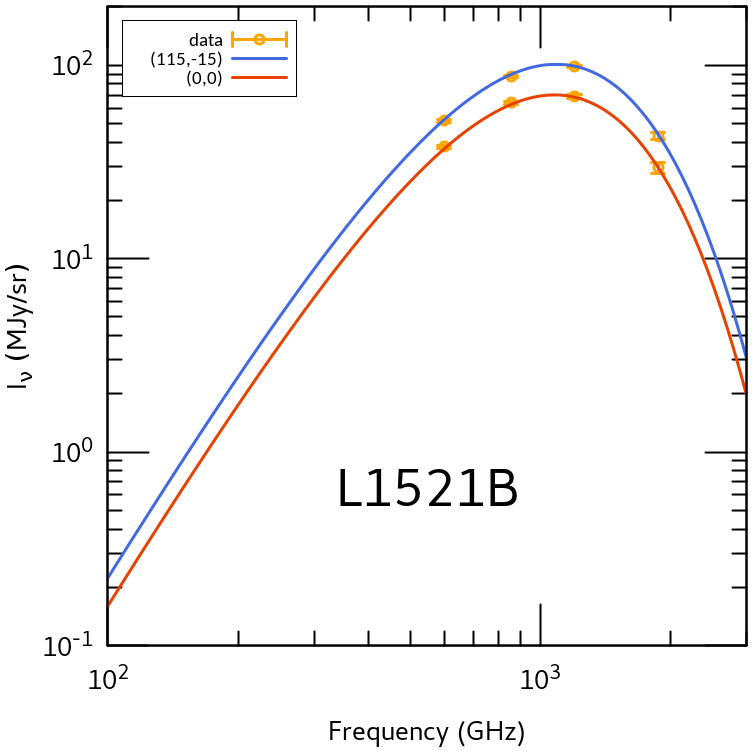}
        \caption{Determination of \Nhh\ in L1521B (see Fig.~\ref{fig:NH2_l1521e}).  The ammonia cores (35 and 36) mapped by Seo et al 2015 are shown in the 250\micr\ map of the left panel.}
        \label{fig:NH2_l1521b}
\end{figure*}

For L1521B, Herschel maps at wavelengths 160, 250, 350, and 500 microns have been used. As for L1521E, the source is not detected at 70 \micr\ and only a lower limit was obtained at 850~\micr\ \citep{kirk2005}. The maps, convolved by a 35"-HPBW Gaussian kernel and re-sampled on the 500~\micr\ grid, are shown in Fig.~\ref{fig:NH2_l1521b}. The specific intensity are listed in Table~\ref{tab:sed}. At 160 \micr, we note that our intensity compares well with that reported by \cite{kirk2007} based on Spitzer data.
        
The resulting physical and dust parameters obtained from our likelihood calculations are shown in the middle panel of Fig.~\ref{fig:NH2_l1521b}. In addition to positions observed in this study, with the $(0",0")$ position corresponding to the peak of the carbon-chain molecule CCS emission \cite{hirota2004}, we also determined the \hh\ column density toward the dust emission peak which is located at $(115",-15")$ or RA,\,DEC\,=\,+04:24:23,+26:36:43 (J2000.0) (see Figs.~\ref{fig:maps} and \ref{fig:NH2_l1521b}).

The most likely dust temperatures are $\approx 10-11$K, typical of starless cores, and in good agreement with the results from ammonia line studies indicating that L1521B has a lower gas kinetic temperature than other starless cores in the L1495 region \citep{seo2015}. The spectral index, $\beta=1.8-2.0$, is also typical of the expected value in starless cores. The \hh\ column density is well constrained, 22.10(7)dex at $(0",0")$ offsets and 22.24(5) toward the dust peak at (115.-15). These values are in very good agreement with that derived by \cite{seo2015} toward their cores 35 and 37 (see their Table 1).

\subsection{L1498}
\begin{figure*}
        \centering
        \vfig{6}{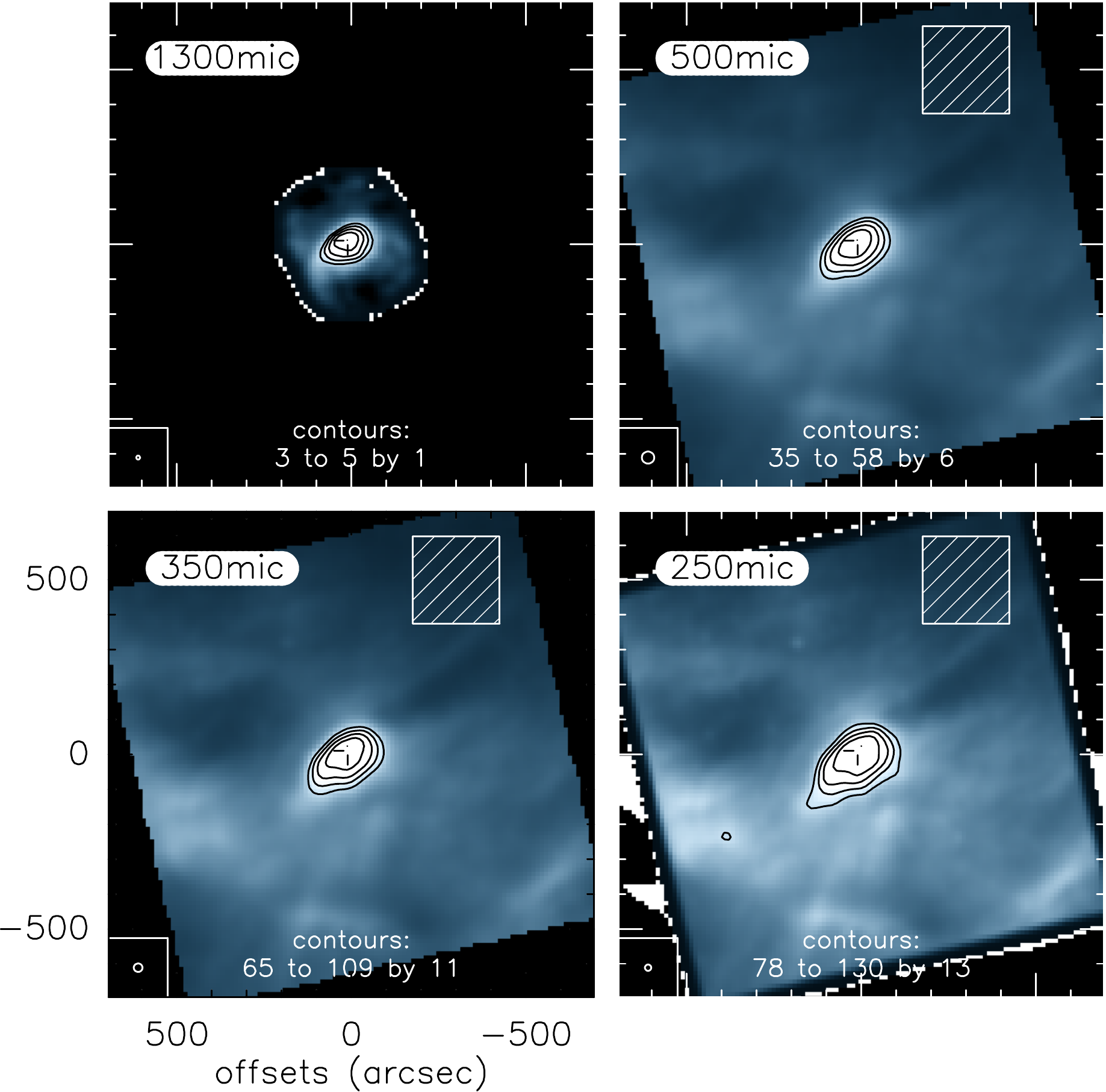}\hfill%
        \vfig{6}{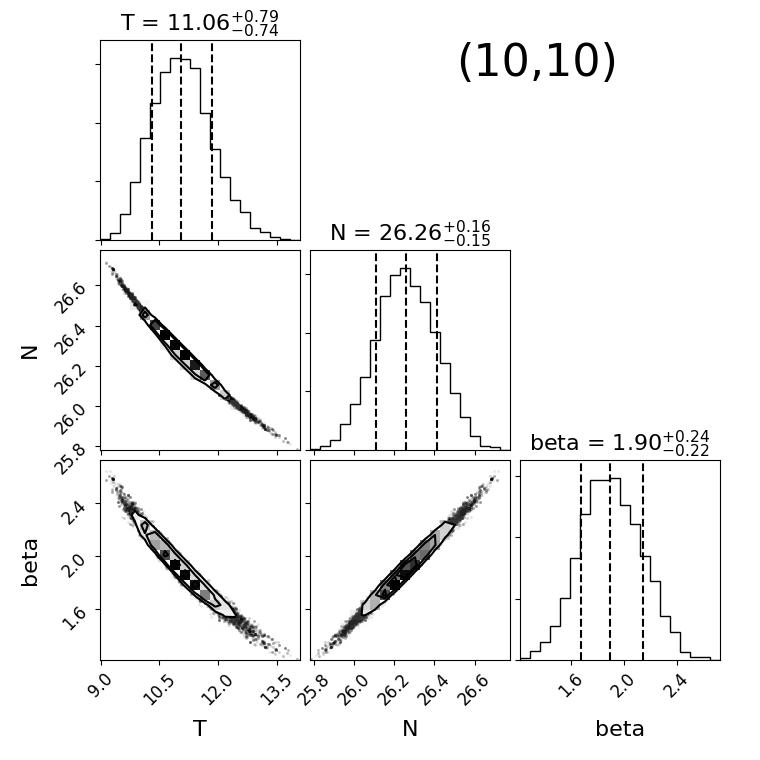}\hfill%
        \vfig{6}{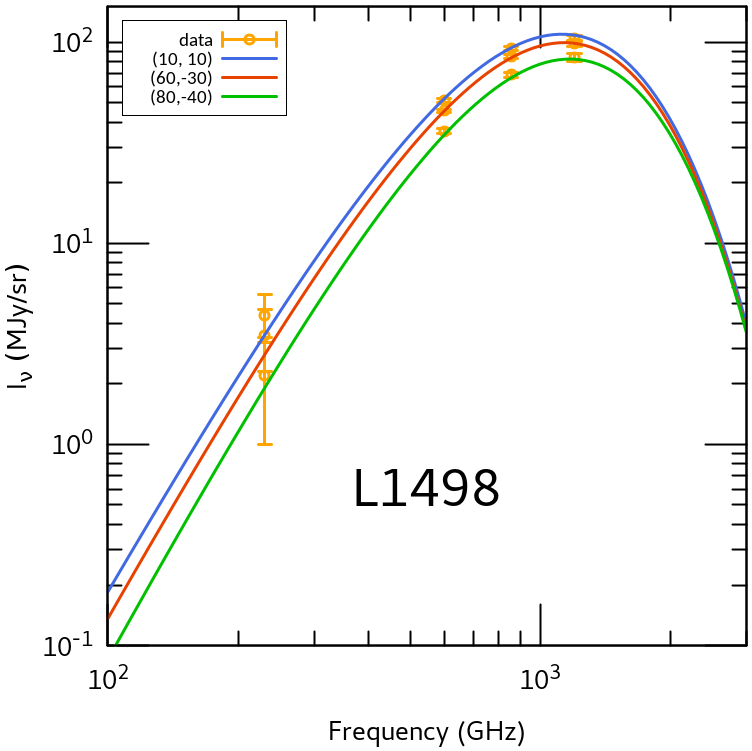}
        \caption{Determination of \Nhh\ in L1498 (see Fig.~\ref{fig:NH2_l1521e}).}
        \label{fig:NH2_l1498}
\end{figure*}
Similarly to L1521E and L1521B, continuum maps made with IRAM/30m and Herschel/SPIRE were used to constrain the \hh\ column density toward several offsets in L1498. In Fig.~\ref{fig:NH2_l1498}, results for three offsets are shown, the dust peak, and two directions for which the fluxes differ by more than 10\% compared to the peak. The resulting dust temperature is 11(1)~K at all positions, in agreement with previous studies \citep{tafalla2004a, magalhaes2018a}. The spectral index is $\approx 2$. The total 35"-averaged \hh\ column density are constrained with $\approx30-40$\% accuracy (1$\sigma$), decreasing from 1.8(+8,-6)\tdix{22}\cc\ toward the peak to 1.4(+8,-5)\tdix{22}\cc, toward $(+80",-40")$. It is worth mentioning that background subtraction changes the minimized values only within 1$\sigma$.
        
We note that the 35"-averaged peak column density is thus 1.5 times smaller than the 26"-beam average value derived from the \hh\ density profile of \cite{magalhaes2018a}, 2.8\tdix{22}\cc {, and a factor of two below the value from \cite{tafalla2004a} based on 1.3mm data only.} Nevertheless, we consider this as a good agreement, and will adopt the here derived column density, taking into account that the profile obtained by Magalhaes et al is based on azimuth averaging.

\subsection{L1544}
\begin{figure*}
        \centering
        \vfig{6}{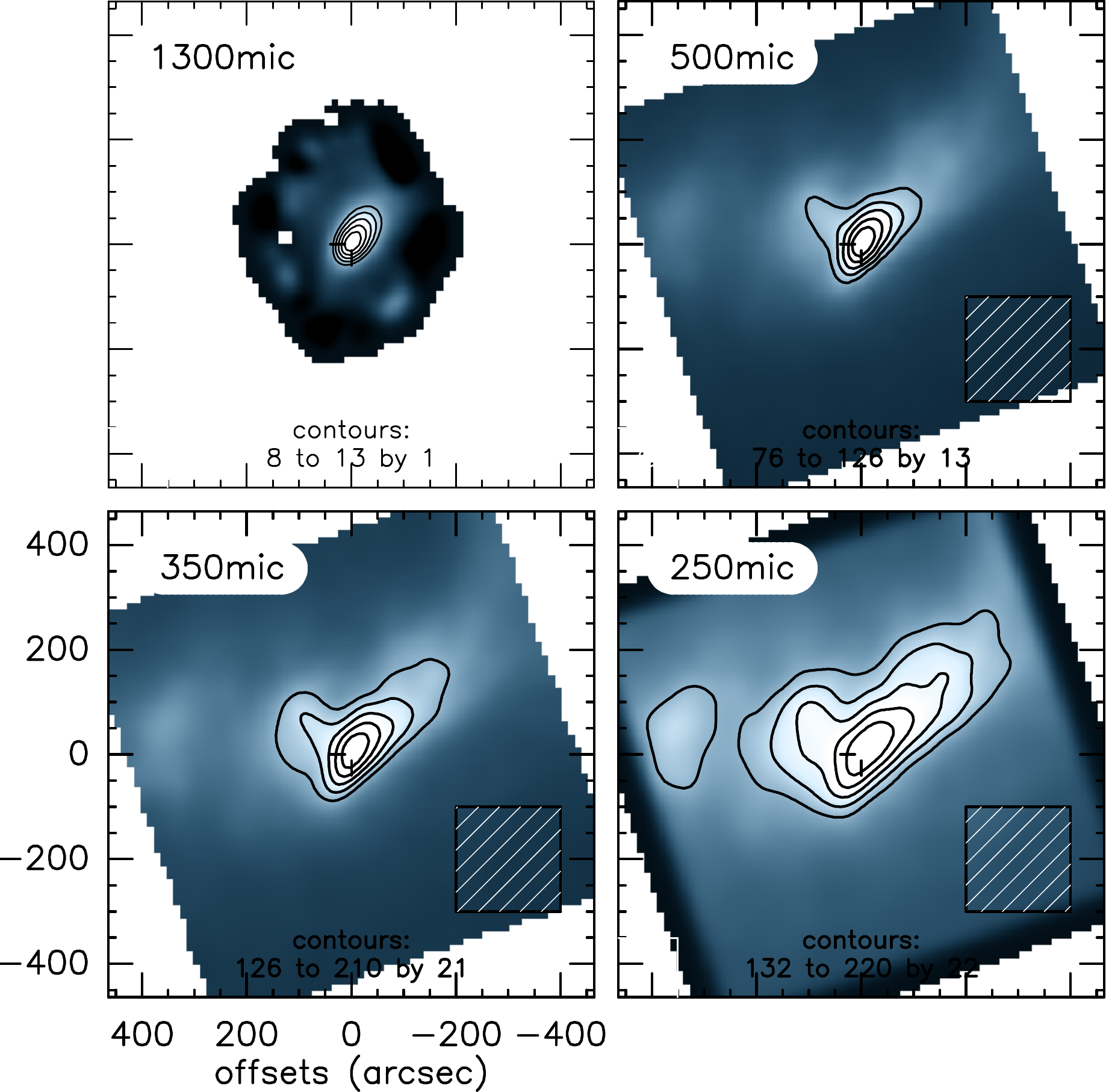}\hfill%
        \vfig{6}{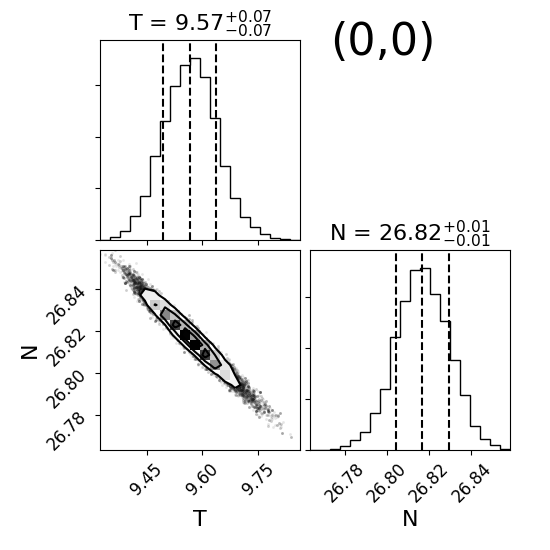}\hfill%
        \vfig{6}{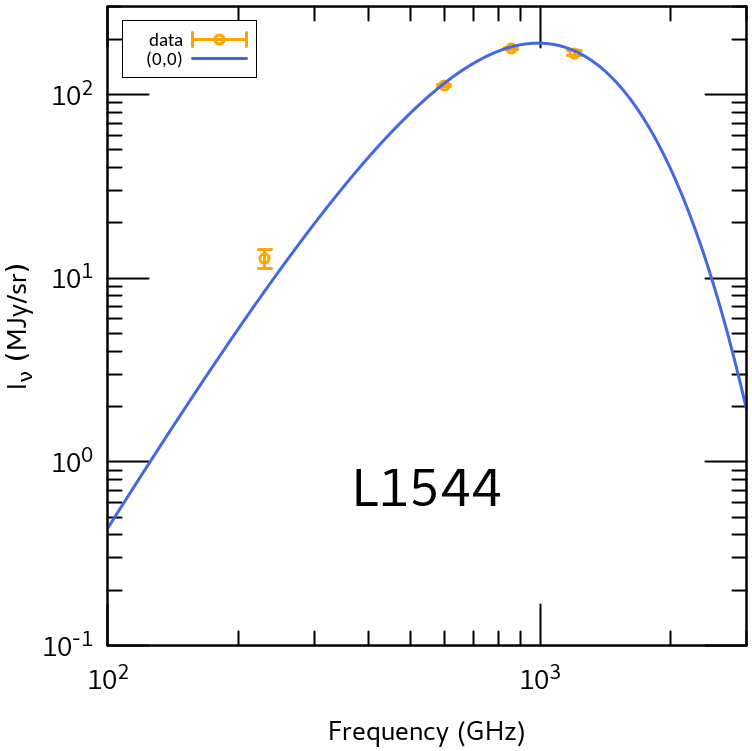}
        \caption{Determination of \Nhh\ in L1544 (see Fig.~\ref{fig:NH2_l1521e}). The spectral index was held fixed with $\beta=2$.}
        \label{fig:NH2_l1544}
\end{figure*}

Computing the \hh\ column density toward L1544 could in principle be done from the source model used by \cite{bizzocchi2013}, itself adopted from \cite{keto2010b}. This density profile has a similar form as for L1521E and L1498, with $n_0=2\tdix{7}\ccc$, $r_0=270$~au (or 1.9", based on Fig. 1 of Keto et al, although no value is provided in these references), and $\alpha=2$. The 26"-beam averaged \hh\ column density is thus 8.9\tdix{22}\cc, which is larger than the value of 6.6\tdix{22}\cc\ reported by Bizzocchi et al. However, a more observation-based determination of \Nhh\ can be obtained as for the other three sources. In contrast with the other sources, the spectral index was kept fixed, $\beta=2$. If not, the most likely index is $\approx 1.6$ and the dust temperature is  $\approx 11$~K: since at long wavelength, $I_\nu \sim \nu^{2+\beta}$, this is likely a result of the 1.3mm intensity being overestimated (no background estimate). Instead, keeping $\beta$ equal to 2 leads to dust temperatures $\approx 10$~K. The 35"-beam--averaged column densities are summarized in Table~\ref{tab:sed}, with a peak value of 6.6\tdix{22}~\cc, in remarkable agreement with the value from Bizzocchi et al. Varying the spectral index by $\pm0.2$, the dust temperature varies by 0.5~K, between 10.1 and 9.1 for $\beta=1.8$ and 2.2 respectively, with corresponding column densities 22.68(1) and 22.95(1) {dex}. We thus adopt a peak column density $\Nhh = 6.6^{+2.3}_{-1.8} \tdix{22}\cc$.

\section{Analysis of NS lines}
\label{app:gfit}

The hyperfine lines of each manifold were fitted independently, with no constraints on their relative velocity or line width (FWHM). In all cases, single-component Gaussian fits were used, except toward offsets $(0",0")$ and $(-20",20")$ in L1544 for which two-component Gaussian fits have been applied. The results, for each source, are shown in Fig.~\ref{fig:gfit_l1521e}--\ref{fig:gfit_l1544} and the line properties are provided in Tables~\ref{tab:l1521e_gfit}--\ref{tab:l1544_gfit}.

\def\wh{0.65\hsize}
\begin{figure}
        \centering
        \hfig{0.45}{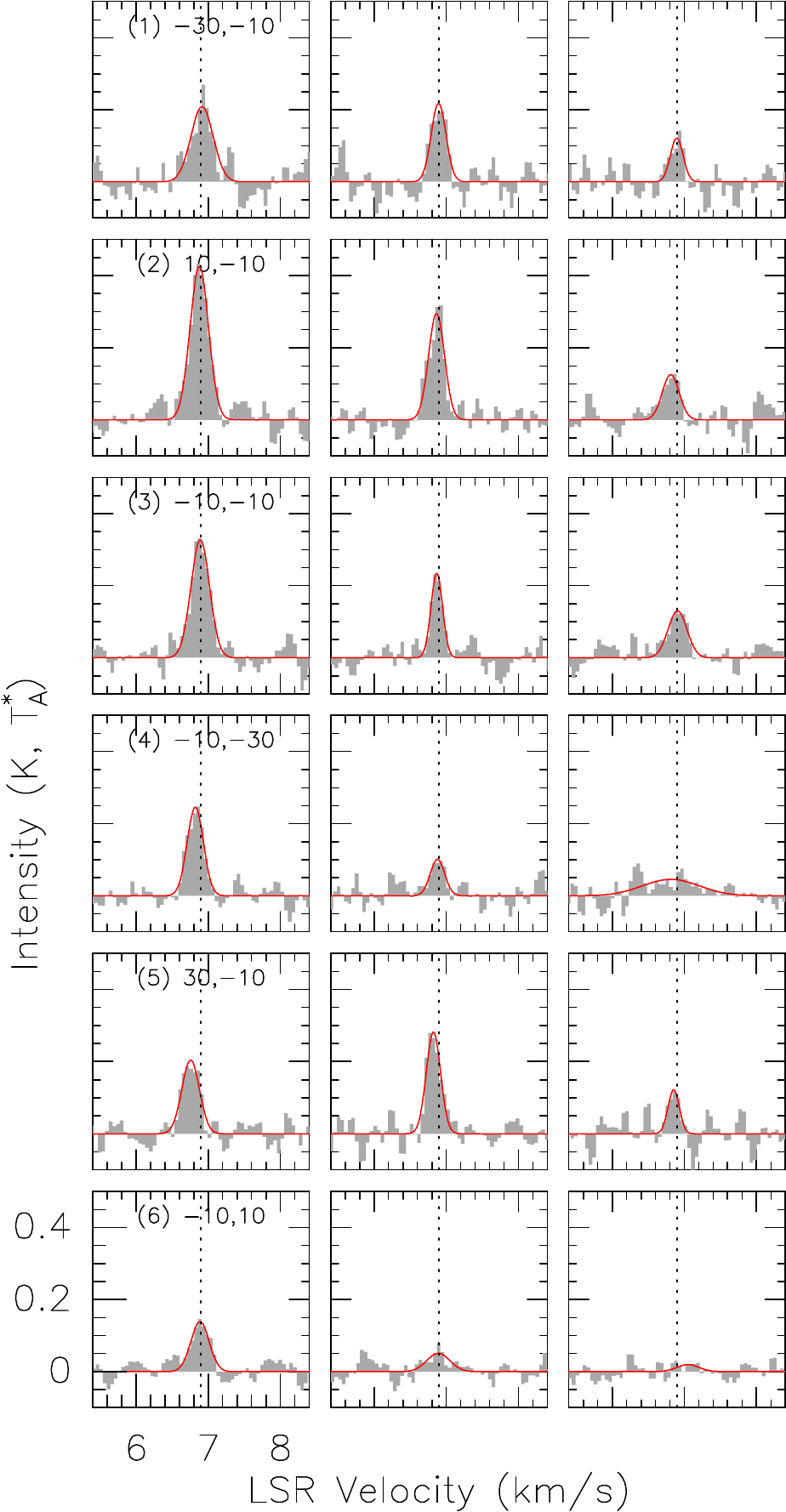}\hfill%
        \hfig{0.45}{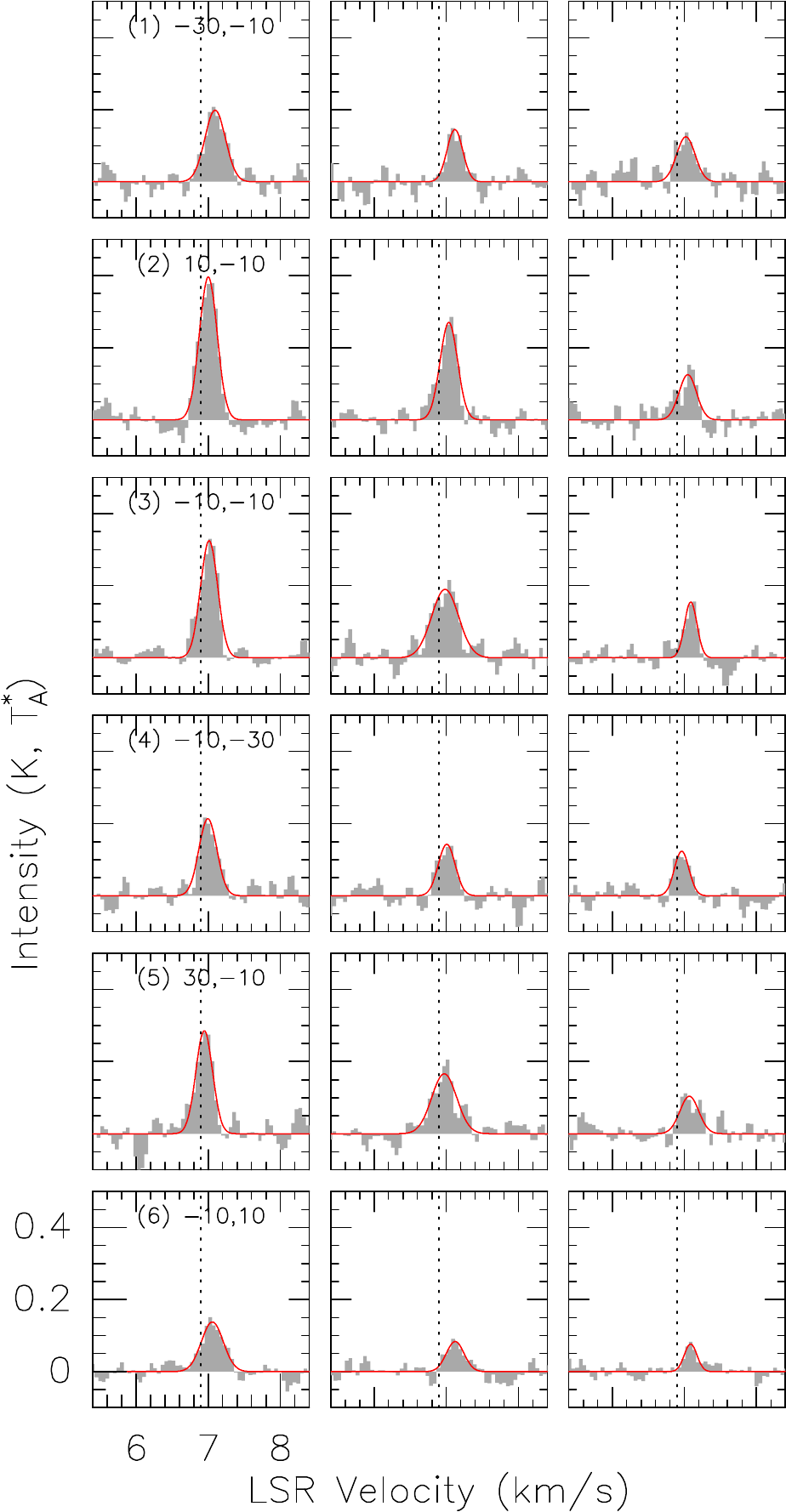}\bigskip\\
        \hfig{0.45}{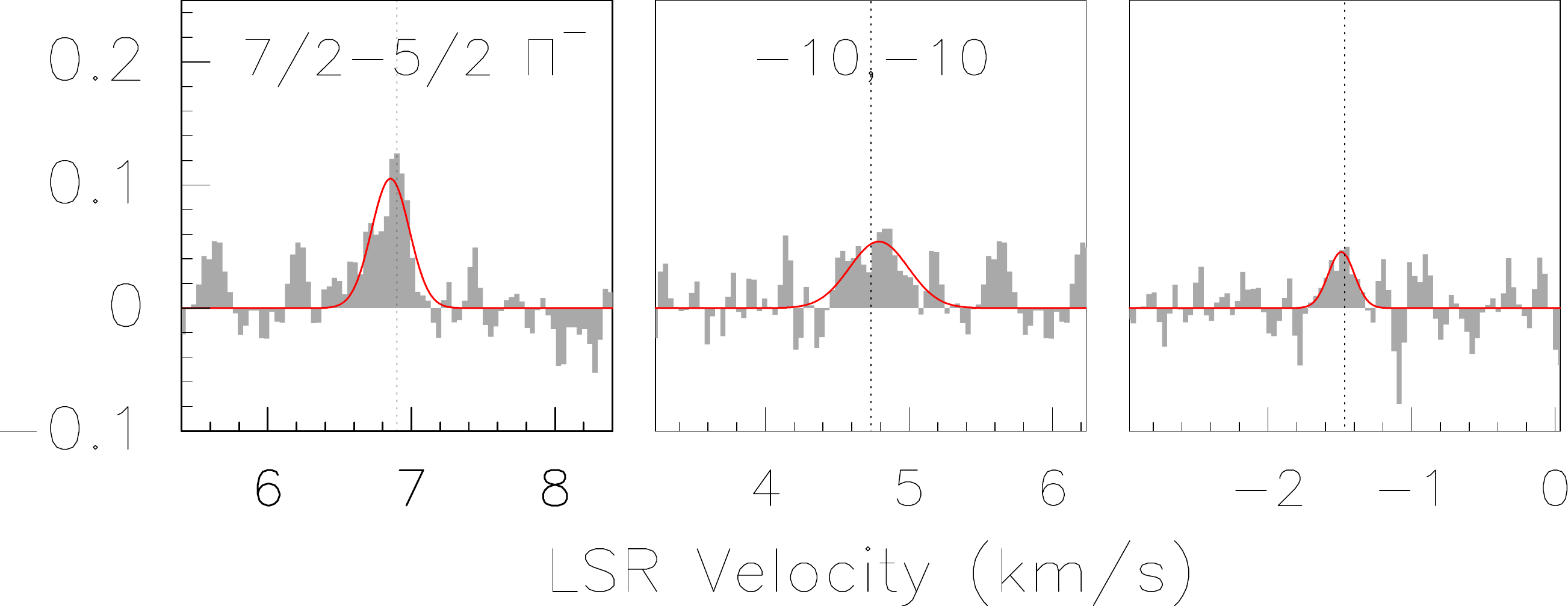}\hfill%
        \hfig{0.45}{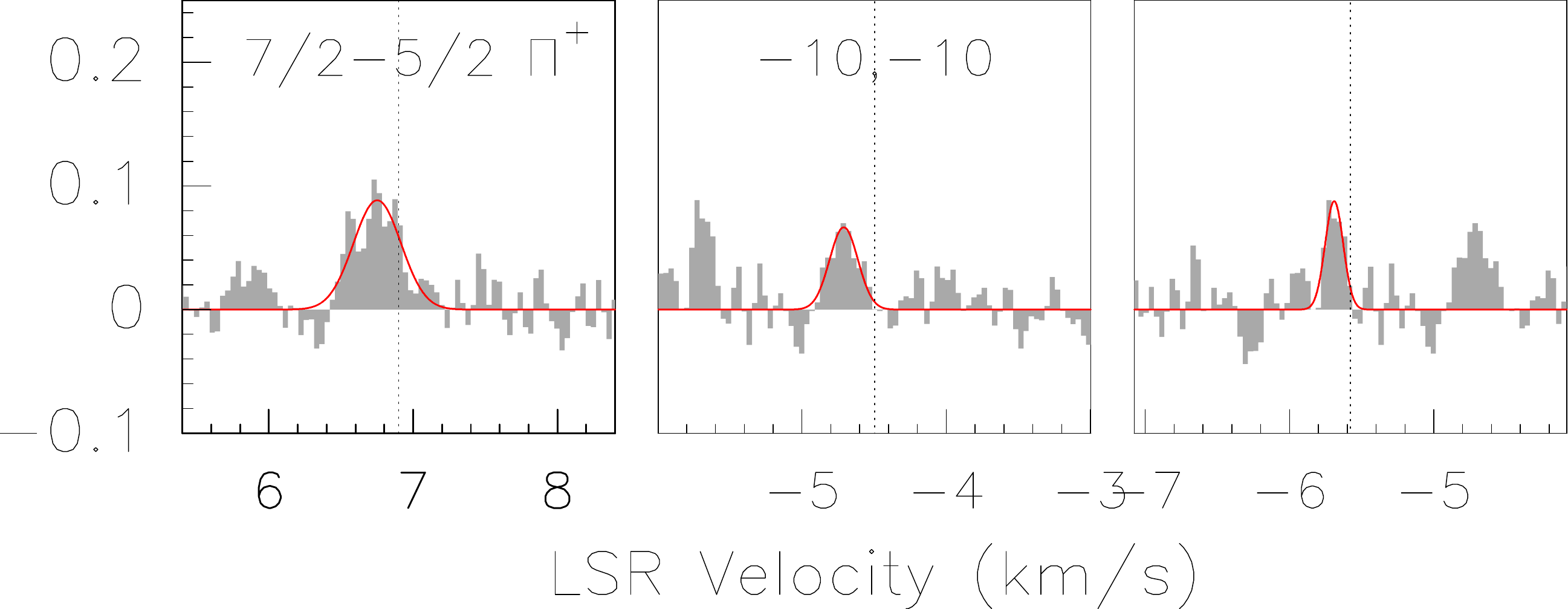}
        \caption{Results of the Gaussian fitting to the $J=5/2-3/2$ $\Pi^-$ ({top} left) and $\Pi^+$ ({top} right) {and $J=7/2-5/2$ $\Pi^-$ ({bottom} left) and $\Pi^+$ ({bottom} right)} emission lines of NS toward L1521E (\tant\ scale). For each multi-fold, the three hyperfine lines (see Fig.~\ref{fig:spectro}) are shown separately. In each panel, the vertical dashed line indicates the systemic velocity of 6.9\kms, and the Gaussian fit is shown in red.}
        \label{fig:gfit_l1521e}
\end{figure}
\begin{figure}
        \centering
        \hfig{0.45}{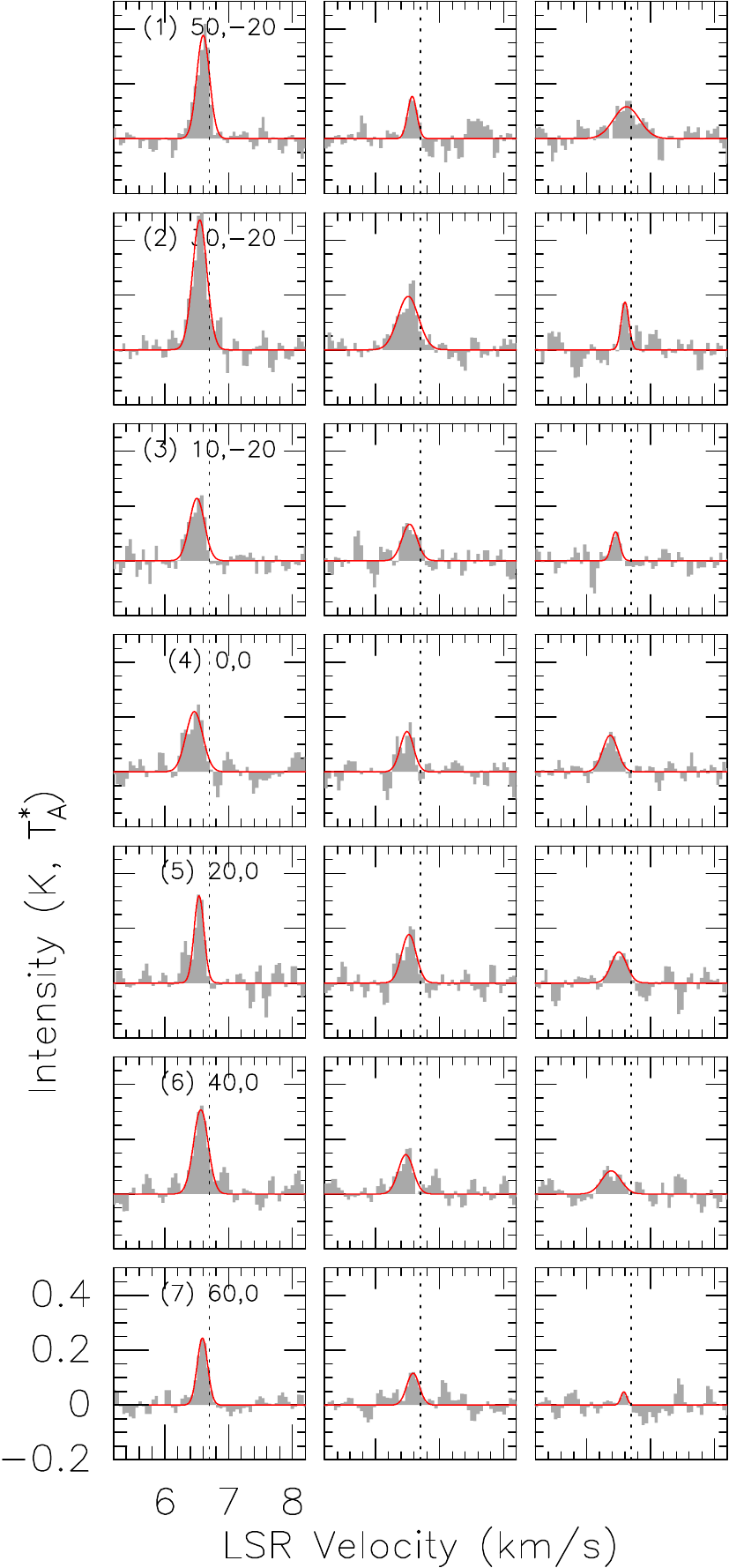}\hfill%
        \hfig{0.45}{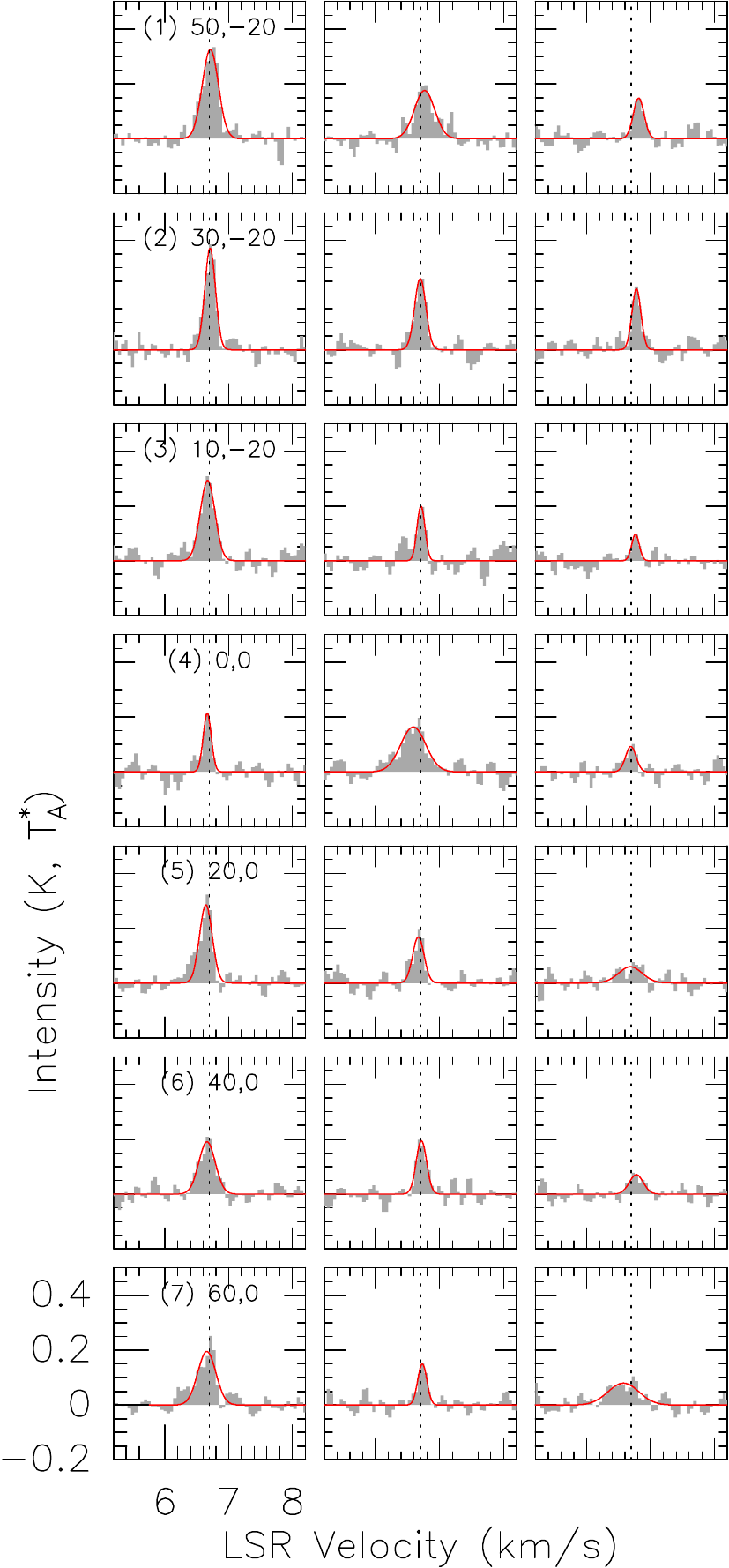}
        \caption{Results of the Gaussian fitting for the $J=5/2-3/2$ $\Pi^-$ (left) and $\Pi^+$ (right) emission lines of NS toward L1521B (see Fig.~\ref{fig:gfit_l1521e}) . The systemic velocity is 6.7 \kms.}
        \label{fig:gfit_l1521b}
\end{figure}
\begin{figure*}
        \centering
        \vfig{9}{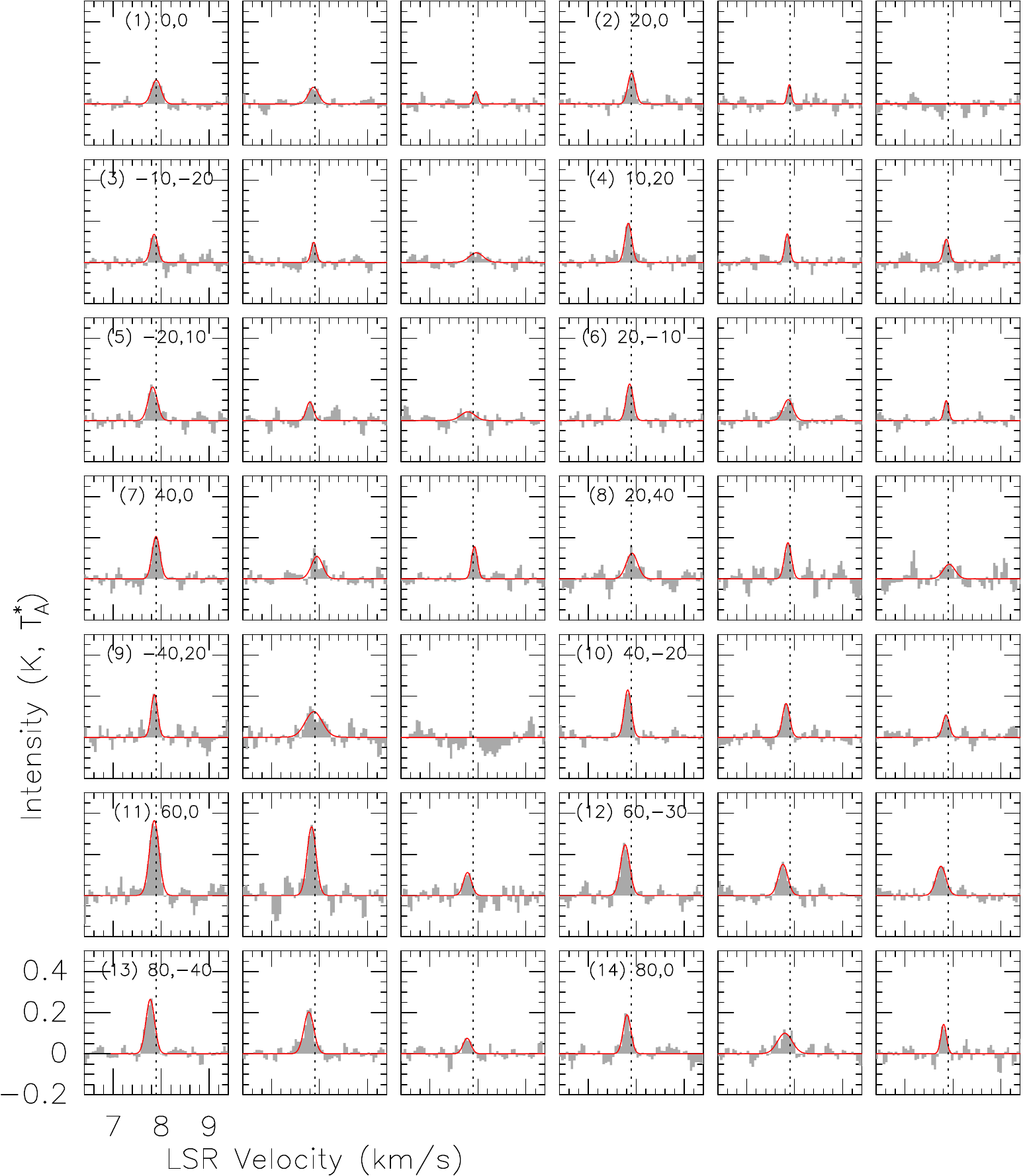}\hspace{1cm}
        \vfig{9}{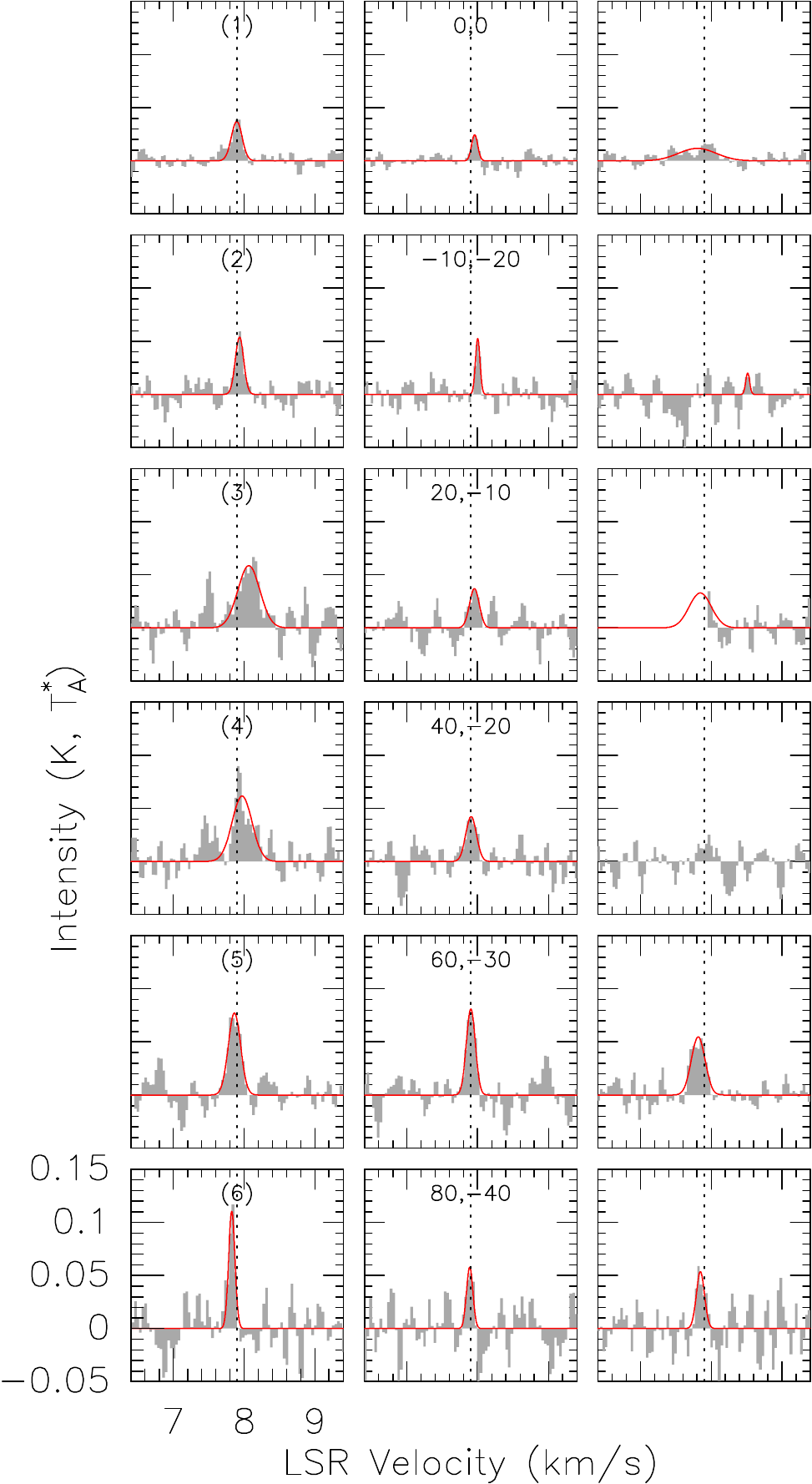}\bigskip\\
        \vfig{9}{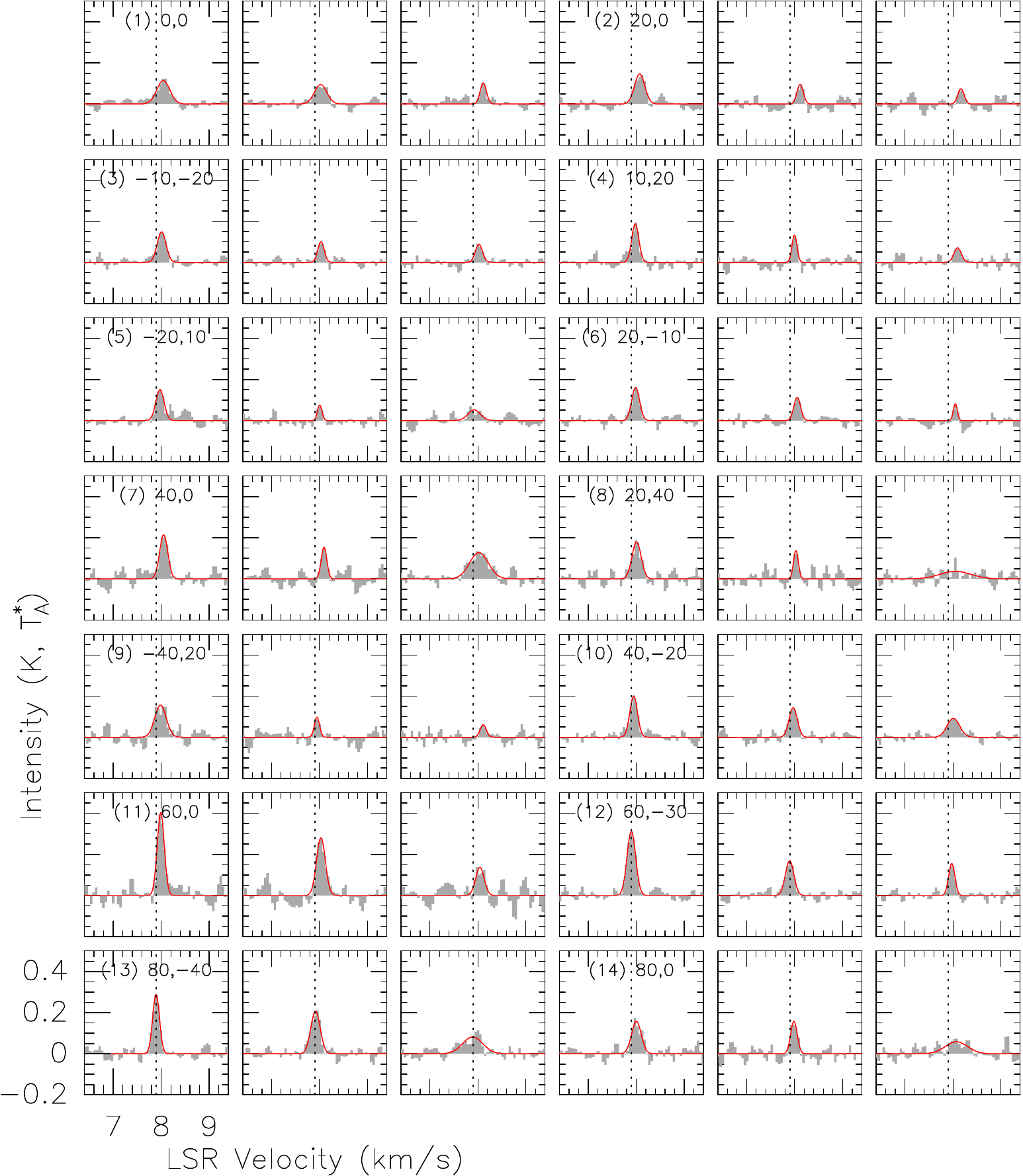}\hspace{1cm}
        \vfig{9}{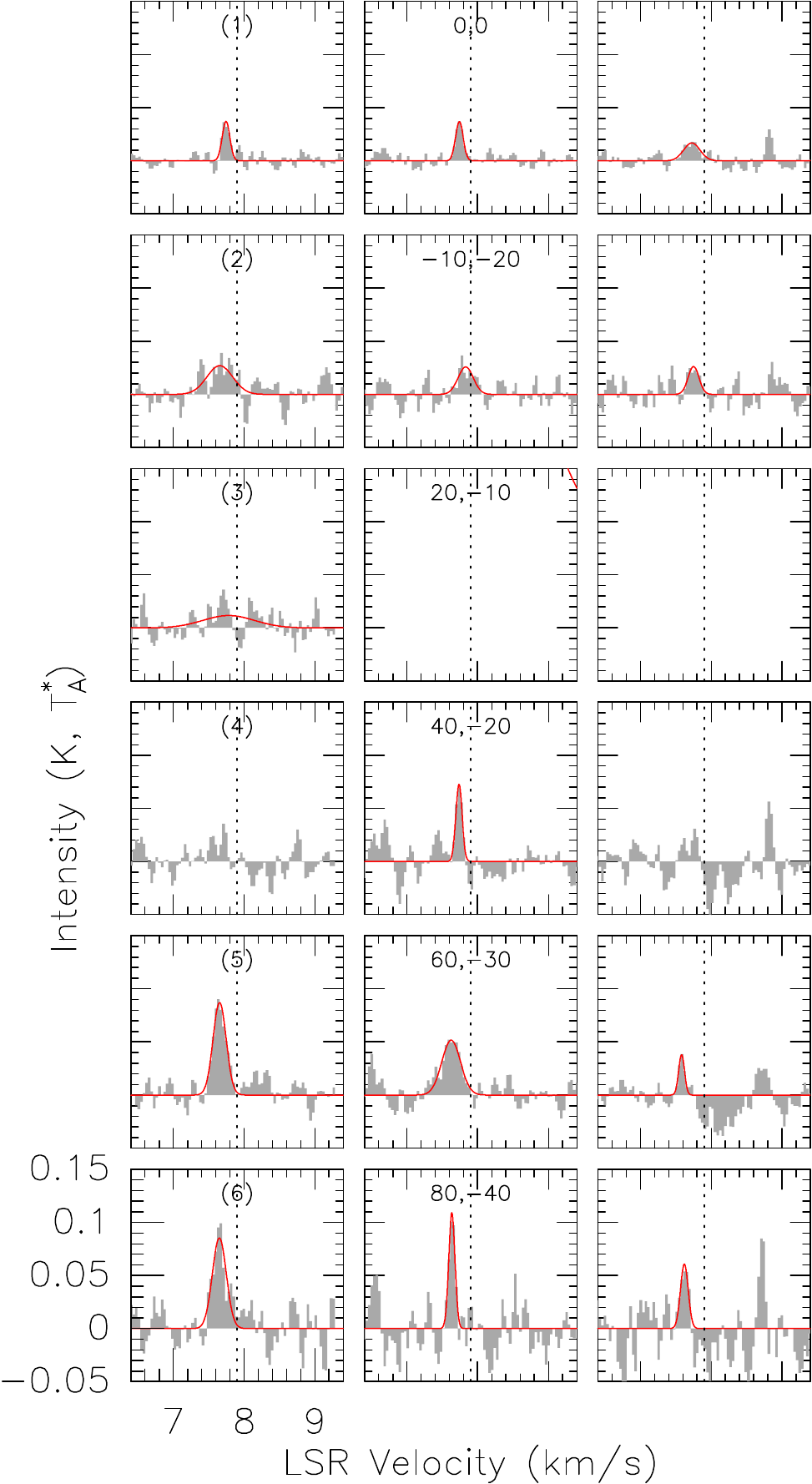}\bigskip\\
        \caption{Results of the Gaussian fitting for the NS lines toward L1498. \textit{Top:}  $J=5/2-3/2$ (left) and $7/2-5/2$ (right) $\Pi^-$. \textit{Bottom:} same as top row for the $5/2-3/2$ and $7/2-5/2$ $\Pi^+$ manifolds. The systemic velocity is 7.9\kms.}
        \label{fig:gfit_l1498}
\end{figure*}
\def\wa{0.65\hsize}
\begin{figure}
        \centering
        \hfig{0.45}{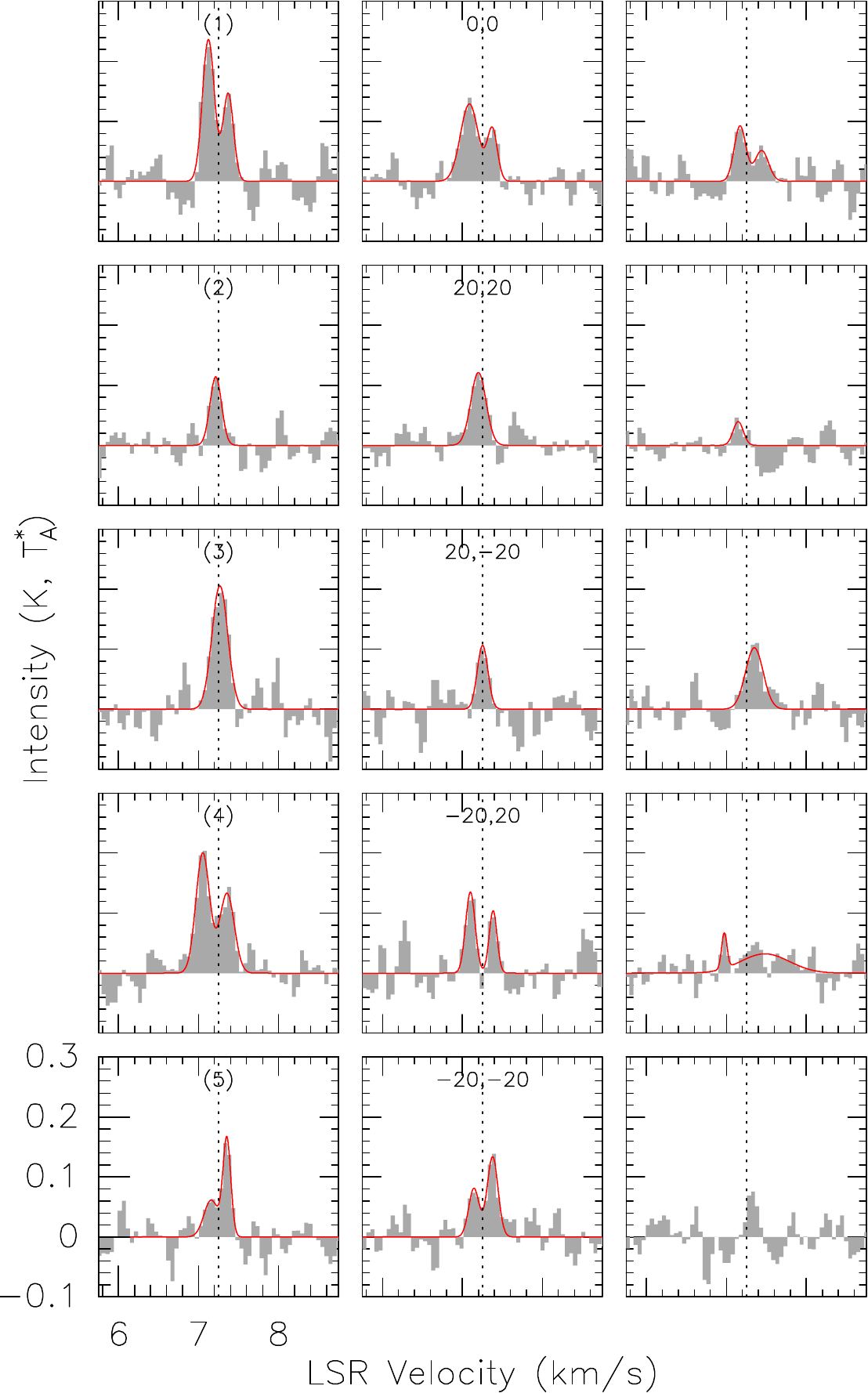}\hfill%
        \hfig{0.45}{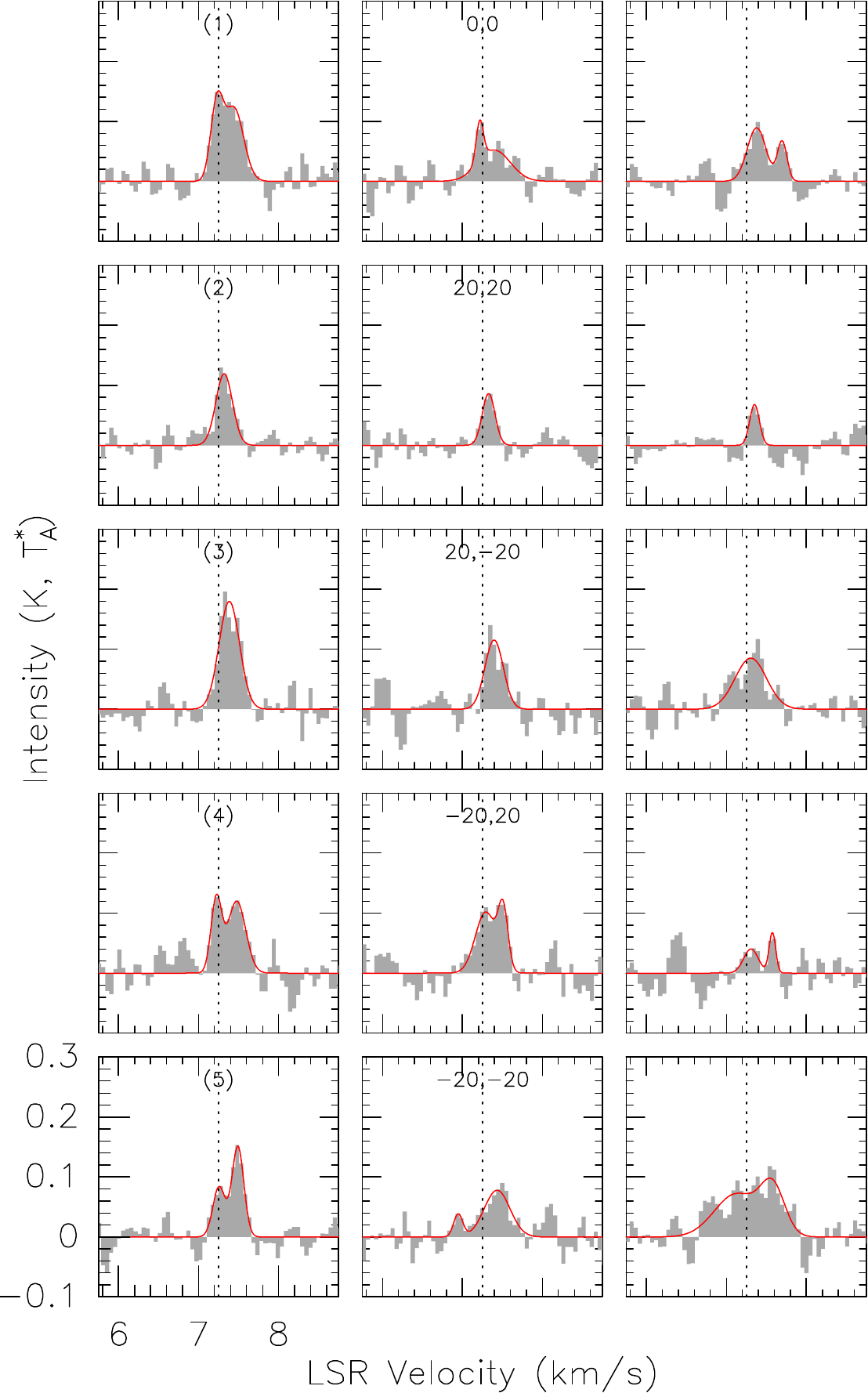}
        \caption{Results of the Gaussian fitting for the $J=5/2-3/2$ $\Pi^-$ (left) and $\Pi^+$ (right) emission lines of NS in L1544 (see Fig.~\ref{fig:gfit_l1521e}). A 2-components Gaussian profile was adopted. The systemic velocity is 7.25\,\kms.}
        \label{fig:gfit_l1544}
\end{figure}

\begin{table*}[htbp]
  \centering
  \scriptsize
  \caption{\label{tab:l1521e_gfit} Line properties derived from
    Gaussian fits to the spectra toward L1521E. At each offsets, the
    three hf lines of each manifold are given separately.}
  \centering
  \begin{tabular}{rrrrr rrrr}
    \toprule
    Offsets & $W$     & \vlsr & FWHM & $T$ & $W$     & \vlsr & FWHM & $T$\\
        "," & mK \kms & \kms  & \kms & mK  & mK \kms & \kms  & \kms & mK \\
    \midrule
        & \mc{4}{$J=5/2-3/2, \Pi^-$} & \mc{4}{$J=5/2-3/2, \Pi^+$}\\
-30,-10 & 97(11) &   6.91(2) &   0.36(6) &  251(50) & 86(7) &   7.09(1) &   0.34(3) &  239(36)  \\ 
-30,-10 &  68(7) &   6.89(1) &   0.24(2) &  262(47) & 48(6) &   7.12(2) &   0.26(4) &  175(37)  \\ 
-30,-10 &  32(7) &   6.89(2) &   0.21(5) &  146(47) & 48(7) &   7.02(2) &   0.30(5) &  151(36)  \\ 
 10,-10 & 165(8) &   6.88(1) &   0.30(2) &  512(41) & 147(7) &   7.00(1) &   0.29(1) &  477(38) \\ 
 10,-10 &  98(7) &   6.86(1) &   0.26(2) &  355(39) & 99(6) &   7.04(1) &   0.29(2) &  326(30)  \\ 
 10,-10 &  44(9) &   6.81(3) &   0.28(6) &  151(48) & 46(7) &   7.05(3) &   0.29(6) &  151(34)  \\ 
-10,-10 & 128(7) &   6.89(1) &   0.30(2) &  394(39) & 117(5) &   7.01(1) &   0.28(2) &  391(30) \\ 
-10,-10 &  57(5) &   6.87(1) &   0.19(2) &  282(37) & 107(8) &   6.98(2) &   0.44(4) &  229(34) \\ 
-10,-10 &  52(8) &   6.91(2) &   0.31(6) &  157(39) & 40(5) &   7.09(1) &   0.20(3) &  186(30)  \\ 
-10,-30 &  84(5) &   6.82(1) &   0.27(2) &  296(32) & 80(6) &   7.00(1) &   0.29(3) &  257(32)  \\ 
-10,-30 &  31(6) &   6.88(2) &   0.24(6) &  122(36) & 50(6) &   7.01(2) &   0.28(3) &  172(36)  \\ 
-10,-30 & 54(12) &  6.81(11) &  0.92(19) &   54(39) & 38(4) &   6.96(1) &   0.24(3) &  149(29)  \\ 
 30,-10 &  75(7) &   6.75(1) &   0.29(3) &  246(37) & 100(8) &   6.94(1) &   0.27(3) &  344(42) \\ 
 30,-10 &  84(7) &   6.82(1) &   0.23(2) &  340(43) & 87(8) &   6.97(2) &   0.41(5) &  200(34)  \\ 
 30,-10 &  29(7) &   6.85(2) &   0.18(4) &  147(52) & 41(7) &   7.07(3) &   0.31(5) &  125(37)  \\ 
-10, 10 &  53(6) &   6.88(2) &   0.30(4) &  167(34) & 63(5) &   7.06(1) &   0.36(3) &  166(24)  \\ 
-10, 10 &  25(6) &   6.89(5) &  0.38(12) &   61(27) & 31(4) &   7.12(2) &   0.29(4) &  101(24)  \\ 
-10, 10 &   9(5) &  7.05(12) &  0.36(20) &   23(29) & 20(4) &   7.09(2) &   0.21(7) &   92(22)  \\ 
\smallskip\\
        & \mc{4}{$J=7/2-3/2, \Pi^-$}& \mc{4}{$J=7/2-3/2, \Pi^+$}\\
-10,-10 & 46(5) &   6.86(2) &   0.31(5) &  141(31)  & 49(5) &   6.75(2) &   0.39(4) &  118(30) \\ 
-10,-10 & 37(6) &   4.79(4) &   0.48(10) &   72(31) & 22(4) &  -4.71(2) &   0.23(4) &   89(30) \\ 
-10,-10 & 14(4) &  -1.49(3) &   0.21(7) &   60(31)  & 18(3) &  -5.69(1) &   0.15(3) &  117(30) \\ 
\bottomrule
\end{tabular}
\tabnote 1$\sigma$ uncertainty (in brackets) are given in units of the last digit.
\end{table*}

\begin{table*}[htbp]
  \centering
  \scriptsize
  \caption{\label{tab:l1521b_gfit}Gaussian fit results for L1521B (see Table~\ref{tab:l1521e_gfit}).}
  \centering
  \begin{tabular}{rrrrr rrrr}
    \toprule
    Offsets & $W$     & \vlsr & FWHM & $T$ & $W$     & \vlsr & FWHM & $T$\\
        "," & mK \kms & \kms  & \kms & mK  & mK \kms & \kms  & \kms & mK \\
    \midrule
        & \mc{4}{$J=5/2-3/2, \Pi^-$} & \mc{4}{$J=5/2-3/2, \Pi^+$}\\
 50,-20 & 114(8) &   6.60(1) &   0.24(2) &  454(45) &124(8) &   6.71(1) &   0.30(2) &  392(39) \\ 
 50,-20 & 34(7) &   6.57(2) &   0.17(3) &  186(49)  & 84(9) &   6.76(2) &   0.37(5) &  212(36) \\ 
 50,-20 & 72(10) &   6.62(3) &   0.48(8) &  140(44) & 38(5) &   6.81(1) &   0.20(3) &  178(33) \\ 
 30,-20 & 162(8) &   6.55(1) &   0.27(2) &  570(43) & 93(6) &   6.71(1) &   0.20(2) &  446(37) \\ 
 30,-20 & 96(8) &   6.51(2) &   0.39(4) &  235(37)  & 69(6) &   6.70(1) &   0.21(2) &  312(37) \\ 
 30,-20 & 31(7) &   6.60(1) &   0.14(4) &  209(47)  & 48(6) &   6.78(1) &   0.17(3) &  268(37) \\ 
 10,-20 & 81(7) &   6.50(1) &   0.28(3) &  275(40)  &102(7) &   6.67(1) &   0.27(2) &  353(39) \\ 
 10,-20 & 48(7) &   6.53(2) &   0.28(4) &  161(42)  & 37(6) &   6.71(1) &   0.15(3) &  240(44) \\ 
 10,-20 & 21(6) &   6.45(2) &   0.15(5) &  128(49)  & 18(5) &   6.76(2) &   0.14(4) &  117(36) \\ 
  0,  0 & 88(9) &   6.46(2) &   0.31(3) &  265(50)  & 41(5) &   6.67(1) &   0.15(2) &  257(35) \\ 
  0,  0 & 47(6) &   6.49(2) &   0.25(3) &  177(41)  & 97(8) &   6.59(2) &   0.46(5) &  197(33) \\ 
  0,  0 & 47(7) &   6.37(2) &   0.28(5) &  161(42)  & 23(6) &   6.69(3) &   0.19(7) &  111(33) \\ 
 20,  0 & 74(9) &   6.54(1) &   0.18(3) &  383(50)  & 86(7) &   6.64(1) &   0.24(3) &  343(36) \\ 
 20,  0 & 61(8) &   6.52(2) &   0.27(4) &  213(47)  & 45(6) &   6.67(2) &   0.21(4) &  202(35) \\ 
 20,  0 & 41(8) &   6.50(3) &   0.28(5) &  136(48)  & 33(7) &   6.68(4) &   0.43(9) &   71(30) \\ 
 40,  0 &107(9) &   6.57(1) &   0.27(3) &  370(47)  & 74(6) &   6.66(1) &   0.30(3) &  229(33) \\ 
 40,  0 & 51(8) &   6.47(2) &   0.28(7) &  173(40)  & 48(5) &   6.72(1) &   0.19(3) &  234(36) \\ 
 40,  0 & 38(7) &   6.38(4) &   0.35(7) &  102(37)  & 23(5) &   6.78(3) &   0.26(5) &   84(29) \\ 
 60,  0 & 60(5) &   6.59(1) &   0.19(2) &  294(34)  & 83(8) &   6.66(1) &   0.33(5) &  236(31) \\ 
 60,  0 & 35(8) &   6.58(2) &   0.23(8) &  141(41)  & 32(4) &   6.73(1) &   0.17(3) &  180(31) \\ 
 60,  0 &  7(4) &   6.58(4) &   0.11(7) &   57(39)  & 58(9) &   6.58(4) &  0.57(10) &   96(33) \\ 
    \bottomrule
  \end{tabular}
\tabnote 1$\sigma$ uncertainty (in brackets) are given in units of the last digit.
\end{table*}

\begin{table*}[htbp]
  \centering
  \scriptsize
  \caption{\label{tab:l1498_gfit_53}Gaussian fit results for L1498 (see Table~\ref{tab:l1521e_gfit}).}
  \centering
  \begin{tabular}{rrrrr rrrr}
    \toprule
    Offsets & $W$     & \vlsr & FWHM & $T$ & $W$     & \vlsr & FWHM & $T$\\
        "," & mK \kms & \kms  & \kms & mK  & mK \kms & \kms  & \kms & mK \\
    \midrule
    & \mc{4}{$J=5/2-3/2, \Pi^-$} & \mc{4}{$J=5/2-3/2, \Pi^+$}\\
  0,  0 & 35(4) &   7.90(1) &   0.24(2) &  141(23) & 47(4) &   8.04(1) &   0.32(4) &  137(21) \\
  0,  0 & 26(4) &   7.88(2) &   0.25(4) &   98(24) & 37(3) &   8.02(1) &   0.30(3) &  114(17) \\
  0,  0 &  9(3) &   7.96(2) &   0.11(3) &   73(24) & 20(3) &   8.11(1) &   0.15(4) &  123(24) \\
-10,-20 & 30(4) &   7.86(1) &   0.17(2) &  164(30) & 41(3) &   8.01(1) &   0.22(2) &  178(19) \\
-10,-20 & 16(3) &   7.88(1) &   0.13(3) &  117(25) & 20(2) &   8.03(1) &   0.16(2) &  120(19) \\
-10,-20 & 20(4) &   7.97(4) &   0.34(7) &   55(24) & 20(3) &   8.02(1) &   0.17(3) &  106(19) \\
 10, 20 & 42(5) &   7.83(1) &   0.17(3) &  229(30) & 42(4) &   7.98(1) &   0.18(2) &  226(25) \\
 10, 20 & 22(3) &   7.85(1) &   0.12(2) &  167(27) & 21(3) &   8.00(1) &   0.12(2) &  160(28) \\
 10, 20 & 21(4) &   7.86(1) &   0.14(2) &  135(31) & 16(4) &   8.10(2) &   0.18(5) &   84(29) \\
-20, 10 & 48(6) &   7.82(1) &   0.23(3) &  197(35) & 39(5) &   7.98(1) &   0.20(4) &  181(29) \\
-20, 10 & 20(5) &   7.80(2) &   0.17(5) &  110(36) & 11(3) &   8.00(1) &   0.11(2) &   91(25) \\
-20, 10 & 19(6) &   7.79(6) &   0.35(8) &   51(31) & 21(5) &   7.92(4) &   0.31(7) &   62(26) \\
 20, 40 &45(10) &   7.91(3) &   0.28(8) &  150(50) & 52(8) &   8.01(2) &   0.22(4) &  216(50) \\
 20, 40 & 39(8) &   7.86(2) &   0.17(5) &  212(56) & 21(4) &   8.03(1) &   0.12(3) &  165(38) \\
 20, 40 & 27(9) &   7.92(5) &   0.29(9) &   85(51) &36(11) &  8.05(12) &  0.76(24) &   44(37) \\
-40, 20 & 43(6) &   7.86(1) &   0.16(3) &  250(46) & 56(7) &   7.99(2) &   0.28(3) &  188(40) \\
-40, 20 &66(10) &   7.88(3) &   0.42(6) &  149(47) & 16(4) &   7.95(2) &   0.13(3) &  117(35) \\
-40, 20 &  0(0) &   7.91(0) & 0.29(368) &    0(52) & 12(4) &   8.11(3) &   0.15(6) &   73(31) \\
 20,  0 & 36(4) &   7.90(1) &   0.18(3) &  181(26) & 47(5) &   8.07(1) &   0.25(3) &  176(31) \\
 20,  0 & 12(3) &   7.89(1) &   0.10(3) &  114(33) & 19(3) &   8.12(2) &   0.16(3) &  116(27) \\
 20,  0 &  0(0) &   8.29(0) &   1.73(0) &    0(36) & 15(4) &   8.17(2) &   0.15(4) &   89(32) \\
 40,  0 & 51(5) &   7.89(1) &   0.19(2) &  249(36) & 55(5) &   8.05(1) &   0.20(2) &  259(35) \\
 40,  0 & 39(6) &   7.95(2) &   0.28(5) &  132(33) & 27(4) &   8.09(1) &   0.14(2) &  186(36) \\
 40,  0 & 29(4) &   7.93(1) &   0.14(3) &  189(33) & 76(7) &   8.02(2) &   0.46(4) &  154(29) \\
 60,  0 &110(9) &   7.86(1) &   0.24(2) &  439(55) & 90(7) &   7.99(1) &   0.17(2) &  486(48) \\
 60,  0 &96(10) &   7.84(1) &   0.22(3) &  403(58) & 78(8) &   8.03(1) &   0.22(2) &  338(49) \\
 60,  0 & 28(8) &   7.78(3) &   0.20(6) &  135(53) & 37(7) &   8.04(2) &   0.21(4) &  165(50) \\
 80,  0 & 46(6) &   7.81(1) &   0.19(3) &  229(39) & 48(5) &   8.00(1) &   0.24(3) &  190(34) \\
 80,  0 & 46(8) &   7.80(3) &   0.36(7) &  120(39) & 32(4) &   7.99(1) &   0.16(2) &  190(34) \\
 80,  0 & 25(5) &   7.81(1) &   0.14(2) &  172(39) & 41(8) &   8.07(6) &  0.55(14) &   70(32) \\
 20,-10 & 40(4) &   7.86(1) &   0.17(2) &  215(27) & 40(3) &   7.99(1) &   0.19(2) &  194(23) \\
 20,-10 & 32(4) &   7.87(2) &   0.24(4) &  123(24) & 24(3) &   8.05(1) &   0.16(2) &  136(19) \\
 20,-10 & 15(3) &   7.86(1) &   0.12(2) &  117(25) & 10(3) &   8.05(1) &   0.10(2) &   97(26) \\
 40,-20 & 52(4) &   7.82(1) &   0.18(1) &  277(27) & 48(3) &   7.95(1) &   0.19(2) &  241(21) \\
 40,-20 & 37(5) &   7.82(1) &   0.18(2) &  197(33) & 37(3) &   7.97(1) &   0.20(2) &  173(22) \\
 40,-20 & 22(5) &   7.86(2) &   0.16(4) &  130(36) & 36(4) &   8.01(2) &   0.31(4) &  111(21) \\
 60,-30 & 72(6) &   7.77(1) &   0.23(2) &  297(37) & 85(3) &   7.90(0) &   0.21(1) &  376(22) \\
 60,-30 & 46(6) &   7.76(1) &   0.24(4) &  183(34) & 45(4) &   7.90(1) &   0.21(2) &  204(26) \\
 60,-30 & 45(6) &   7.75(2) &   0.25(4) &  171(34) & 30(3) &   7.98(1) &   0.15(2) &  187(24) \\
 80,-40 & 73(4) &   7.77(1) &   0.21(1) &  319(28) & 62(4) &   7.89(1) &   0.17(1) &  348(29) \\
 80,-40 & 64(6) &   7.78(1) &   0.24(4) &  245(32) & 61(5) &   7.92(1) &   0.23(2) &  250(29) \\
 80,-40 & 19(5) &   7.78(2) &   0.19(6) &   91(31) & 52(6) &   7.89(3) &   0.49(6) &   98(24) \\
    \bottomrule
  \end{tabular}
\tabnote 1$\sigma$ uncertainty (in brackets) are given in units of the last digit.
\end{table*}

\begin{table*}[htbp]
  \centering
  \scriptsize
  \caption{\label{tab:l1498_gfit_75}Gaussian fit results for the $J=7/2-5/2$ manifolds toward L1498 (see Table~\ref{tab:l1498_gfit_53}).}
  \centering
  \begin{tabular}{rrrrr rrrr}
    \toprule
    Offsets & $W$     & \vlsr & FWHM & $T$ & $W$     & \vlsr & FWHM & $T$\\
        "," & mK \kms & \kms  & \kms & mK  & mK \kms & \kms  & \kms & mK \\
    \midrule
        & \mc{4}{$J=7/2-3/2, \Pi^-$}& \mc{4}{$J=7/2-3/2, \Pi^+$}\\
  0,  0 & 9(1) &   7.89(1) &   0.18(3) &   48(8)       & 7(1) &   7.74(1) &   0.13(1) &   48(7)      \\ 
  0,  0 & 4(1) &   7.96(1) &   0.10(2) &   32(7)       & 7(1) &   7.74(1) &   0.13(2) &   48(7)      \\ 
  0,  0 & 10(1) &   7.80(5) &   0.62(9) &   15(7)      & 6(1) &   7.73(3) &   0.27(8) &   22(9)      \\ 
-10,-20 & 11(2) &   7.93(1) &   0.14(3) &   70(14)     & 16(3) &   7.66(4) &   0.43(7) &   35(14)    \\ 
-10,-20 & 6(1) &   8.00(1) &   0.08(1) &   68(13)      & 9(2) &   7.83(3) &   0.25(5) &   33(13)     \\ 
-10,-20 & 2(1) &   8.52(2) &   0.07(6) &   26(18)      & 7(2) &   7.75(2) &   0.18(4) &   34(14)     \\ 
 20,-10 & 30(4) &   8.07(2) &   0.37(6) &   76(22)     & 14(4) &   7.78(15) &   0.86(26) &   15(16)  \\ 
 20,-10 & 8(2) &   7.95(2) &   0.16(4) &   48(18)      &   -- &     -- &     -- &      --      \\ 
 20,-10 & 17(5) &   7.84(3) &   0.37(18) &   --  & --   &   --   &   --   &      --      \\ 
 40,-20 & 28(5) &   7.97(2) &   0.33(9) &   80(20)     & 0(0) &   7.86(89) &   1.73(274) &    0(17)  \\ 
 40,-20 & 11(2) &   7.91(2) &   0.18(4) &   55(18)     & 11(2) &   7.74(1) &   0.10(2) &   95(19)    \\ 
 40,-20 & 0(0) &   8.84(62) &   0.52(153) &    0(19)   & 0(0) &   6.71(0) &   0.04(283) &    0(23)   \\ 
 60,-30 & 22(2) &   7.86(1) &   0.21(2) &  101(17)     & 27(2) &   7.65(1) &   0.22(2) &  114(13)    \\ 
 60,-30 & 17(2) &   7.90(1) &   0.15(2) &  106(19)     & 22(2) &   7.63(2) &   0.31(4) &   67(15)    \\ 
 60,-30 & 17(2) &   7.82(1) &   0.23(2) &   71(12)     & 5(2) &   7.58(2) &   0.10(4) &   49(19)     \\ 
 80,-40 & 17(2) &   7.83(1) &   0.11(2) &  148(24)     & 28(3) &   7.65(1) &   0.23(4) &  114(23)    \\ 
 80,-40 & 9(2) &   7.89(2) &   0.11(3) &   77(29)      & 16(3) &   7.63(1) &   0.10(2) &  146(28)    \\ 
 80,-40 & 10(3) &   7.85(2) &   0.14(3) &   71(26)     & 11(3) &   7.62(2) &   0.13(4) &   81(35)    \\ 
-40,-80 & 8(2) &   7.97(4) &   0.24(10) &   32(15)     & 0(0) &   8.03(67) &   0.34(288) &    0(14)  \\ 
-40,-80 & 0(0) &   8.24(0) &   2.12(163) &    0(14)    & 15(2) &   8.09(4) &   0.40(6) &   35(15)    \\ 
-40,-80 & 9(3) &   7.91(7) &   0.49(18) &   17(13)     & 6(1) &   8.13(1) &   0.10(2) &   57(16)     \\ 
-30,-60 & 9(5) &   7.13(8) &   0.25(9) &   35(41)      & 43(8) &   7.96(7) &   0.70(14) &   57(35)   \\ 
-30,-60 & 41(10) &   7.95(11) &   0.78(20) &   48(43)  & 7(3) &   8.09(2) &   0.08(4) &   86(46)     \\ 
-30,-60 & 0(0) &   8.14(72) &   2.23(0) &    0(40)     & 7(3) &   8.52(2) &   0.09(4) &   71(42)     \\ 
-20,-40 & 0(0) &   8.14(0) &   2.33(0) &    0(14)      & 6(1) &   7.76(2) &   0.16(4) &   36(13)     \\ 
-20,-40 & 17(3) &   7.93(9) &   0.83(16) &   19(13)    & 0(0) &   9.06(62) &   0.70(0) &    0(14)    \\ 
-20,-40 & 0(0) &   7.86(0) &   3.04(0) &    0(12)      & 0(0) &   7.91(119) &   0.20(274) &    0(14) \\ 
    \bottomrule
  \end{tabular}
\tabnote 1$\sigma$ uncertainty (in brackets) are given in units of the last digit.
\end{table*}

% build_gfit_textable.sh l1544-ns115M-gfit-1comp-2comp.dat
% build_gfit_textable.sh l1544-ns115P-gfit-1comp-2comp.dat
\begin{table*}[htbp]
  \centering
  \scriptsize
  \caption{\label{tab:l1544_gfit}Gaussian fit results for L1544 (see also Table~\ref{tab:l1521e_gfit}).}
  \centering
  \begin{tabular}{rrrrr rrrr}
    \toprule
    Offsets & $W$     & \vlsr & FWHM & $T$ & $W$     & \vlsr & FWHM & $T$\\
        "," & mK \kms & \kms  & \kms & mK  & mK \kms & \kms  & \kms & mK \\
    \midrule
    & \mc{4}{$J=5/2-3/2, \Pi^-$} & \mc{4}{$J=5/2-3/2, \Pi^+$}\\
  0,  0 & 54(6) &   7.12(1) &   0.18(2) &  285(36)  &29(21) &   7.23(4) &   0.18(7) &  147(23)\\ 
  0,  0 & 29(6) &   7.37(1) &   0.16(3) &  176(36)  &45(21) &   7.44(7) &  0.29(10) &  147(23)\\ 
  0,  0 & 44(7) &   7.09(2) &   0.26(6) &  156(30)  &  9(5) &   7.22(2) &   0.11(5) &   80(25)\\ 
  0,  0 & 16(6) &   7.37(3) &   0.15(5) &  103(30)  & 32(9) &   7.40(7) &  0.48(11) &   63(25)\\ 
  0,  0 & 20(6) &   7.17(3) &   0.17(5) &  111(35)  & 32(5) &   7.38(2) &   0.28(5) &  108(25)\\ 
  0,  0 & 13(7) &   7.44(5) &   0.20(14)&   61(35)  & 12(3) &   7.70(2) &   0.14(4) &   78(25)\\ 
 20, 20 & 25(4) &   7.21(1) &   0.17(3) &  138(27)  & 38(3) &   7.32(1) &   0.25(3) &  144(17)\\ 
 20, 20 & 35(4) &   7.20(1) &   0.22(3) &  146(25)  & 21(3) &   7.33(1) &   0.19(3) &  104(21)\\ 
 20, 20 &  8(4) &   7.15(4) &   0.15(9) &   47(26)  & 12(3) &   7.35(2) &   0.14(3) &   81(23)\\ 
 20,-20 & 62(6) &   7.27(1) &   0.23(2) &  248(36)  & 70(5) &   7.38(1) &   0.30(2) &  216(26)\\ 
 20,-20 & 21(4) &   7.25(2) &   0.16(3) &  128(34)  & 40(6) &   7.40(2) &   0.27(4) &  139(32)\\ 
 20,-20 & 34(6) &   7.35(2) &   0.26(6) &  123(33)  & 51(5) &   7.31(2) &   0.46(5) &  103(23)\\ 
-20, 20 & 52(7) &   7.05(1) &   0.20(3) &  241(29)  & 25(9) &   7.22(2) &   0.16(5) &  147(36)\\ 
-20, 20 & 37(7) &   7.36(2) &   0.22(5) &  160(29)  &41(10) &   7.48(3) &   0.26(7) &  145(36)\\ 
-20, 20 & 25(4) &   7.10(1) &   0.14(3) &  163(35)  &40(11) &   7.29(4) &   0.31(9) &  123(32)\\ 
-20, 20 & 16(4) &   7.38(2) &   0.12(3) &  125(35)  & 18(9) &   7.51(2) &   0.14(4) &  116(32)\\ 
-20, 20 &  6(4) &   6.97(3) &   0.08(4) &   71(39)  & 11(5) &   7.31(5) &   0.22(9) &   48(31)\\ 
-20, 20 &30(11) &  7.48(14) &  0.72(24) &   38(39)  &  8(3) &   7.58(2) &   0.10(4) &   80(31)\\ 
-20,-20 &17(11) &   7.16(5) &  0.21(18) &   74(32)  & 19(5) &   7.26(3) &   0.18(5) &  100(27)\\ 
-20,-20 & 24(8) &   7.35(1) &   0.12(3) &  194(32)  & 34(5) &   7.49(1) &   0.17(3) &  182(27)\\ 
-20,-20 & 16(5) &   7.14(3) &   0.15(5) &   97(34)  &  7(4) &   6.95(3) &   0.13(9) &   45(27)\\ 
-20,-20 & 27(7) &   7.38(1) &   0.16(5) &  161(34)  & 38(6) &   7.43(3) &   0.38(7) &   93(27)\\ 
-20,-20 &  0(0) &  6.22(13) &  0.05(86) &    2(36)  &60(21) &  7.13(13) &  0.65(16) &   86(32)\\ 
-20,-20 &  5(6) &   8.26(3) &   0.09(8) &   51(36)  &35(21) &   7.58(6) &  0.36(11) &   93(32)\\ 
%   0,  0 &29(21) &   7.23(4) &   0.18(7) &  147(23) \\ 
%   0,  0 &45(21) &   7.44(7) &  0.29(10) &  147(23) \\ 
%   0,  0 &  9(5) &   7.22(2) &   0.11(5) &   80(25) \\ 
%   0,  0 & 32(9) &   7.40(7) &  0.48(11) &   63(25) \\ 
%   0,  0 & 32(5) &   7.38(2) &   0.28(5) &  108(25) \\ 
%   0,  0 & 12(3) &   7.70(2) &   0.14(4) &   78(25) \\ 
%  20, 20 & 38(3) &   7.32(1) &   0.25(3) &  144(17) \\ 
%  20, 20 & 21(3) &   7.33(1) &   0.19(3) &  104(21) \\ 
%  20, 20 & 12(3) &   7.35(2) &   0.14(3) &   81(23) \\ 
%  20,-20 & 70(5) &   7.38(1) &   0.30(2) &  216(26) \\ 
%  20,-20 & 40(6) &   7.40(2) &   0.27(4) &  139(32) \\ 
%  20,-20 & 51(5) &   7.31(2) &   0.46(5) &  103(23) \\ 
% -20, 20 & 25(9) &   7.22(2) &   0.16(5) &  147(36) \\ 
% -20, 20 &41(10) &   7.48(3) &   0.26(7) &  145(36) \\ 
% -20, 20 &40(11) &   7.29(4) &   0.31(9) &  123(32) \\ 
% -20, 20 & 18(9) &   7.51(2) &   0.14(4) &  116(32) \\ 
% -20, 20 & 11(5) &   7.31(5) &   0.22(9) &   48(31) \\ 
% -20, 20 &  8(3) &   7.58(2) &   0.10(4) &   80(31) \\ 
% -20,-20 & 19(5) &   7.26(3) &   0.18(5) &  100(27) \\ 
% -20,-20 & 34(5) &   7.49(1) &   0.17(3) &  182(27) \\ 
% -20,-20 &  7(4) &   6.95(3) &   0.13(9) &   45(27) \\ 
% -20,-20 & 38(6) &   7.43(3) &   0.38(7) &   93(27) \\ 
% -20,-20 &60(21) &  7.13(13) &  0.65(16) &   86(32) \\ 
% -20,-20 &35(21) &   7.58(6) &  0.36(11) &   93(32) \\ 
\bottomrule
  \end{tabular}
  \tabnotes 1$\sigma$ uncertainty (in brackets) are given in units of the last digit. Toward offsets $(0",0")$, $(-20",20")$, and $(-20",-20")$, a 2-components fit was applied: the results are thus given separately for each component. Toward offsets $(20",20")$ and $(20",-20")$, a single component was fitted.
\end{table*}

%%% Local Variables:
%%% mode: latex
%%% TeX-master: t
%%% End:

\section{NS column density}
\label{app:minimize}

\subsection{Collisional excitation of NS}
\label{app:coll}

The energy level diagram of the NS radical is shown in \rfig{spectro} and the details on the hyperfine transitions of the $J=5/2-3/2$ and $J=7/2-5/2$ rotational transitions are summarized in \rtab{spectro}.

In the radiative transfer calculations used in this work, we used new collision rate coefficients for the NS molecule that will be presented in a forthcoming paper. Nevertheless, we provide here a brief description of the calculations of these new NS--He data. NS--He rate coefficients were computed from new rigid-rotor NS--He interaction potentials determined using the spin-unrestricted single and double excitation coupled cluster approach with non-iterative perturbation treatment of triples [UCCSD(T)] \citep{knowles:93, knowles:00}. The NS bond distance was kept fixed at the equilibrium distance $r_e$= 2.82~bohr \citep{amano69}. The basis set for these electronic structure calculations was composed from augmented correlation-consistent polarized valence triple-zeta (aug-cc-pVTZ) basis set supplemented with additional mid-bond functions \citep{dunning:89}. Close-coupling calculations of the fine structure resolved rate coefficients \citep{alexander1985a} were performed between NS energy levels up to $J=13/2$ in the ground spin orbit manifold. Hyperfine-structure--resolved rate coefficients were then derived using a re-coupling technique \citep{alexander1985b}. Finally, the NS--He rate coefficients were scaled by a factor of 1.4 \citep{roueff2013molecular} to obtain data for NS in collision with \hh.

%Figure~\ref{fig:ns161} shows some $J=7/2-5/2$ obtained during our observing run. Given the low signal-to-noise ratio, and paucity, of these 161~GHz spectra, they have not been used in the present study.

\begin{figure}
        \centering
        \includegraphics[width=0.8\hsize]{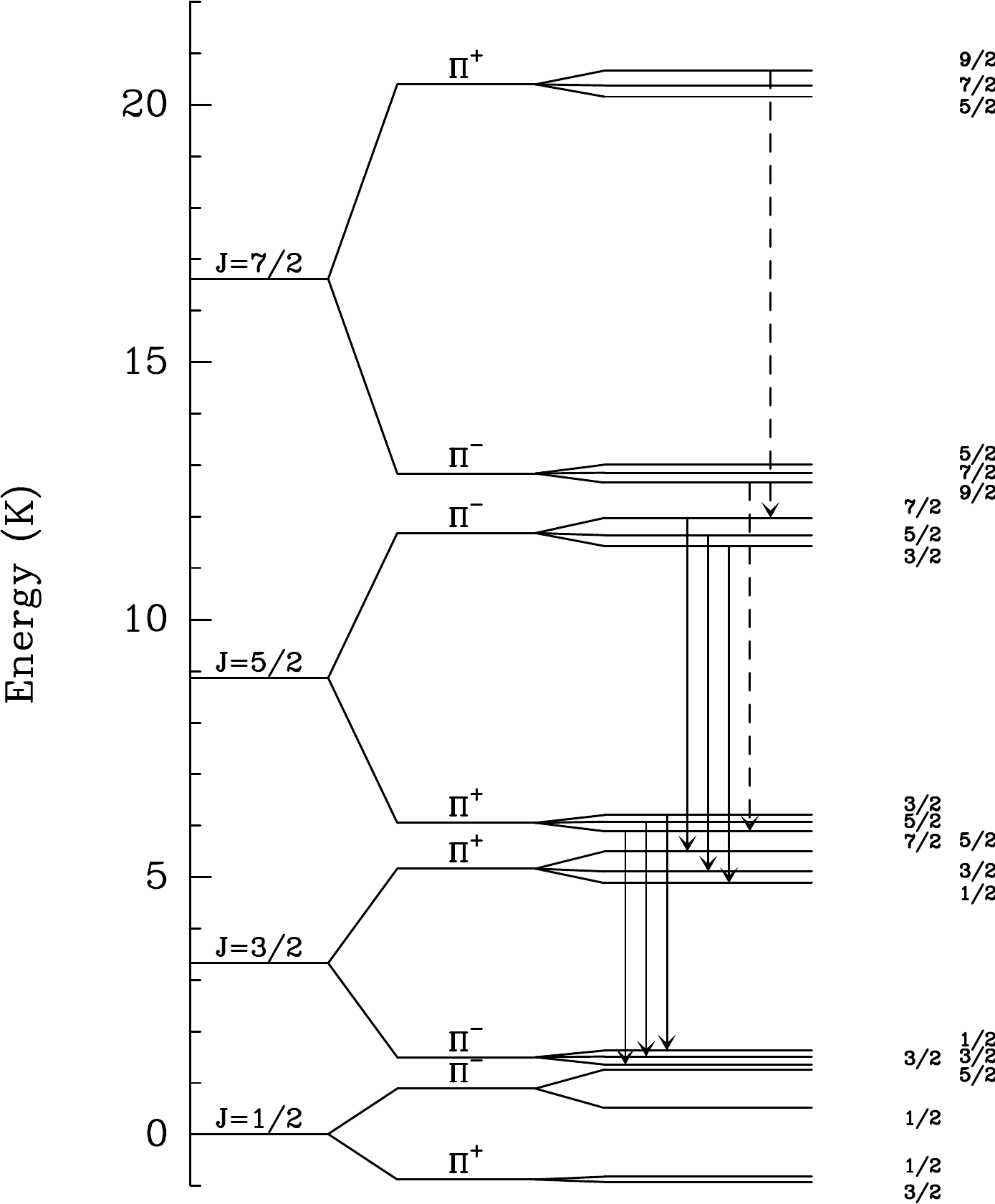}
        \caption{Energy level diagram (in K) for the first four rotational levels $J$ of the $^2\Pi_{1/2}$ spin sub-level of NS. The energy shift of the $\Lambda$-doubling states and the hyperfine levels with respect to the rotational level has been multiplied by 100 for the sake of visibility. The hyperfine quantum number $F$ is indicated on the right of each hf level, where $\mathbf{F = I+J}$ with $\mathbf I$ the nuclear spin of the nitrogen atom ($I=1$) and $\mathbf J$ the orbital angular momentum. Allowed transitions are restricted by the selection rules $\Delta J=\pm1$, $\Delta F=0,\pm1$, and $+\leftrightarrow -$. The transitions shown in Figs.~\ref{fig:ns115} and \ref{fig:gfit_l1521e}--\ref{fig:gfit_l1544}, are outlined with arrows, where full lines indicate the 115~GHz lines, and the dashed lines those at 161~GHz.}
        \label{fig:spectro}
\end{figure}

\begin{table}
  \centering
  \scriptsize
  \caption{\label{tab:spectro}Spectroscopic parameters of the NS
    $J=5/2-3/2$ and $7/2-5/2$ rotational transitions.}
  \begin{tabular}{ccccccccc}
    \toprule
    $J_u$-$J_l$ &$E_u^\ddag$ & $F_u$-$F_l$ & $g_u$ & Parity$^\dag$
     & Frequency & $A_{ul}$ & R.I.$^\S$ \\
     &K&&&&&MHz&\pers\\
    \midrule
    5/2 - 3/2 & 8.8
         & 7/2 - 5/2 & 8 & + & 115153.935 &0.233E-04 &0.444\\
       & & 5/2 - 3/2 & 6 & + & 115156.812 &0.196E-04 &0.280\\
       & & 3/2 - 1/2 & 4 & + & 115162.982 &0.175E-04 &0.167\\
       & & 3/2 - 3/2 & 4 & + & 115185.336 &0.559E-05 &0.053\\
       & & 5/2 - 5/2 & 6 & + & 115191.456 &0.373E-05 &0.053\\
       & & 3/2 - 5/2 & 4 & + & 115219.975 &0.233E-06 &0.002\\ 
       & & 3/2 - 5/2 & 4 & - & 115443.234 &0.235E-06 &0.002\\
       & & 5/2 - 5/2 & 6 & - & 115489.412 &0.376E-05 &0.053\\
       & & 3/2 - 3/2 & 4 & - & 115524.603 &0.565E-05 &0.053\\
       & & 7/2 - 5/2 & 8 & - & 115556.253 &0.235E-04 &0.444\\
       & & 5/2 - 3/2 & 6 & - & 115570.763 &0.198E-04 &0.280\\
       & & 3/2 - 1/2 & 4 & - & 115571.954 &0.177E-04 &0.167\\
      \midrule
      7/2 - 5/2 & 16.6
        & 9/2 - 7/2 &10 & - & 161297.246 & 0.686E-04 &0.417\\
      & & 7/2 - 5/2 & 8 & - & 161298.411 & 0.630E-04 &0.306\\
      & & 5/2 - 3/2 & 6 & - & 161301.747 & 0.610E-04 &0.222\\
      & & 5/2 - 5/2 & 6 & - & 161330.290 & 0.747E-05 &0.027\\
      & & 7/2 - 7/2 & 8 & - & 161335.952 & 0.560E-05 &0.027\\
      & & 5/2 - 7/2 & 6 & - & 161367.831 & 0.156E-06 &0.001\\
      & & 5/2 - 7/2 & 6 & + & 161590.959 & 0.157E-06 &0.001\\
      & & 7/2 - 7/2 & 8 & + & 161636.517 & 0.752E-05 &0.036\\
      & & 5/2 - 5/2 & 6 & + & 161657.816 & 0.564E-05 &0.020\\
      & & 9/2 - 7/2 &10 & + & 161697.257 & 0.691E-04 &0.416\\
      & & 7/2 - 5/2 & 8 & + & 161703.404 & 0.635E-04 &0.306\\
      & & 5/2 - 3/2 & 6 & + & 161703.987 & 0.614E-04 &0.222\\
      \bottomrule
    \end{tabular}
        \tabnotes Frequencies are quoted from the CDMS database. $\ddag$ Rotational energy of the upper level. $\dag$ Parity of the upper level. $\S$ Relative intensity.
\end{table}

\subsection{Minimization}
\label{app:cdens}
{The results of the MCMC minimization using our NS--He collision rate coefficients are shown in Figs.~\ref{fig:l1521e_cross}-\ref{fig:l1544_cross}.}

\begin{figure*}
        \centering
        \includegraphics[width=.48\hsize]{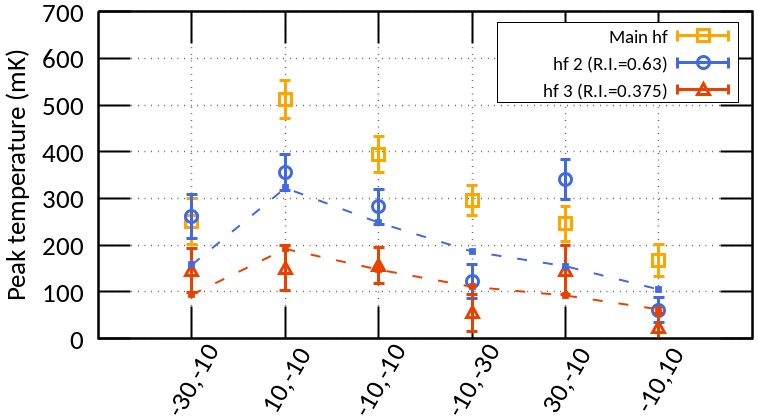}\hfill%
        \includegraphics[width=.48\hsize]{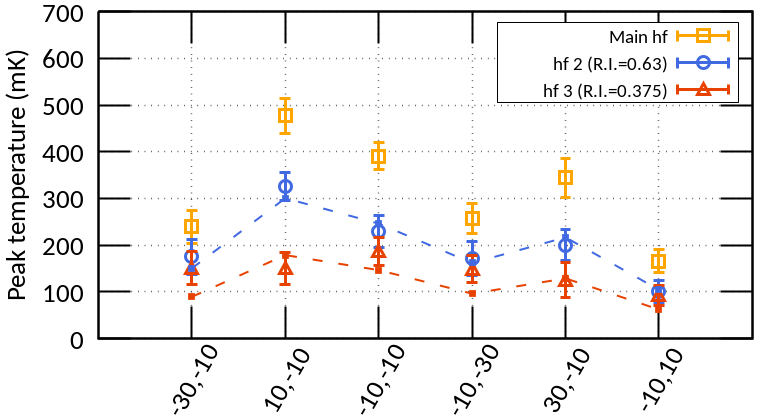}\bigskip\\
        \includegraphics[width=.48\hsize]{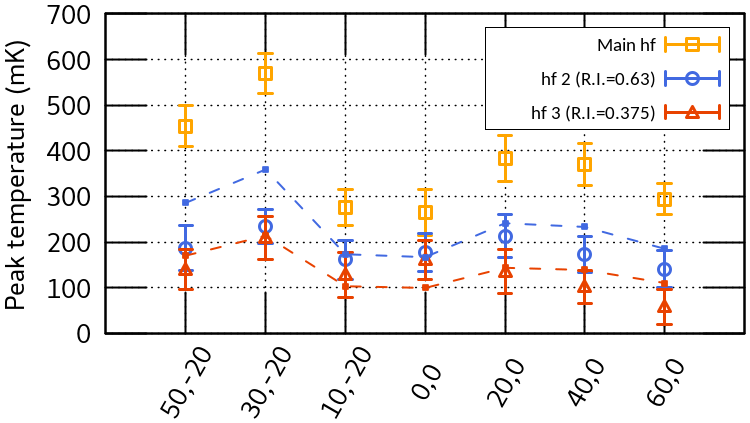}\hfill%
        \includegraphics[width=.48\hsize]{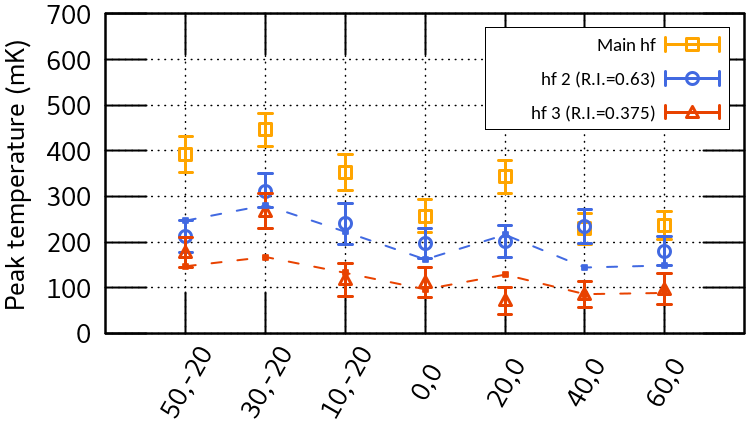}\bigskip\\
        \includegraphics[width=.48\hsize]{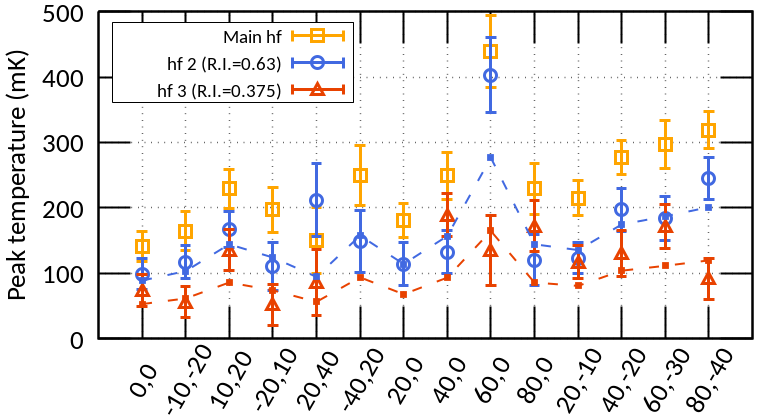}\hfill%
        \includegraphics[width=.48\hsize]{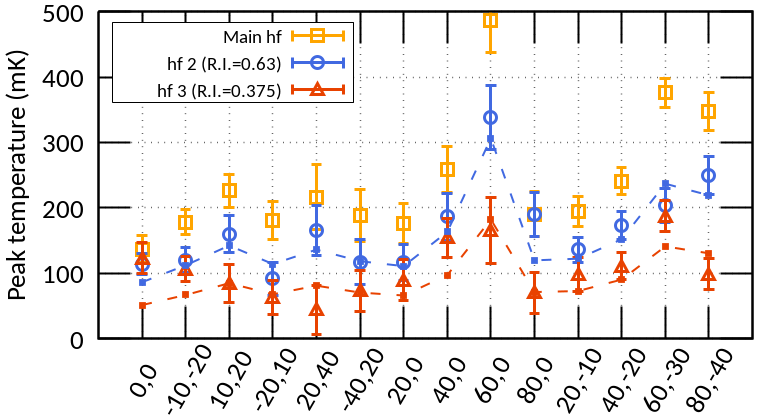}\bigskip\\
        \includegraphics[width=.48\hsize]{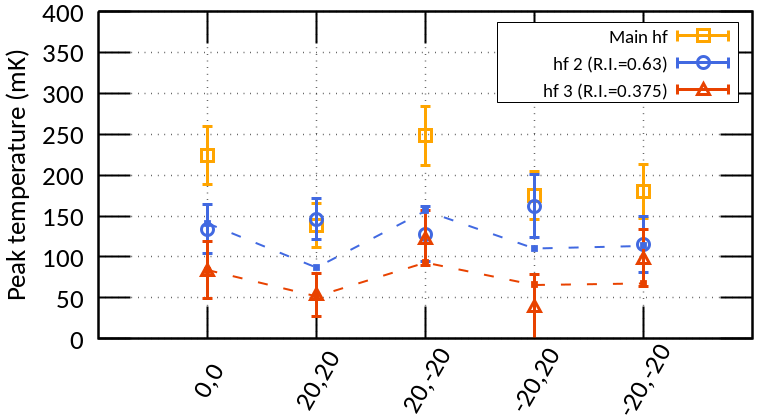}\hfill%
        \includegraphics[width=.48\hsize]{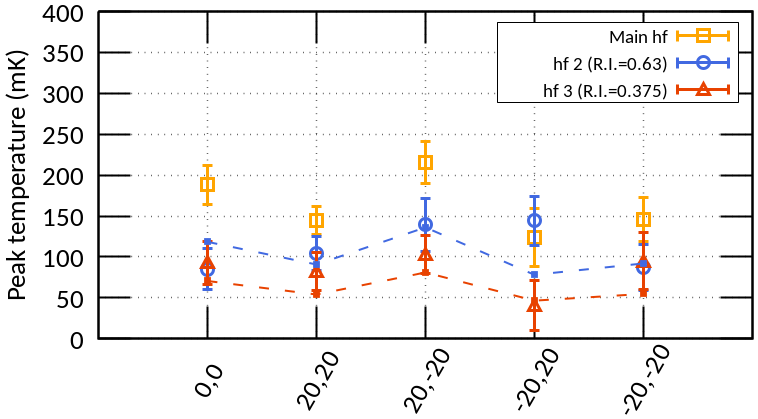}
        \caption{\new Peak temperature (in mK) of the $J=5/2-3/2$ $\Pi^-$ and $\Pi^+$ manifolds (left and right columns resp.) toward L1521E, L1521B, L1498, and L1544 (from top to bottom) as obtained from Gaussian fits to the spectra. In each panel, the x-axis gives the position offsets (in \arcsec). Open symbols show the observations with 1$\sigma$ error bars while the dashed lines and filled squares locate the theoretical peak intensity if optically thin lines and Boltzmann equilibrium assumptions hold.}
        \label{fig:gfit_tpeak}
\end{figure*}

\def\wh{0.33\hsize}
\def\wv{3cm}
\begin{figure*}
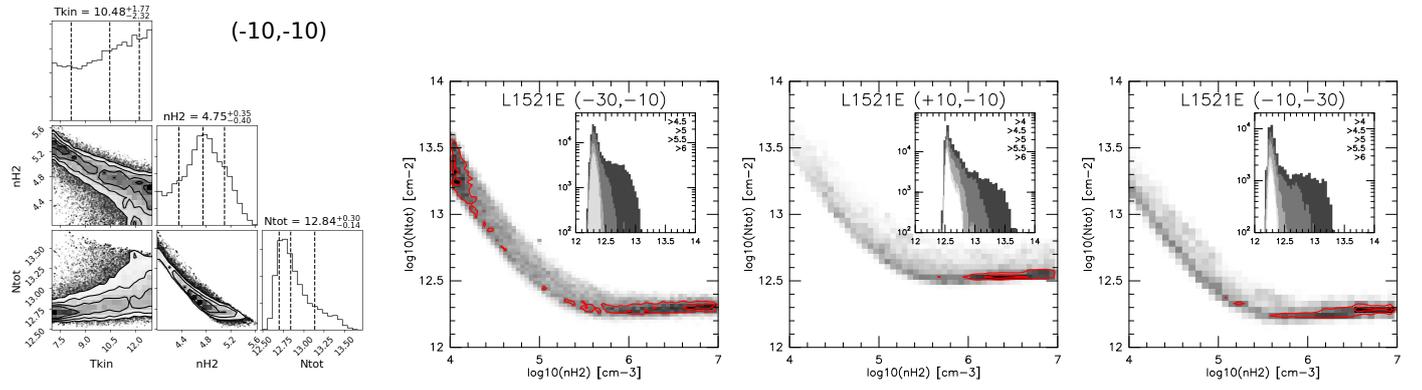

        \centering
        \vfig{5}{l1521e/m10m10-TR_W16_S8192_1229_corner.png}\hfill%
        \vfig{4}{l1521e/m30m10_crosshisto.pdf}\hfill%
        \vfig{4}{l1521e/p10m10_crosshisto.pdf}\hfill%
        \vfig{4}{l1521e/m10m30_crosshisto.pdf}
        \caption{Results of the RADEX-MCMC minimization of NS lines toward L1521E. \textit{Left:} Results at offsets $(-10",-10")$. Corner plot showing the distribution of \tkin, $\logd \nhh$, and $\logd N$ within the prior intervals, [7:13], [4:7], and [11.5:14], respectively. \textit{Other panels:} Cross-histograms toward position offsets $(-30",-10")$, $(+10",-10")$, and $(-10",-30")$, from left to right. The inset shows the histograms of the column density for $\logd\nhh>4$ by step of 0.5. The red contours show 50\% and 80\% of the maximum.}
        \label{fig:l1521e_cross}
\end{figure*}

\def\wa{0.4\hsize}
\def\wa{4cm}
\begin{figure*}
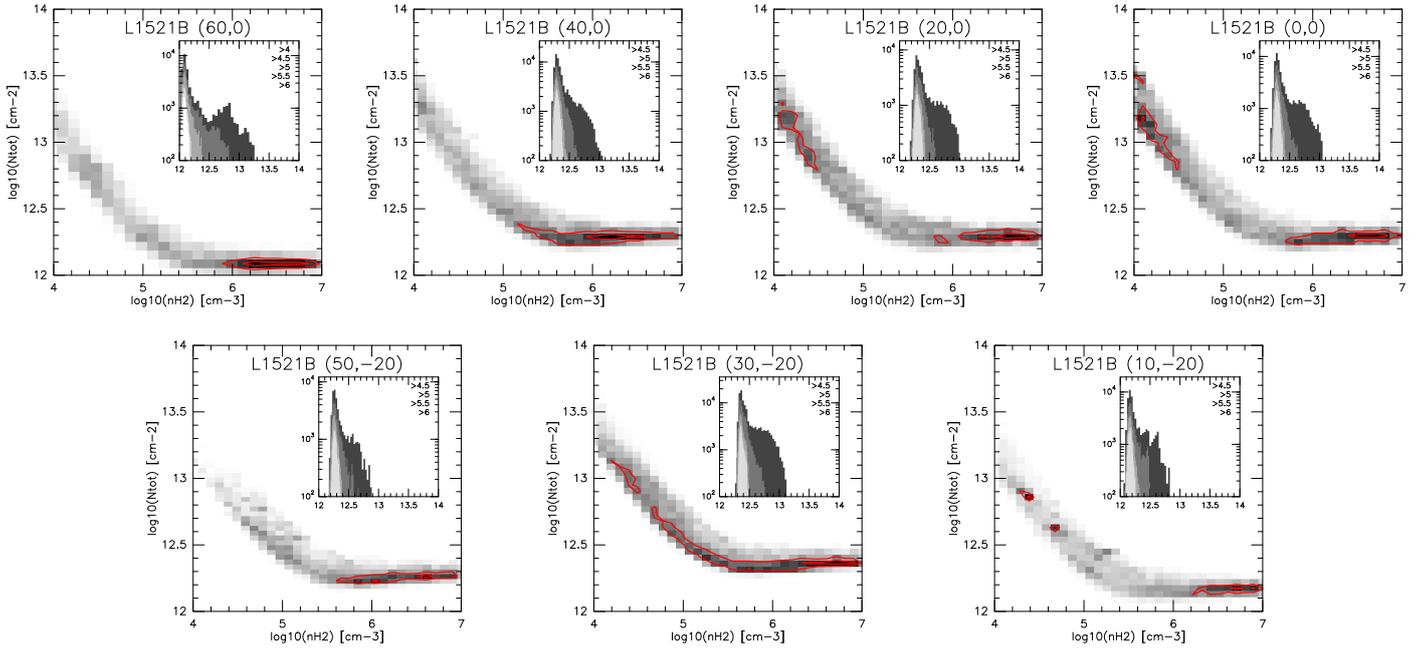

        \centering
        \vfig{4}{l1521b/p60m0_crosshisto}\hfill%
        \vfig{4}{l1521b/p40m0_crosshisto}\hfill%
        \vfig{4}{l1521b/p20m0_crosshisto}\hfill%
        \vfig{4}{l1521b/p0m0_crosshisto}\bigskip\\
        \vfig{4}{l1521b/p50m20_crosshisto}\hspace{1cm}
        \vfig{4}{l1521b/p30m20_crosshisto}\hspace{1cm}
        \vfig{4}{l1521b/p10m20_crosshisto}\bigskip\\
        \caption{\new  Cross-histograms toward L1521B (see Fig.~\ref{fig:l1521e_cross}).}
        \label{fig:l1521b_cross}
\end{figure*}

\def\wa{0.4\hsize}
\def\wh{0.33\hsize}
\def\wv{4.5cm}
\def\hx{\hspace{0.3cm}}
\begin{figure*}
        \vfig{4.4}{l1498/p0m0-TR_3_W8_S8192_1000_corner.png}\hx
        \vfig{3.5}{l1498/p0m0-TR_Tkin8_V1_W8_S2048_1000_corner.png}\hx
        \vfig{3.5}{l1498/p0m0-TR-Tkin10_V1_W8_S2048_1000_corner.png}\bigskip\\
        \vfig{4.4}{l1498/m10m20-TR_V1_W8_S16384_1000_corner.png}\hx
        \vfig{4.4}{l1498/p20m10-TR_Tkin10_V3_W16_S4096_1000_corner.png}\hx
        \vfig{4.2}{l1498/p10p20_crosshisto}\hx
        \vfig{4.4}{l1498/m20p10-TR_2_W8_S8192_1000_corner.png}\bigskip\\
        \vfig{4.4}{l1498/p20p40-TR_1_W8_S8192_1000_corner.png}\hx
        \vfig{4.4}{l1498/m40p20-TR_1_W8_S8192_1000_corner.png}\hx
        \vfig{4.4}{l1498/p60m30-TR_tkin10_V3_W16_S8192_1000_corner.png}\hx
        \vfig{4.4}{l1498/p80m40-TR_2_W8_S8192_1000_corner.png}
        \caption{Results of RADEX-MCMC minimization for L1498 (see also Fig.~\ref{fig:l1521e_cross}). For offsets (0",0"): the first panel has three free parameters and the next two panels correspond to \tkin\ held fixed at 8~K and 10~K. Priors on \tkin, \logd\nhh, and $\logd N$ taken in [7:13], [4:7], and [11:14], respectively. Toward (20",-10"), (10",20"), and (60",-30"), the kinetic temperature was fixed and equal to 10~K. For offsets (+10",+20"), the results are shown in a similar way as for L1521E.}
        \label{fig:l1498_cross}
\end{figure*}

\def\hx{\hspace{0.9cm}}
\begin{figure*}
        \centering
        \vfig{3.7}{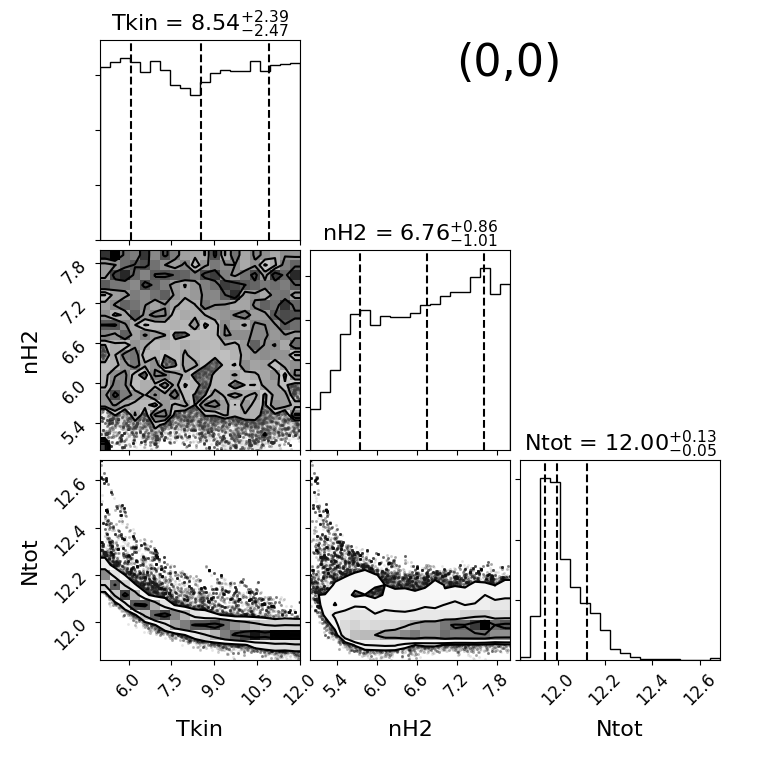}\hx
        \vfig{3.7}{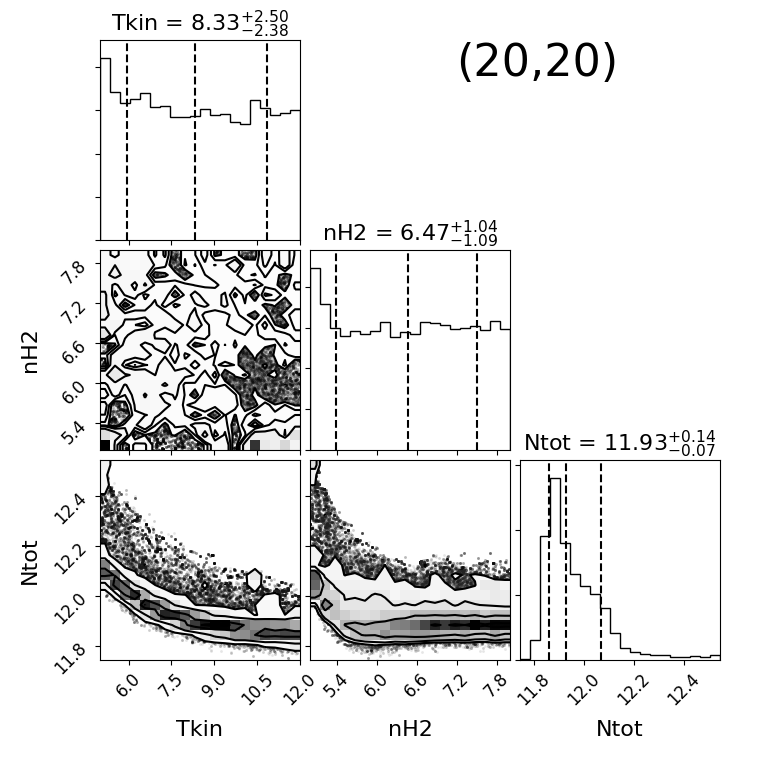}\hx
        \vfig{3.7}{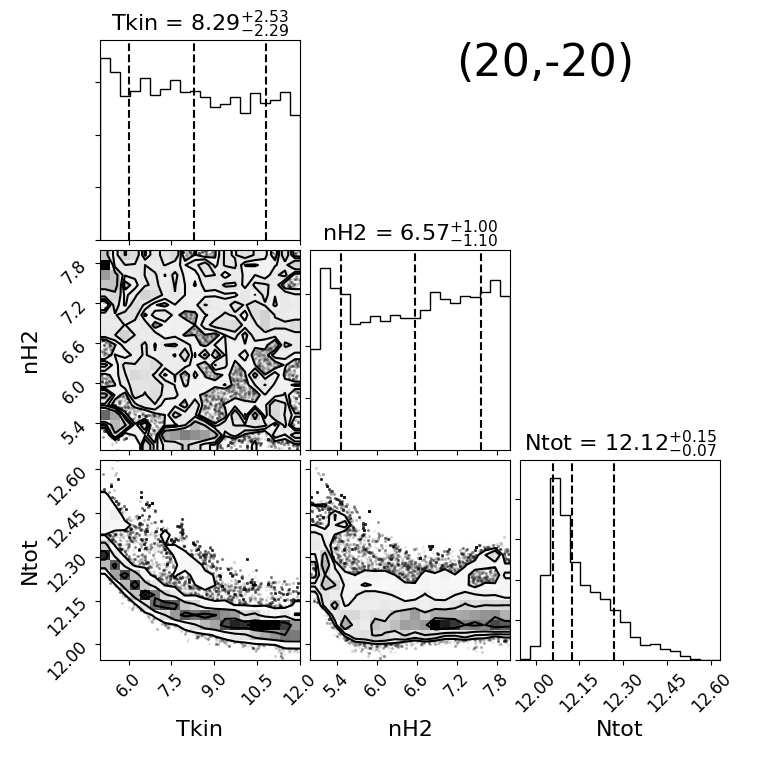}\bigskip\\
        \vfig{3.7}{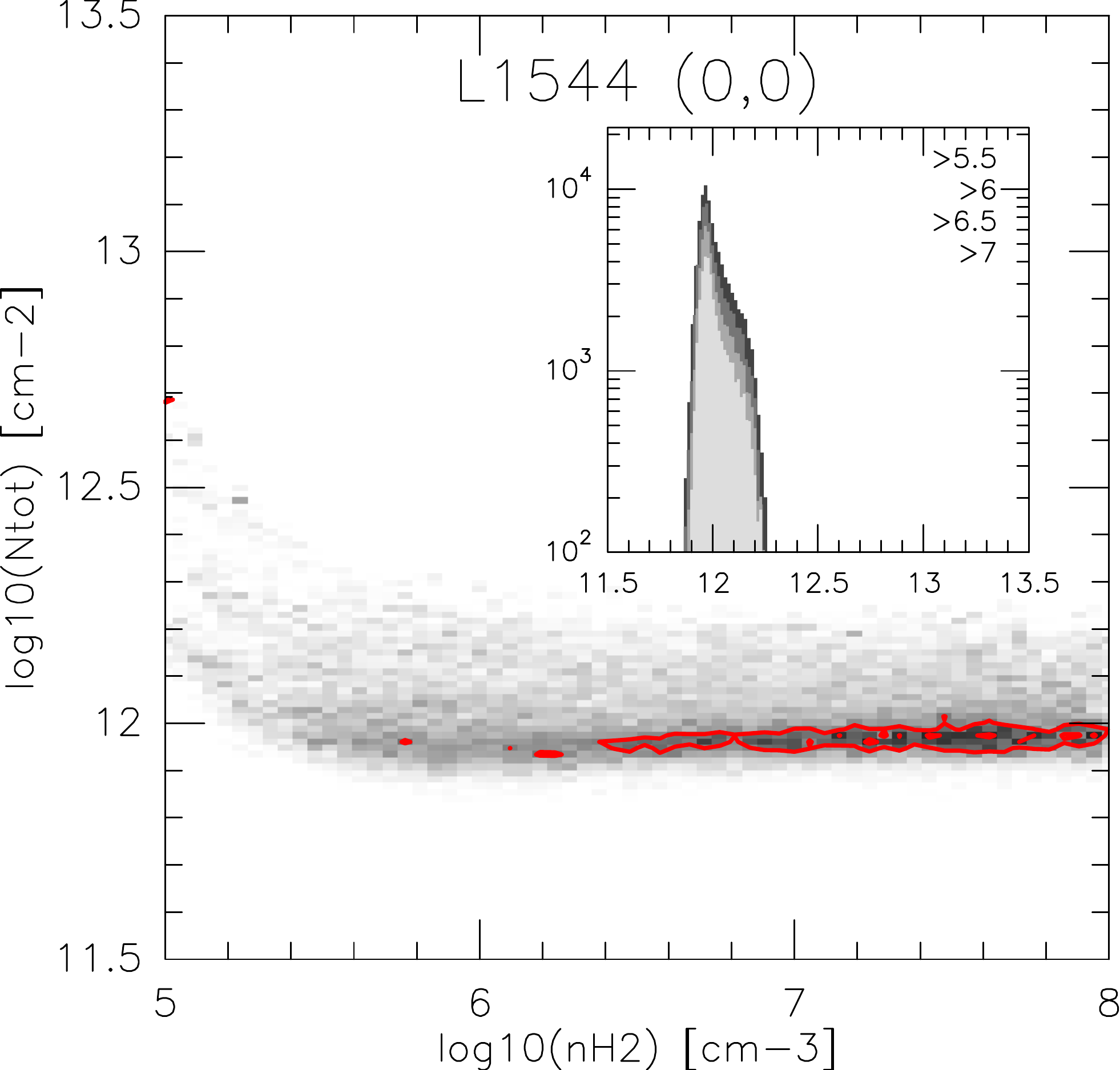}\hx
        \vfig{3.7}{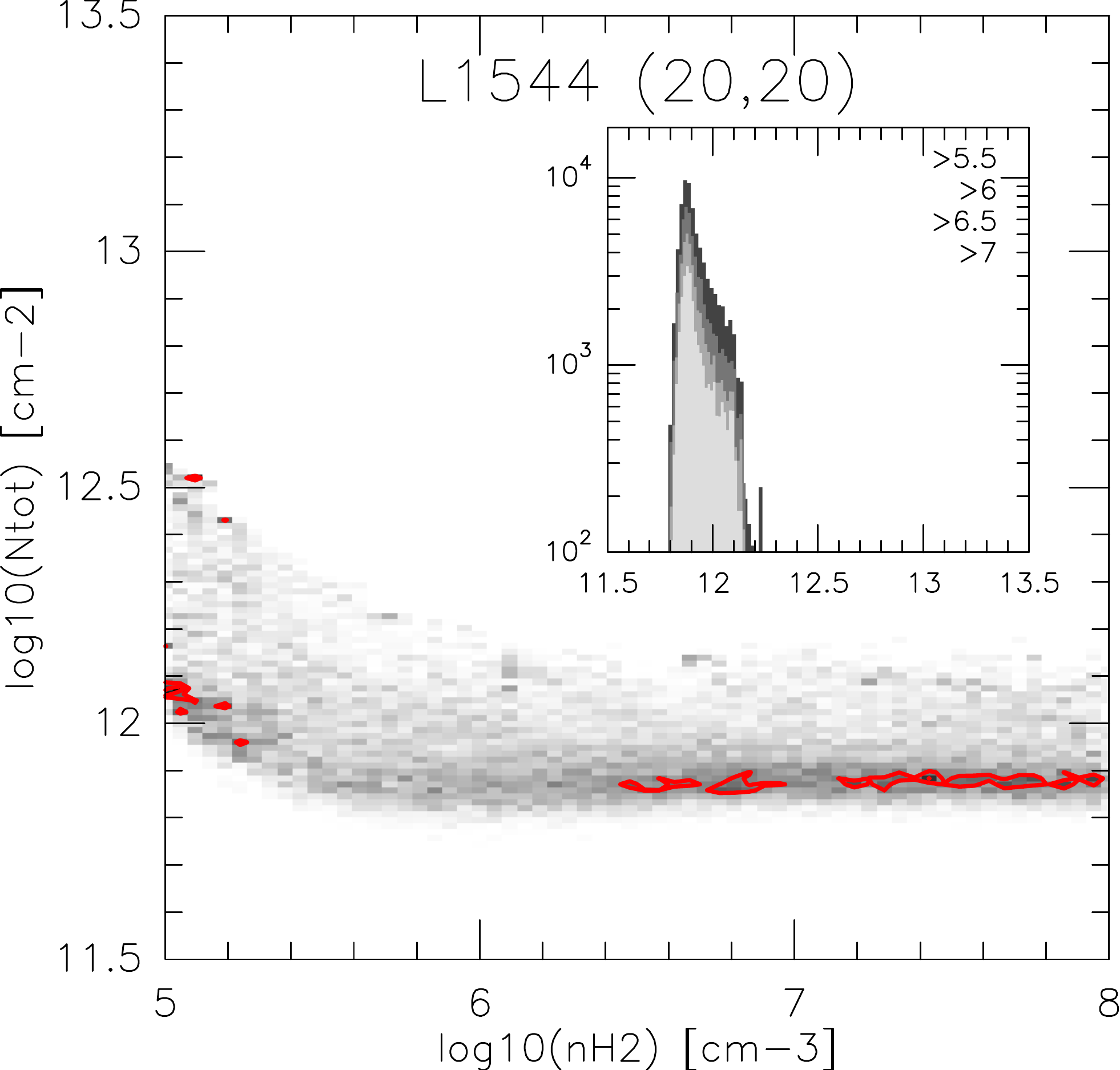}\hx
        \vfig{3.7}{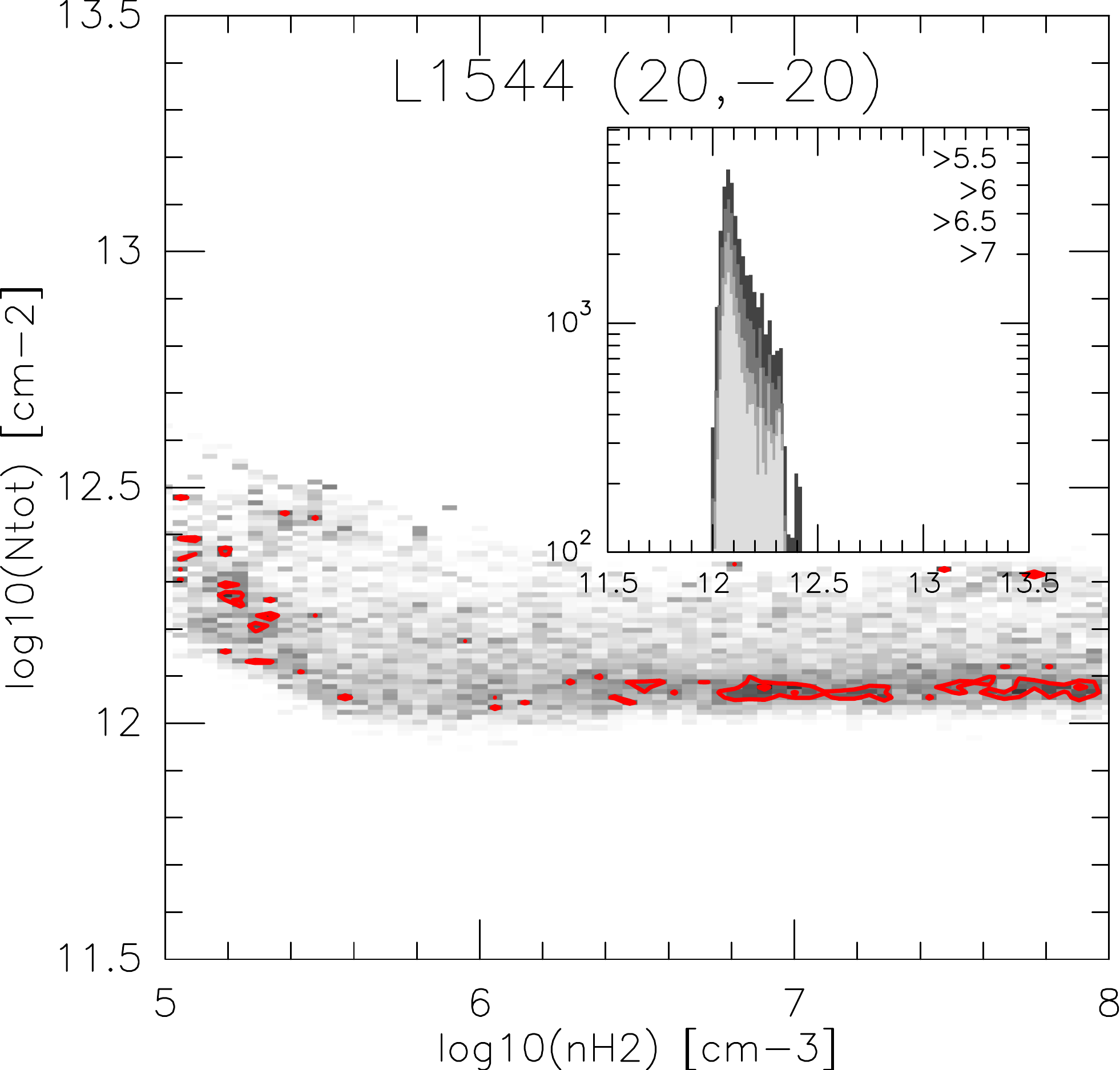}
        \caption{Results of MCMC parameter space exploration (see Fig.~\ref{fig:l1521e_cross}) for L1544 with priors on \tkin, $\logd\nhh$, and $\logd N$ taken in [5:12], [5:8], and [11:15] respectively. The bottom row shows the results in a similar way as for L1521E.}
        \label{fig:l1544_cross}
\end{figure*}
\def\hx{\hspace{0.3cm}}

\section{Chemical models}
\label{app:network}

The list of the updated and new reactions included to the UGAN network are provided in Table~\ref{tab:chemistry}. {The predicted steady-state abundances at density $\nh=\dix{4}$, \dix{5}, and \dix{6}\ccc\ are listed in Tables~\ref{tab:abundances_n4} to \ref{tab:abundances_n6}, respectively.}

\begin{table*}[t]
        \centering
%       \input{chemistry}
% Total: 31
\caption{\label{tab:chemistry}List of updated and new reactions added to the UGAN network.}
\footnotesize
\begin{tabular}{r lllll rrr lc}
\toprule
Num & \multicolumn{5}{c}{Reaction} & $\alpha$ & $\beta$ & $\gamma$ & Status & Comment \\
&&&&&&s$^{-1}$ & & K \\
\midrule
  1 & NH    &   S        &  NS     &  H     &       &                 1.000E-10 &  0.000E+00 &  0.000E+00 & New & \\
  2 & SH    &   N        &  NS     &  H     &       &                 1.000E-10 &  0.000E+00 &  0.000E+00 & New & \\
  3 & NS    &   N        &  N2     &  S     &       &                 4.000E-11 & -2.000E-01 &  2.000E+01 & New & \\
  4 & NS    &   O        &  NO     &  S     &       &                 3.000E-11 &  0.000E+00 &  0.000E+00 & New & \\
  5 & NS    &   C        &  CN     &  S     &       &                 2.000E-10 &  0.000E+00 &  0.000E+00 & New & \\
  6 & NS    &   C+       &  CS+    &  N     &       &                 1.300E-09 & -5.000E-01 &  0.000E+00 & New & (3)\\
  7 & NS    &   C+       &  NS+    &  C     &       &                 1.300E-09 & -5.000E-01 &  0.000E+00 & New & (3)\\
  8 & NS    &   H+       &  NS+    &  H     &       &                 8.000E-09 & -5.000E-01 &  0.000E+00 & New & (3)\\
  9 & NS    &   He+      &  N+     &  S     &   He  &                 2.900E-09 & -5.000E-01 &  0.000E+00 & New & (3)\\
 10 & NS    &   He+      &  S+     &  N     &   He  &                 2.900E-09 & -5.000E-01 &  0.000E+00 & New & (3)\\
 11 & NS    &   H3+      &  HNS+   &  H2    &       &                 4.800E-09 & -5.000E-01 &  0.000E+00 & New & (3)\\
 12 & NS    &   HCO+     &  HNS+   &  CO    &       &                 2.000E-09 & -5.000E-01 &  0.000E+00 & New & (3)\\
 13 & NS    &   SECPHO$^\ddagger$   &  N      &  S     &       &                 3.000E+02 &  0.000E+00 &  0.000E+00 & New & (1) \\
 14 & N     &   SH+      &  H      &  NS+   &       &                 7.400E-10 &  0.000E+00 &  0.000E+00 & New & \\
 15 & N     &   SO+      &  O      &  NS+   &       &                 5.000E-11 &  0.000E+00 &  0.000E+00 & New & \\
 16 & N     &   H2S+     &  H2     &  NS+   &       &                 7.900E-10 &  0.000E+00 &  0.000E+00 & New & \\
 17 & S     &   H3+      &  SH+    &  H2    &       &                 2.000E-09 &  0.000E+00 &  0.000E+00 & Updated & \\
 18 & S     &   H2O+     &  OH     &  SH+   &       &                 4.000E-10 &  0.000E+00 &  0.000E+00 & New & \\
 19 & S     &   NH+      &  N      &  SH+   &       &                 6.900E-10 &  0.000E+00 &  0.000E+00 & New & (3)\\
 20 & H2S   &   H+       &  H2     &  SH+   &       &                 7.100E-10 & -5.000E-01 &  0.000E+00 & New & \\
 21 & H2S   &   H+       &  H      &  H2S+  &       &                 4.000E-09 & -5.000E-01 &  0.000E+00 & Updated & \\
 22 & H2O   &   SH+      &  H3O+   &  S     &       &                 2.300E-09 & -5.000E-01 &  0.000E+00 & Updated & \\
 23 & SH+   &   H2       &  H3S+   &  PHOTON&       &                 1.400E-16 & -6.000E-01 &  0.000E+00 & Updated & \\
 24 & HNS+  &   ELECTR   &  NS     &  H     &       &                 3.000E-07 & -5.000E-01 &  0.000E+00 & New & (3)\\
 25 & HNS+  &   Gr$^\dag$       &  NS     &  Gr+   &   H   &                 2.400E-07 &  5.000E-01 &  0.000E+00 & New & (2)\\
 26 & NS+   &   O        &  NO+    &  S     &       &                 6.100E-10 &  0.000E+00 &  0.000E+00 & New & \\
 27 & NS+   &   C        &  CN     &  S+    &       &                 6.000E-10 &  0.000E+00 &  0.000E+00 & New & \\
 28 & NS+   &   N        &  N2     &  S+    &       &                 6.000E-10 &  0.000E+00 &  0.000E+00 & New & \\
 29 & NS+   &   ELECTR   &  N      &  S     &       &                 2.000E-07 & -5.000E-01 &  0.000E+00 & New & (3)\\
 30 & S     &   NH+      &  H      &  NS+   &       &                 6.900E-10 &  0.000E+00 &  0.000E+00 & New & (3)\\
 31 & S     &   NH+      &  N      &  SH+   &       &                 6.900E-10 &  0.000E+00 &  0.000E+00 & New & (3)\\
\bottomrule
\end{tabular}
\tabnotes  All reaction rates were taken from KIDA2018 unless specified. The reaction rates are computed as $\alpha\,(T/300)^\beta\,\exp(-\gamma/T)$\cccs. $^\dag$ {\new The species denoted \verb|Gr| and \verb|Gr+| are not involved in gas-grain chemistry but only in determining the charge of the gas. $^\ddagger$ \verb|SECPHO| are secondary photons; (1) guess based on \ce{NO + SECPHO} \citep{heays2017}; (2) guess based on \ce{HCS+ + Gr}; (3) Not validated by the KIDA committee \citep{wakelam2015}.}
\end{table*}

\newcolumntype{d}[1]{D{.}{.}{#1}}
\newcolumntype{.}{D{.}{.}{-1}}
\begin{table*}
        \centering
        \caption{Steady-state abundance of the key species involved in the formation and destruction of NS, as predicted by our model for varying conditions.}
        \label{tab:abundances_n4}
        \scriptsize
        \begin{tabular}%
                {cccccccccccccccccc}
                \toprule
                %               $\zeta_{17}$&
                \stot&  \otot&    C/O& \ab{O} & \ab{C} & \ab{S} & \ab{N} & \ab{NH} &     \ab{SH} & \ab{NS} & \ab{S+}& \ab{H+} & \ab{m+}$^\dagger$&   \ab{N2H+}&  \ab{e-}
                & \ab{S}/\sgas\,$^\ddagger$\\
                %$\zeta_{17}$&  \stot&  \ctot&  \otot&    C/O&      O&     O2&      S&     NS&     H+&    H3+&   HCO+&   H3O+&     S+&     e-\\
                \midrule
%{ccccccccccccc}
%\toprule
%  \stot&  \otot&    C/O&      O&      C&      S&      N&     NH&     SH&     NS&     S+&     H+&     m+&   N2H+&     e-&[S]/\stot\\
%\midrule
&&&&&&&&$\zeta_{17}=    1.0$ \\
   0.14&  207.5&    0.4& 4.0E-5& 2.8E-8& 1.4E-7& 5.7E-6& 8.6E-10& 1.3E-12& 9.7E-12& 4.2E-10& 2.2E-10& 1.3E-8& 5.3E-10& 1.7E-8&98.5\\
   0.14&  138.3&    0.6& 2.3E-5& 6.1E-8& 1.4E-7& 7.5E-6& 1.6E-9& 2.8E-12& 2.9E-11& 8.5E-10& 5.1E-10& 1.3E-8& 5.8E-10& 1.8E-8&98.2\\
   0.14&  103.8&    0.8& 1.5E-5& 2.4E-7& 1.4E-7& 1.3E-5& 2.0E-9& 2.8E-12& 3.8E-11& 2.0E-9& 2.0E-9& 1.1E-8& 5.1E-10& 2.3E-8&97.1\\
   0.14&   83.0&    1.0& 5.3E-6& 3.7E-6& 1.2E-7& 1.5E-5& 2.6E-9& 5.5E-12& 2.9E-11& 3.2E-9& 2.4E-9& 7.7E-9& 4.6E-10& 3.0E-8&82.6\\
   0.14&   69.2&    1.2& 1.6E-6& 8.5E-6& 8.1E-8& 6.2E-6& 3.0E-9& 9.4E-12& 1.6E-11& 2.2E-9& 6.3E-10& 8.5E-9& 7.0E-10& 2.2E-8&57.9\\
  0.443&  207.5&    0.4& 4.0E-5& 2.8E-8& 4.3E-7& 5.8E-6& 8.5E-10& 5.6E-12& 3.1E-11& 1.3E-9& 2.2E-10& 1.3E-8& 5.2E-10& 1.8E-8&98.1\\
  0.443&  138.3&    0.6& 2.3E-5& 6.2E-8& 4.3E-7& 7.6E-6& 1.5E-9& 1.2E-11& 9.1E-11& 2.6E-9& 5.2E-10& 1.2E-8& 5.6E-10& 2.0E-8&97.9\\
  0.443&  103.8&    0.8& 1.5E-5& 2.5E-7& 4.3E-7& 1.4E-5& 1.9E-9& 1.2E-11& 1.2E-10& 6.2E-9& 2.0E-9& 9.4E-9& 4.6E-10& 2.6E-8&97.0\\
  0.443&   83.0&    1.0& 5.5E-6& 4.1E-6& 3.7E-7& 1.6E-5& 2.4E-9& 1.8E-11& 7.9E-11& 9.6E-9& 2.7E-9& 6.6E-9& 3.9E-10& 3.7E-8&83.6\\
  0.443&   69.2&    1.2& 1.8E-6& 9.4E-6& 2.7E-7& 6.7E-6& 2.7E-9& 2.9E-11& 4.3E-11& 6.7E-9& 7.2E-10& 7.5E-9& 6.2E-10& 2.6E-8&60.0\\
    1.4&  207.5&    0.4& 4.1E-5& 2.9E-8& 1.4E-6& 6.0E-6& 8.1E-10& 3.0E-11& 9.6E-11& 4.1E-9& 2.2E-10& 1.2E-8& 4.8E-10& 2.0E-8&97.4\\
    1.4&  138.3&    0.6& 2.4E-5& 6.4E-8& 1.4E-6& 8.0E-6& 1.5E-9& 5.9E-11& 2.9E-10& 8.1E-9& 5.3E-10& 1.0E-8& 4.9E-10& 2.3E-8&97.5\\
    1.4&  103.8&    0.8& 1.6E-5& 2.9E-7& 1.4E-6& 1.5E-5& 1.7E-9& 5.0E-11& 3.5E-10& 1.9E-8& 2.3E-9& 7.1E-9& 3.5E-10& 3.7E-8&97.1\\
    1.4&   83.0&    1.0& 6.1E-6& 5.1E-6& 1.2E-6& 1.8E-5& 1.9E-9& 5.4E-11& 1.8E-10& 2.8E-8& 3.2E-9& 4.8E-9& 2.6E-10& 5.7E-8&86.0\\
    1.4&   69.2&    1.2& 2.3E-6& 1.2E-5& 9.3E-7& 8.2E-6& 2.2E-9& 8.6E-11& 1.0E-10& 2.1E-8& 9.9E-10& 5.4E-9& 4.3E-10& 4.1E-8&66.3\\
   4.43&  207.5&    0.4& 4.5E-5& 3.0E-8& 4.3E-6& 6.6E-6& 7.1E-10& 1.6E-10& 2.8E-10& 1.3E-8& 2.2E-10& 9.1E-9& 3.9E-10& 2.6E-8&96.0\\
   4.43&  138.3&    0.6& 2.6E-5& 6.9E-8& 4.3E-6& 9.1E-6& 1.2E-9& 2.6E-10& 7.8E-10& 2.5E-8& 5.6E-10& 6.8E-9& 3.4E-10& 3.7E-8&96.9\\
   4.43&  103.8&    0.8& 1.8E-5& 4.3E-7& 4.3E-6& 1.9E-5& 1.2E-9& 1.6E-10& 7.6E-10& 5.9E-8& 3.0E-9& 4.1E-9& 1.7E-10& 7.7E-8&97.2\\
   4.43&   83.0&    1.0& 7.5E-6& 6.7E-6& 4.0E-6& 2.1E-5& 1.1E-9& 1.5E-10& 3.3E-10& 8.3E-8& 3.2E-9& 3.1E-9& 1.2E-10& 1.2E-7&89.4\\
   4.43&   69.2&    1.2& 3.4E-6& 1.5E-5& 3.5E-6& 1.2E-5& 1.3E-9& 2.2E-10& 2.1E-10& 6.9E-8& 1.4E-9& 3.2E-9& 1.8E-10& 9.5E-8&78.0\\
     14&  207.5&    0.4& 5.6E-5& 3.3E-8& 1.3E-5& 8.5E-6& 4.7E-10& 5.2E-10& 5.9E-10& 4.1E-8& 2.5E-10& 5.0E-9& 2.0E-10& 5.0E-8&95.3\\
     14&  138.3&    0.6& 3.5E-5& 9.3E-8& 1.4E-5& 1.3E-5& 6.4E-10& 6.1E-10& 1.3E-9& 8.7E-8& 7.1E-10& 3.2E-9& 1.3E-10& 9.6E-8&97.2\\
     14&  103.8&    0.8& 2.1E-5& 1.4E-6& 1.4E-5& 3.0E-5& 3.7E-10& 2.8E-10& 7.9E-10& 2.2E-7& 4.4E-9& 2.0E-9& 3.6E-11& 2.5E-7&96.9\\
     14&   83.0&    1.0& 1.1E-5& 9.7E-6& 1.3E-5& 2.7E-5& 3.9E-10& 3.2E-10& 4.4E-10& 2.5E-7& 2.5E-9& 1.8E-9& 3.5E-11& 2.8E-7&92.8\\
     14&   69.2&    1.2& 6.2E-6& 1.8E-5& 1.2E-5& 1.8E-5& 4.8E-10& 4.2E-10& 3.0E-10& 2.3E-7& 1.4E-9& 1.8E-9& 4.7E-11& 2.7E-7&88.4\\
&&&&&&&&$\zeta_{17}=    3.0$ \\
   0.14&  207.5&    0.4& 4.1E-5& 6.7E-8& 1.4E-7& 7.9E-6& 2.9E-9& 2.1E-12& 3.0E-11& 9.5E-10& 6.4E-10& 2.4E-8& 1.1E-9& 3.1E-8&98.6\\
   0.14&  138.3&    0.6& 2.4E-5& 1.4E-7& 1.4E-7& 1.1E-5& 5.0E-9& 3.6E-12& 7.1E-11& 1.7E-9& 1.5E-9& 2.3E-8& 1.1E-9& 3.4E-8&97.9\\
   0.14&  103.8&    0.8& 1.6E-5& 6.4E-7& 1.3E-7& 2.1E-5& 5.0E-9& 2.8E-12& 5.1E-11& 3.9E-9& 7.0E-9& 1.7E-8& 7.6E-10& 5.0E-8&95.8\\
   0.14&   83.0&    1.0& 9.3E-6& 9.0E-6& 1.2E-7& 2.7E-5& 3.6E-9& 2.9E-12& 1.1E-11& 6.1E-9& 2.3E-8& 9.6E-9& 3.2E-10& 1.2E-7&86.1\\
   0.14&   69.2&    1.2& 5.1E-6& 1.7E-5& 1.0E-7& 1.8E-5& 4.0E-9& 4.2E-12&10.0E-12& 5.5E-9& 8.1E-9& 1.1E-8& 5.5E-10& 8.8E-8&73.7\\
  0.443&  207.5&    0.4& 4.2E-5& 6.7E-8& 4.4E-7& 8.0E-6& 2.9E-9& 8.7E-12& 9.2E-11& 3.0E-9& 6.4E-10& 2.3E-8& 1.0E-9& 3.3E-8&98.3\\
  0.443&  138.3&    0.6& 2.4E-5& 1.4E-7& 4.3E-7& 1.1E-5& 4.8E-9& 1.5E-11& 2.2E-10& 5.4E-9& 1.6E-9& 2.1E-8& 1.1E-9& 3.7E-8&97.8\\
  0.443&  103.8&    0.8& 1.7E-5& 6.9E-7& 4.2E-7& 2.2E-5& 4.6E-9& 1.1E-11& 1.5E-10& 1.2E-8& 7.4E-9& 1.6E-8& 6.6E-10& 5.8E-8&95.8\\
  0.443&   83.0&    1.0& 9.6E-6& 9.3E-6& 3.8E-7& 2.8E-5& 3.2E-9& 9.3E-12& 3.2E-11& 1.9E-8& 2.2E-8& 9.0E-9& 2.8E-10& 1.3E-7&86.5\\
  0.443&   69.2&    1.2& 5.7E-6& 1.8E-5& 3.4E-7& 2.0E-5& 3.5E-9& 1.2E-11& 2.6E-11& 1.8E-8& 9.9E-9& 9.3E-9& 4.2E-10& 1.1E-7&76.3\\
    1.4&  207.5&    0.4& 4.3E-5& 6.9E-8& 1.4E-6& 8.3E-6& 2.7E-9& 4.3E-11& 2.8E-10& 9.3E-9& 6.5E-10& 2.1E-8& 9.3E-10& 3.7E-8&98.1\\
    1.4&  138.3&    0.6& 2.5E-5& 1.5E-7& 1.4E-6& 1.2E-5& 4.3E-9& 6.8E-11& 6.3E-10& 1.7E-8& 1.6E-9& 1.8E-8& 8.8E-10& 4.5E-8&97.7\\
    1.4&  103.8&    0.8& 1.8E-5& 8.5E-7& 1.3E-6& 2.4E-5& 3.5E-9& 4.1E-11& 3.4E-10& 4.0E-8& 8.9E-9& 1.2E-8& 4.3E-10& 8.6E-8&95.9\\
    1.4&   83.0&    1.0& 1.1E-5& 1.0E-5& 1.2E-6& 2.9E-5& 2.3E-9& 2.9E-11& 7.8E-11& 6.2E-8& 2.0E-8& 7.7E-9& 1.9E-10& 1.8E-7&87.5\\
    1.4&   69.2&    1.2& 6.8E-6& 1.9E-5& 1.1E-6& 2.3E-5& 2.5E-9& 3.4E-11& 5.9E-11& 5.7E-8& 1.1E-8& 7.6E-9& 2.5E-10& 1.6E-7&80.3\\
   4.43&  207.5&    0.4& 4.7E-5& 7.4E-8& 4.3E-6& 9.4E-6& 2.2E-9& 2.1E-10& 7.3E-10& 2.9E-8& 6.7E-10& 1.6E-8& 6.8E-10& 5.1E-8&97.5\\
   4.43&  138.3&    0.6& 2.8E-5& 1.7E-7& 4.3E-6& 1.4E-5& 3.1E-9& 2.8E-10& 1.4E-9& 5.3E-8& 1.7E-9& 1.2E-8& 5.2E-10& 7.6E-8&97.6\\
   4.43&  103.8&    0.8& 2.1E-5& 1.9E-6& 4.2E-6& 3.3E-5& 1.4E-9& 9.5E-11& 3.6E-10& 1.5E-7& 1.5E-8& 6.9E-9& 1.2E-10& 2.1E-7&95.2\\
   4.43&   83.0&    1.0& 1.3E-5& 1.3E-5& 4.0E-6& 3.3E-5& 1.1E-9& 7.9E-11& 1.3E-10& 2.0E-7& 1.5E-8& 5.5E-9& 8.1E-11& 3.1E-7&89.3\\
   4.43&   69.2&    1.2& 9.2E-6& 2.2E-5& 3.8E-6& 2.8E-5& 1.2E-9& 8.4E-11& 1.0E-10& 1.9E-7&10.0E-9& 5.4E-9&10.0E-11& 3.1E-7&85.2\\
     14&  207.5&    0.4& 6.0E-5& 9.1E-8& 1.4E-5& 1.3E-5& 1.2E-9& 6.5E-10& 1.3E-9& 9.8E-8& 7.7E-10& 8.5E-9& 2.8E-10& 1.1E-7&97.4\\
     14&  138.3&    0.6& 4.0E-5& 3.1E-7& 1.4E-5& 2.3E-5& 1.1E-9& 4.7E-10& 1.4E-9& 2.2E-7& 2.8E-9& 5.8E-9& 1.2E-10& 2.4E-7&97.7\\
     14&  103.8&    0.8& 2.8E-5& 7.4E-6& 1.3E-5& 4.4E-5& 2.7E-10& 1.4E-10& 2.1E-10& 5.6E-7& 1.5E-8& 3.4E-9& 1.8E-11& 6.6E-7&94.1\\
     14&   83.0&    1.0& 1.9E-5& 1.9E-5& 1.3E-5& 4.0E-5& 3.1E-10& 1.5E-10& 1.5E-10& 5.8E-7& 9.9E-9& 3.2E-9& 2.0E-11& 7.1E-7&91.6\\
     14&   69.2&    1.2& 1.5E-5& 2.8E-5& 1.3E-5& 3.6E-5& 3.6E-10& 1.6E-10& 1.2E-10& 5.7E-7& 7.5E-9& 3.2E-9& 2.3E-11& 7.1E-7&90.1\\
&&&&&&&&$\zeta_{17}=   10.0$ \\
   0.14&  207.5&    0.4& 4.4E-5& 1.6E-7& 1.4E-7& 1.2E-5& 9.3E-9& 2.8E-12& 7.7E-11& 2.2E-9& 2.1E-9& 4.6E-8& 1.9E-9& 6.3E-8&97.9\\
   0.14&  138.3&    0.6& 2.6E-5& 3.5E-7& 1.4E-7& 1.7E-5& 1.4E-8& 3.5E-12& 1.3E-10& 3.5E-9& 5.2E-9& 4.4E-8& 1.8E-9& 7.2E-8&96.8\\
   0.14&  103.8&    0.8& 2.0E-5& 2.3E-6& 1.3E-7& 3.5E-5& 7.9E-9& 1.8E-12& 2.6E-11& 8.4E-9& 3.5E-8& 2.9E-8& 5.8E-10& 1.6E-7&92.6\\
   0.14&   83.0&    1.0& 2.0E-5& 2.0E-5& 1.2E-7& 4.5E-5& 2.0E-9& 1.0E-12& 1.8E-12& 1.5E-8& 1.4E-7& 1.4E-8& 8.4E-11& 5.2E-7&84.5\\
   0.14&   69.2&    1.2& 1.8E-5& 3.2E-5& 1.2E-7& 4.5E-5& 1.7E-9& 8.7E-13& 1.2E-12& 1.5E-8& 1.5E-7& 1.2E-8& 7.0E-11& 6.1E-7&82.5\\
  0.443&  207.5&    0.4& 4.4E-5& 1.6E-7& 4.3E-7& 1.2E-5& 9.0E-9& 1.1E-11& 2.3E-10& 6.9E-9& 2.1E-9& 4.5E-8& 1.8E-9& 6.6E-8&97.8\\
  0.443&  138.3&    0.6& 2.6E-5& 3.6E-7& 4.3E-7& 1.7E-5& 1.3E-8& 1.3E-11& 3.8E-10& 1.1E-8& 5.3E-9& 4.1E-8& 1.7E-9& 7.8E-8&96.7\\
  0.443&  103.8&    0.8& 2.1E-5& 2.6E-6& 4.1E-7& 3.7E-5& 6.3E-9& 6.0E-12& 6.0E-11& 2.8E-8& 4.0E-8& 2.6E-8& 4.4E-10& 1.9E-7&92.2\\
  0.443&   83.0&    1.0& 2.1E-5& 2.0E-5& 3.7E-7& 4.6E-5& 1.8E-9& 3.3E-12& 5.2E-12& 4.8E-8& 1.3E-7& 1.3E-8& 7.7E-11& 5.4E-7&84.4\\
  0.443&   69.2&    1.2& 1.9E-5& 3.2E-5& 3.7E-7& 4.5E-5& 1.5E-9& 2.9E-12& 3.6E-12& 4.9E-8& 1.3E-7& 1.2E-8& 6.6E-11& 6.3E-7&82.4\\
    1.4&  207.5&    0.4& 4.6E-5& 1.7E-7& 1.4E-6& 1.3E-5& 8.1E-9& 4.7E-11& 6.6E-10& 2.2E-8& 2.2E-9& 4.0E-8& 1.6E-9& 7.6E-8&97.9\\
    1.4&  138.3&    0.6& 2.8E-5& 3.8E-7& 1.4E-6& 1.9E-5& 1.1E-8& 5.6E-11& 9.6E-10& 3.6E-8& 5.5E-9& 3.6E-8& 1.3E-9& 9.7E-8&96.9\\
    1.4&  103.8&    0.8& 2.4E-5& 4.7E-6& 1.3E-6& 4.3E-5& 2.9E-9& 1.6E-11& 6.3E-11& 1.1E-7& 6.1E-8& 1.8E-8& 1.7E-10& 3.3E-7&90.6\\
    1.4&   83.0&    1.0& 2.2E-5& 2.2E-5& 1.2E-6& 4.7E-5& 1.3E-9& 1.0E-11& 1.3E-11& 1.5E-7& 1.1E-7& 1.2E-8& 5.6E-11& 6.3E-7&84.5\\
    1.4&   69.2&    1.2& 2.0E-5& 3.3E-5& 1.2E-6& 4.6E-5& 1.2E-9& 9.2E-12& 9.6E-12& 1.5E-7& 1.0E-7& 1.1E-8& 5.2E-11& 7.1E-7&82.7\\
   4.43&  207.5&    0.4& 5.1E-5& 1.9E-7& 4.3E-6& 1.5E-5& 5.9E-9& 2.1E-10& 1.5E-9& 6.9E-8& 2.3E-9& 3.0E-8& 9.9E-10& 1.1E-7&97.7\\
   4.43&  138.3&    0.6& 3.3E-5& 4.9E-7& 4.3E-6& 2.3E-5& 5.9E-9& 1.9E-10& 1.6E-9& 1.2E-7& 6.6E-9& 2.4E-8& 6.1E-10& 1.8E-7&96.7\\
   4.43&  103.8&    0.8& 3.3E-5& 1.2E-5& 3.9E-6& 5.0E-5& 6.6E-10& 3.1E-11& 3.7E-11& 4.4E-7& 8.0E-8& 1.0E-8& 3.2E-11& 7.8E-7&88.1\\
   4.43&   83.0&    1.0& 2.8E-5& 2.8E-5& 3.8E-6& 5.0E-5& 5.4E-10& 2.7E-11& 2.1E-11& 4.7E-7& 7.6E-8& 8.9E-9& 2.4E-11& 9.6E-7&85.3\\
   4.43&   69.2&    1.2& 2.5E-5& 3.8E-5& 3.7E-6& 5.0E-5& 5.4E-10& 2.4E-11& 1.8E-11& 4.8E-7& 6.8E-8& 8.5E-9& 2.4E-11& 1.0E-6&84.2\\
     14&  207.5&    0.4& 6.9E-5& 2.8E-7& 1.4E-5& 2.2E-5& 2.3E-9& 5.0E-10& 1.6E-9& 2.6E-7& 2.9E-9& 1.6E-8& 2.8E-10& 2.9E-7&97.6\\
     14&  138.3&    0.6& 6.2E-5& 7.8E-6& 1.3E-5& 5.4E-5& 1.8E-10& 6.3E-11& 6.3E-11& 1.2E-6& 5.3E-8& 6.4E-9& 8.8E-12& 1.4E-6&90.9\\
     14&  103.8&    0.8& 4.9E-5& 2.7E-5& 1.2E-5& 5.6E-5& 1.3E-10& 5.3E-11& 2.9E-11& 1.3E-6& 5.6E-8& 5.4E-9& 5.5E-12& 1.8E-6&88.4\\
     14&   83.0&    1.0& 4.1E-5& 4.0E-5& 1.2E-5& 5.6E-5& 1.5E-10& 5.0E-11& 2.4E-11& 1.3E-6& 4.8E-8& 5.2E-9& 5.6E-12& 1.9E-6&87.6\\
     14&   69.2&    1.2& 3.6E-5& 4.9E-5& 1.2E-5& 5.5E-5& 1.7E-10& 4.8E-11& 2.2E-11& 1.3E-6& 4.5E-8& 5.1E-9& 5.8E-12& 1.9E-6&87.3\\
            \bottomrule
    \end{tabular}
        \tabnotes  The density is $\nh=\dix{4}\ccc$ in these models. The total gas-phase abundances \stot\ and \otot\ are given in ppm relative to H nuclei (\ctot=83 ppm). $^\dagger$ \ab{m+} = \ab{H3+} + \ab{HCO+} + \ab{H3O+}; $^\ddagger$ {Ratio in \%}.
\end{table*}

\begin{table*}
        \centering
        \caption{Steady-state abundance as predicted by our model for a density of $10^5$\ccc (see Table \ref{tab:abundances_n4}).}
        \label{tab:abundances_n5}
        \scriptsize
        \begin{tabular}%
                {ccccccccccccccccccc}
                \toprule
                %$\zeta_{17}$&
                \stot&  \otot&    C/O& \ab{O} & \ab{C} & \ab{S} & \ab{N} & \ab{NH} &     \ab{SH} & \ab{NS} & \ab{S+}& \ab{H+} & \ab{m+}&   \ab{N2H+}&  \ab{e-}
                & \ab{S}/\sgas\\
                \midrule
%               \input{models/V1/grid5_analysis_paper}
%{ccccccccccccc}
%\toprule
%  \stot&  \otot&    C/O&      O&      C&      S&      N&     NH&     SH&     NS&     S+&     H+&     m+&   N2H+&     e-&[S]/\stot\\
%\midrule
&&&&&&&&$\zeta_{17}=    1.0$ \\
   0.14&  207.5&    0.4& 3.9E-5& 3.5E-9& 1.4E-7& 3.7E-6& 5.0E-11& 3.7E-13& 6.8E-13& 7.8E-11& 2.2E-11& 3.8E-9& 8.3E-11& 5.7E-9&97.2\\
   0.14&  138.3&    0.6& 2.1E-5& 8.6E-9& 1.4E-7& 4.1E-6& 9.8E-11& 1.0E-12& 2.5E-12& 1.7E-10& 5.2E-11& 3.7E-9& 9.4E-11& 5.9E-9&97.1\\
   0.14&  103.8&    0.8& 1.3E-5& 3.3E-8& 1.4E-7& 6.0E-6& 1.5E-10& 1.6E-12& 6.2E-12& 4.4E-10& 1.7E-10& 3.4E-9& 9.7E-11& 6.3E-9&97.9\\
   0.14&   83.0&    1.0& 2.7E-6& 7.0E-7& 1.2E-7& 6.0E-6& 4.0E-10& 5.9E-12& 2.7E-11& 6.8E-10& 1.8E-10& 2.9E-9& 1.1E-10& 6.9E-9&85.6\\
   0.14&   69.2&    1.2& 1.4E-7& 2.4E-6& 5.8E-8& 5.6E-7& 6.8E-10& 2.8E-11& 1.1E-11& 1.8E-10& 3.3E-11& 2.9E-9& 1.3E-10& 6.0E-9&41.3\\
  0.443&  207.5&    0.4& 3.9E-5& 3.5E-9& 4.3E-7& 3.7E-6& 5.0E-11& 1.7E-12& 2.3E-12& 2.4E-10& 2.2E-11& 3.7E-9& 8.2E-11& 5.9E-9&96.3\\
  0.443&  138.3&    0.6& 2.2E-5& 8.6E-9& 4.3E-7& 4.1E-6& 9.7E-11& 4.6E-12& 8.7E-12& 5.4E-10& 5.2E-11& 3.6E-9& 9.2E-11& 6.1E-9&96.3\\
  0.443&  103.8&    0.8& 1.3E-5& 3.4E-8& 4.3E-7& 6.0E-6& 1.5E-10& 7.4E-12& 2.3E-11& 1.3E-9& 1.7E-10& 3.1E-9& 9.4E-11& 6.9E-9&97.4\\
  0.443&   83.0&    1.0& 2.8E-6& 7.8E-7& 3.8E-7& 6.1E-6& 3.9E-10& 2.3E-11& 8.7E-11& 2.0E-9& 1.8E-10& 2.5E-9& 1.0E-10& 7.9E-9&85.4\\
  0.443&   69.2&    1.2& 1.5E-7& 2.5E-6& 1.8E-7& 5.9E-7& 6.3E-10& 9.7E-11& 3.4E-11& 6.0E-10& 3.3E-11& 2.7E-9& 1.3E-10& 6.3E-9&41.6\\
    1.4&  207.5&    0.4& 4.0E-5& 3.5E-9& 1.3E-6& 3.7E-6& 4.8E-11& 9.0E-12& 7.8E-12& 7.4E-10& 2.2E-11& 3.5E-9& 7.9E-11& 6.2E-9&94.1\\
    1.4&  138.3&    0.6& 2.2E-5& 8.6E-9& 1.3E-6& 4.1E-6& 9.5E-11& 2.4E-11& 3.1E-11& 1.6E-9& 5.3E-11& 3.1E-9& 8.8E-11& 6.9E-9&94.3\\
    1.4&  103.8&    0.8& 1.4E-5& 3.4E-8& 1.4E-6& 6.0E-6& 1.6E-10& 3.8E-11& 8.8E-11& 3.8E-9& 1.7E-10& 2.4E-9& 8.6E-11& 8.8E-9&96.5\\
    1.4&   83.0&    1.0& 2.9E-6& 9.8E-7& 1.2E-6& 6.7E-6& 3.5E-10& 8.9E-11& 2.7E-10& 5.5E-9& 1.8E-10& 1.8E-9& 8.9E-11& 1.1E-8&85.2\\
    1.4&   69.2&    1.2& 1.8E-7& 3.0E-6& 6.0E-7& 6.7E-7& 5.2E-10& 3.5E-10& 9.0E-11& 2.1E-9& 3.3E-11& 2.3E-9& 1.2E-10& 7.3E-9&43.0\\
   4.43&  207.5&    0.4& 4.3E-5& 3.4E-9& 4.0E-6& 3.9E-6& 4.3E-11& 4.6E-11& 2.6E-11& 2.2E-9& 2.3E-11& 2.8E-9& 7.0E-11& 7.3E-9&90.1\\
   4.43&  138.3&    0.6& 2.5E-5& 8.2E-9& 4.0E-6& 4.3E-6& 8.6E-11& 1.1E-10& 1.0E-10& 4.8E-9& 5.5E-11& 2.2E-9& 7.5E-11& 9.3E-9&91.3\\
   4.43&  103.8&    0.8& 1.5E-5& 3.5E-8& 4.2E-6& 6.4E-6& 1.5E-10& 1.6E-10& 3.1E-10& 1.1E-8& 1.8E-10& 1.5E-9& 6.6E-11& 1.5E-8&95.2\\
   4.43&   83.0&    1.0& 3.4E-6& 1.4E-6& 3.8E-6& 8.0E-6& 2.6E-10& 2.8E-10& 6.5E-10& 1.5E-8& 1.6E-10& 1.0E-9& 6.4E-11& 1.9E-8&86.3\\
   4.43&   69.2&    1.2& 2.9E-7& 4.5E-6& 2.2E-6& 1.0E-6& 3.5E-10& 9.9E-10& 1.9E-10& 7.8E-9& 3.7E-11& 1.3E-9& 9.9E-11& 1.2E-8&49.7\\
     14&  207.5&    0.4& 5.5E-5& 3.1E-9& 1.2E-5& 4.5E-6& 3.0E-11& 1.5E-10& 6.2E-11& 6.7E-9& 2.5E-11& 1.8E-9& 5.0E-11& 1.1E-8&87.7\\
     14&  138.3&    0.6& 3.2E-5& 7.2E-9& 1.3E-5& 5.4E-6& 6.3E-11& 2.9E-10& 2.3E-10& 1.5E-8& 6.2E-11& 1.1E-9& 4.6E-11& 1.9E-8&90.8\\
     14&  103.8&    0.8& 1.7E-5& 3.6E-8& 1.3E-5& 7.9E-6& 1.1E-10& 4.1E-10& 7.4E-10& 3.2E-8& 1.6E-10& 6.4E-10& 3.5E-11& 3.5E-8&94.6\\
     14&   83.0&    1.0& 4.3E-6& 2.1E-6& 1.3E-5&10.0E-6& 1.5E-10& 6.9E-10& 1.3E-9& 4.1E-8& 1.2E-10& 4.9E-10& 3.1E-11& 4.5E-8&89.5\\
     14&   69.2&    1.2& 7.2E-7& 7.1E-6& 9.7E-6& 2.2E-6& 1.9E-10& 2.0E-9& 4.1E-10& 3.1E-8& 4.1E-11& 5.2E-10& 4.7E-11& 3.4E-8&69.1\\
&&&&&&&&$\zeta_{17}=    3.0$ \\
   0.14&  207.5&    0.4& 3.9E-5& 9.8E-9& 1.4E-7& 4.3E-6& 2.0E-10& 6.9E-13& 2.5E-12& 1.7E-10& 6.5E-11& 7.0E-9& 2.1E-10& 9.5E-9&98.1\\
   0.14&  138.3&    0.6& 2.2E-5& 2.3E-8& 1.4E-7& 5.3E-6& 3.8E-10& 1.7E-12& 8.4E-12& 3.7E-10& 1.5E-10& 6.8E-9& 2.4E-10& 9.9E-9&97.9\\
   0.14&  103.8&    0.8& 1.4E-5& 8.6E-8& 1.4E-7& 8.5E-6& 5.5E-10& 2.3E-12& 1.7E-11& 9.3E-10& 5.3E-10& 6.0E-9& 2.3E-10& 1.1E-8&97.8\\
   0.14&   83.0&    1.0& 3.6E-6& 1.5E-6& 1.2E-7& 8.8E-6& 1.1E-9& 6.1E-12& 3.2E-11& 1.5E-9& 5.4E-10& 4.8E-9& 2.5E-10& 1.3E-8&83.5\\
   0.14&   69.2&    1.2& 4.7E-7& 4.3E-6& 6.8E-8& 1.8E-6& 1.5E-9& 1.6E-11& 1.4E-11& 6.3E-10& 1.2E-10& 5.0E-9& 3.3E-10& 1.1E-8&48.7\\
  0.443&  207.5&    0.4& 3.9E-5& 9.8E-9& 4.3E-7& 4.4E-6& 2.0E-10& 3.1E-12& 8.0E-12& 5.4E-10& 6.5E-11& 6.9E-9& 2.1E-10& 9.8E-9&97.5\\
  0.443&  138.3&    0.6& 2.2E-5& 2.3E-8& 4.3E-7& 5.3E-6& 3.8E-10& 7.7E-12& 2.8E-11& 1.2E-9& 1.6E-10& 6.4E-9& 2.3E-10& 1.0E-8&97.4\\
  0.443&  103.8&    0.8& 1.4E-5& 8.8E-8& 4.3E-7& 8.6E-6& 5.6E-10&10.0E-12& 5.6E-11& 2.8E-9& 5.4E-10& 5.4E-9& 2.2E-10& 1.2E-8&97.6\\
  0.443&   83.0&    1.0& 3.7E-6& 1.7E-6& 3.7E-7& 9.2E-6& 1.0E-9& 2.2E-11& 9.5E-11& 4.3E-9& 5.7E-10& 4.2E-9& 2.3E-10& 1.5E-8&83.9\\
  0.443&   69.2&    1.2& 5.0E-7& 4.6E-6& 2.2E-7& 1.9E-6& 1.4E-9& 5.5E-11& 4.1E-11& 2.0E-9& 1.2E-10& 4.6E-9& 3.1E-10& 1.1E-8&49.5\\
    1.4&  207.5&    0.4& 4.0E-5& 9.9E-9& 1.3E-6& 4.4E-6& 1.9E-10& 1.7E-11& 2.6E-11& 1.7E-9& 6.6E-11& 6.3E-9& 2.0E-10& 1.1E-8&96.0\\
    1.4&  138.3&    0.6& 2.3E-5& 2.3E-8& 1.3E-6& 5.4E-6& 3.7E-10& 4.0E-11& 9.3E-11& 3.5E-9& 1.6E-10& 5.5E-9& 2.1E-10& 1.2E-8&96.2\\
    1.4&  103.8&    0.8& 1.5E-5& 9.6E-8& 1.4E-6& 9.1E-6& 5.5E-10& 4.8E-11& 2.0E-10& 8.2E-9& 5.7E-10& 4.1E-9& 1.9E-10& 1.7E-8&97.2\\
    1.4&   83.0&    1.0& 4.0E-6& 2.2E-6& 1.2E-6& 1.0E-5& 9.1E-10& 7.7E-11& 2.5E-10& 1.2E-8& 6.1E-10& 3.0E-9& 1.8E-10& 2.2E-8&85.1\\
    1.4&   69.2&    1.2& 6.2E-7& 5.8E-6& 7.4E-7& 2.3E-6& 1.1E-9& 1.8E-10& 1.0E-10& 6.5E-9& 1.4E-10& 3.6E-9& 2.7E-10& 1.5E-8&52.6\\
   4.43&  207.5&    0.4& 4.4E-5&10.0E-9& 4.1E-6& 4.7E-6& 1.7E-10& 8.9E-11& 8.2E-11& 5.1E-9& 6.8E-11& 5.0E-9& 1.7E-10& 1.3E-8&93.3\\
   4.43&  138.3&    0.6& 2.5E-5& 2.4E-8& 4.2E-6& 6.0E-6& 3.3E-10& 1.8E-10& 2.9E-10& 1.1E-8& 1.6E-10& 3.8E-9& 1.7E-10& 1.7E-8&94.6\\
   4.43&  103.8&    0.8& 1.6E-5& 1.2E-7& 4.3E-6& 1.1E-5& 4.7E-10& 1.8E-10& 5.9E-10& 2.4E-8& 6.5E-10& 2.4E-9& 1.2E-10& 3.1E-8&96.9\\
   4.43&   83.0&    1.0& 4.7E-6& 3.2E-6& 3.9E-6& 1.3E-5& 6.3E-10& 2.3E-10& 5.3E-10& 3.3E-8& 6.2E-10& 1.7E-9& 1.1E-10& 4.2E-8&87.8\\
   4.43&   69.2&    1.2& 1.0E-6& 8.5E-6& 2.8E-6& 3.4E-6& 7.7E-10& 5.2E-10& 2.2E-10& 2.2E-8& 1.8E-10& 2.0E-9& 1.7E-10& 2.9E-8&63.1\\
     14&  207.5&    0.4& 5.5E-5&10.0E-9& 1.3E-5& 5.8E-6& 1.2E-10& 3.0E-10& 1.9E-10& 1.6E-8& 7.5E-11& 2.9E-9& 1.1E-10& 2.2E-8&91.7\\
     14&  138.3&    0.6& 3.3E-5& 2.6E-8& 1.3E-5& 7.9E-6& 2.2E-10& 4.8E-10& 6.1E-10& 3.5E-8& 1.9E-10& 1.8E-9& 8.6E-11& 3.9E-8&94.6\\
     14&  103.8&    0.8& 1.9E-5& 2.1E-7& 1.4E-5& 1.5E-5& 2.6E-10& 4.2E-10& 1.2E-9& 7.6E-8& 7.6E-10& 1.1E-9& 4.7E-11& 8.3E-8&96.9\\
     14&   83.0&    1.0& 6.4E-6& 4.8E-6& 1.3E-5& 1.7E-5& 2.9E-10& 5.3E-10& 8.6E-10& 9.6E-8& 5.0E-10& 9.1E-10& 4.0E-11& 1.1E-7&91.7\\
     14&   69.2&    1.2& 2.1E-6& 1.2E-5& 1.1E-5& 6.5E-6& 3.7E-10& 1.1E-9& 4.4E-10& 8.0E-8& 2.1E-10& 9.2E-10& 5.9E-11& 8.8E-8&80.6\\
&&&&&&&&$\zeta_{17}=   10.0$ \\
   0.14&  207.5&    0.4& 4.0E-5& 2.8E-8& 1.4E-7& 5.7E-6& 8.6E-10& 1.3E-12& 9.7E-12& 4.2E-10& 2.2E-10& 1.3E-8& 5.3E-10& 1.7E-8&98.5\\
   0.14&  138.3&    0.6& 2.3E-5& 6.1E-8& 1.4E-7& 7.5E-6& 1.6E-9& 2.8E-12& 2.9E-11& 8.5E-10& 5.1E-10& 1.3E-8& 5.8E-10& 1.8E-8&98.2\\
   0.14&  103.8&    0.8& 1.5E-5& 2.4E-7& 1.4E-7& 1.3E-5& 2.0E-9& 2.8E-12& 3.8E-11& 2.0E-9& 2.0E-9& 1.1E-8& 5.1E-10& 2.3E-8&97.1\\
   0.14&   83.0&    1.0& 5.3E-6& 3.7E-6& 1.2E-7& 1.5E-5& 2.6E-9& 5.5E-12& 2.9E-11& 3.2E-9& 2.4E-9& 7.7E-9& 4.6E-10& 3.0E-8&82.6\\
   0.14&   69.2&    1.2& 1.6E-6& 8.5E-6& 8.1E-8& 6.2E-6& 3.0E-9& 9.4E-12& 1.6E-11& 2.2E-9& 6.3E-10& 8.5E-9& 7.0E-10& 2.2E-8&57.9\\
  0.443&  207.5&    0.4& 4.0E-5& 2.8E-8& 4.3E-7& 5.8E-6& 8.5E-10& 5.6E-12& 3.1E-11& 1.3E-9& 2.2E-10& 1.3E-8& 5.2E-10& 1.8E-8&98.2\\
  0.443&  138.3&    0.6& 2.3E-5& 6.2E-8& 4.3E-7& 7.6E-6& 1.5E-9& 1.2E-11& 9.1E-11& 2.6E-9& 5.2E-10& 1.2E-8& 5.6E-10& 2.0E-8&98.0\\
  0.443&  103.8&    0.8& 1.5E-5& 2.5E-7& 4.3E-7& 1.4E-5& 1.9E-9& 1.2E-11& 1.2E-10& 6.2E-9& 2.0E-9& 9.4E-9& 4.6E-10& 2.6E-8&97.1\\
  0.443&   83.0&    1.0& 5.5E-6& 4.1E-6& 3.7E-7& 1.6E-5& 2.4E-9& 1.8E-11& 7.9E-11& 9.6E-9& 2.7E-9& 6.6E-9& 3.9E-10& 3.7E-8&83.7\\
  0.443&   69.2&    1.2& 1.8E-6& 9.4E-6& 2.7E-7& 6.7E-6& 2.7E-9& 2.9E-11& 4.3E-11& 6.7E-9& 7.2E-10& 7.5E-9& 6.2E-10& 2.6E-8&60.0\\
    1.4&  207.5&    0.4& 4.1E-5& 2.9E-8& 1.4E-6& 6.0E-6& 8.1E-10& 3.0E-11& 9.6E-11& 4.1E-9& 2.2E-10& 1.2E-8& 4.8E-10& 2.0E-8&97.4\\
    1.4&  138.3&    0.6& 2.4E-5& 6.4E-8& 1.4E-6& 8.0E-6& 1.5E-9& 5.9E-11& 2.9E-10& 8.1E-9& 5.3E-10& 1.0E-8& 4.9E-10& 2.3E-8&97.5\\
    1.4&  103.8&    0.8& 1.6E-5& 2.9E-7& 1.4E-6& 1.5E-5& 1.7E-9& 5.0E-11& 3.5E-10& 1.9E-8& 2.3E-9& 7.1E-9& 3.5E-10& 3.7E-8&97.1\\
    1.4&   83.0&    1.0& 6.1E-6& 5.1E-6& 1.2E-6& 1.8E-5& 1.9E-9& 5.4E-11& 1.8E-10& 2.8E-8& 3.2E-9& 4.8E-9& 2.6E-10& 5.7E-8&86.0\\
    1.4&   69.2&    1.2& 2.3E-6& 1.2E-5& 9.3E-7& 8.2E-6& 2.2E-9& 8.6E-11& 1.0E-10& 2.1E-8& 9.9E-10& 5.4E-9& 4.3E-10& 4.1E-8&66.3\\
   4.43&  207.5&    0.4& 4.5E-5& 3.0E-8& 4.3E-6& 6.6E-6& 7.1E-10& 1.6E-10& 2.8E-10& 1.3E-8& 2.2E-10& 9.1E-9& 3.9E-10& 2.6E-8&96.0\\
   4.43&  138.3&    0.6& 2.6E-5& 6.9E-8& 4.3E-6& 9.1E-6& 1.2E-9& 2.6E-10& 7.8E-10& 2.5E-8& 5.6E-10& 6.8E-9& 3.4E-10& 3.7E-8&96.9\\
   4.43&  103.8&    0.8& 1.8E-5& 4.3E-7& 4.3E-6& 1.9E-5& 1.2E-9& 1.6E-10& 7.6E-10& 5.9E-8& 3.0E-9& 4.1E-9& 1.7E-10& 7.7E-8&97.2\\
   4.43&   83.0&    1.0& 7.5E-6& 6.7E-6& 4.0E-6& 2.1E-5& 1.1E-9& 1.5E-10& 3.3E-10& 8.3E-8& 3.2E-9& 3.1E-9& 1.2E-10& 1.2E-7&89.4\\
   4.43&   69.2&    1.2& 3.4E-6& 1.5E-5& 3.5E-6& 1.2E-5& 1.3E-9& 2.2E-10& 2.1E-10& 6.9E-8& 1.4E-9& 3.2E-9& 1.8E-10& 9.5E-8&78.0\\
     14&  207.5&    0.4& 5.6E-5& 3.3E-8& 1.3E-5& 8.5E-6& 4.7E-10& 5.2E-10& 5.9E-10& 4.1E-8& 2.5E-10& 5.0E-9& 2.0E-10& 5.0E-8&95.3\\
     14&  138.3&    0.6& 3.5E-5& 9.3E-8& 1.4E-5& 1.3E-5& 6.4E-10& 6.1E-10& 1.3E-9& 8.7E-8& 7.1E-10& 3.2E-9& 1.3E-10& 9.6E-8&97.2\\
     14&  103.8&    0.8& 2.1E-5& 1.4E-6& 1.4E-5& 3.0E-5& 3.7E-10& 2.8E-10& 7.9E-10& 2.2E-7& 4.4E-9& 2.0E-9& 3.6E-11& 2.5E-7&96.9\\
     14&   83.0&    1.0& 1.1E-5& 9.7E-6& 1.3E-5& 2.7E-5& 3.9E-10& 3.2E-10& 4.4E-10& 2.5E-7& 2.5E-9& 1.8E-9& 3.5E-11& 2.8E-7&92.8\\
     14&   69.2&    1.2& 6.2E-6& 1.8E-5& 1.2E-5& 1.8E-5& 4.8E-10& 4.2E-10& 3.0E-10& 2.3E-7& 1.4E-9& 1.8E-9& 4.7E-11& 2.7E-7&88.4\\

                \bottomrule
        \end{tabular}
\end{table*}

\begin{table*}
        \centering
        \caption{Steady-state abundance as predicted by our model for a density of $10^6$\ccc\ (see Table \ref{tab:abundances_n4}).}
        \label{tab:abundances_n6}
        \scriptsize
        \begin{tabular}%
                {ccccccccccccccccccc}
                \toprule
                %$\zeta_{17}$&
                \stot&  \otot&    C/O& \ab{O} & \ab{C} & \ab{S} & \ab{N} & \ab{NH} &     \ab{SH} & \ab{NS} & \ab{S+}& \ab{H+} & \ab{m+}&   \ab{N2H+}&  \ab{e-}
                & \ab{S}/\sgas\\
                \midrule
%               \input{models/V1/grid6_analysis_paper}
%{ccccccccccccc}
%\toprule
%  \stot&  \otot&    C/O&      O&      C&      S&      N&     NH&     SH&     NS&     S+&     H+&     m+&   N2H+&     e-&[S]/\stot\\
%\midrule
&&&&&&&&$\zeta_{17}=    1.0$ \\
   0.14&  207.5&    0.4& 3.8E-5& 3.6E-10& 1.3E-7& 3.3E-6& 2.9E-12& 8.6E-14& 5.7E-14& 1.4E-11& 2.2E-12& 8.9E-10& 9.8E-12& 2.4E-9&94.4\\
   0.14&  138.3&    0.6& 2.1E-5& 9.3E-10& 1.3E-7& 3.3E-6& 5.7E-12& 2.7E-13& 2.5E-13& 3.2E-11& 5.2E-12& 8.8E-10& 1.1E-11& 2.4E-9&94.3\\
   0.14&  103.8&    0.8& 1.3E-5& 4.0E-9& 1.4E-7& 3.8E-6& 9.0E-12& 5.4E-13& 7.7E-13& 8.3E-11& 1.6E-11& 8.4E-10& 1.2E-11& 2.5E-9&96.9\\
   0.14&   83.0&    1.0& 1.9E-6& 2.1E-7& 1.2E-7& 3.5E-6& 3.4E-11& 3.6E-12& 1.2E-11& 1.2E-10& 2.0E-11& 7.4E-10& 1.3E-11& 2.6E-9&89.2\\
   0.14&   69.2&    1.2& 1.2E-8& 8.8E-7& 3.2E-8& 5.1E-8& 1.0E-10& 4.1E-11& 3.0E-12& 1.0E-11& 2.8E-12& 7.1E-10& 1.6E-11& 2.4E-9&23.0\\
  0.443&  207.5&    0.4& 3.8E-5& 3.6E-10& 4.1E-7& 3.2E-6& 2.9E-12& 3.7E-13& 2.0E-13& 4.3E-11& 2.2E-12& 8.7E-10& 9.7E-12& 2.4E-9&92.6\\
  0.443&  138.3&    0.6& 2.1E-5& 9.3E-10& 4.1E-7& 3.2E-6& 5.6E-12& 1.2E-12& 9.1E-13& 9.9E-11& 5.2E-12& 8.4E-10& 1.1E-11& 2.5E-9&92.5\\
  0.443&  103.8&    0.8& 1.3E-5& 3.8E-9& 4.2E-7& 3.6E-6& 9.3E-12& 2.6E-12& 3.1E-12& 2.5E-10& 1.6E-11& 7.9E-10& 1.2E-11& 2.6E-9&95.8\\
  0.443&   83.0&    1.0& 1.8E-6& 2.1E-7& 3.9E-7& 3.2E-6& 3.4E-11& 1.7E-11& 5.1E-11& 3.7E-10& 2.0E-11& 6.8E-10& 1.3E-11& 2.8E-9&87.7\\
  0.443&   69.2&    1.2& 1.3E-8& 9.2E-7& 1.0E-7& 5.3E-8& 9.2E-11& 1.3E-10& 8.8E-12& 3.8E-11& 2.8E-12& 6.7E-10& 1.6E-11& 2.4E-9&23.1\\
    1.4&  207.5&    0.4& 3.9E-5& 3.6E-10& 1.2E-6& 3.1E-6& 2.7E-12& 1.9E-12& 7.8E-13& 1.2E-10& 2.2E-12& 8.2E-10& 9.4E-12& 2.5E-9&88.4\\
    1.4&  138.3&    0.6& 2.1E-5& 8.8E-10& 1.2E-6& 3.0E-6& 5.4E-12& 6.0E-12& 3.6E-12& 2.9E-10& 5.2E-12& 7.6E-10& 1.1E-11& 2.7E-9&88.1\\
    1.4&  103.8&    0.8& 1.2E-5& 3.3E-9& 1.3E-6& 3.1E-6& 9.6E-12& 1.4E-11& 1.4E-11& 7.2E-10& 1.5E-11& 6.6E-10& 1.1E-11& 3.0E-9&92.8\\
    1.4&   83.0&    1.0& 1.8E-6& 2.0E-7& 1.2E-6& 2.9E-6& 3.2E-11& 8.5E-11& 2.2E-10& 1.1E-9& 2.0E-11& 5.4E-10& 1.3E-11& 3.4E-9&84.6\\
    1.4&   69.2&    1.2& 1.6E-8& 1.0E-6& 3.3E-7& 6.1E-8& 7.0E-11& 4.2E-10& 2.3E-11& 1.6E-10& 2.7E-12& 5.7E-10& 1.6E-11& 2.5E-9&23.6\\
   4.43&  207.5&    0.4& 4.3E-5& 3.3E-10& 3.6E-6& 3.1E-6& 2.3E-12& 8.8E-12& 2.7E-12& 3.5E-10& 2.3E-12& 7.0E-10& 8.6E-12& 2.8E-9&81.8\\
   4.43&  138.3&    0.6& 2.4E-5& 7.3E-10& 3.6E-6& 2.8E-6& 4.7E-12& 2.5E-11& 1.2E-11& 8.4E-10& 5.5E-12& 5.9E-10& 9.8E-12& 3.1E-9&81.9\\
   4.43&  103.8&    0.8& 1.3E-5& 2.0E-9& 3.9E-6& 2.8E-6& 9.0E-12& 6.3E-11& 5.0E-11& 2.0E-9& 1.4E-11& 4.5E-10& 1.0E-11& 4.1E-9&87.6\\
   4.43&   83.0&    1.0& 1.9E-6& 1.3E-7& 3.6E-6& 2.6E-6& 2.5E-11& 3.2E-10& 8.2E-10& 3.1E-9& 1.5E-11& 3.3E-10& 1.1E-11& 5.1E-9&80.9\\
   4.43&   69.2&    1.2& 2.9E-8& 1.4E-6& 1.2E-6& 9.8E-8& 4.1E-11& 1.2E-9& 5.9E-11& 7.7E-10& 2.6E-12& 3.7E-10& 1.5E-11& 2.9E-9&27.0\\
     14&  207.5&    0.4& 5.6E-5& 2.6E-10& 1.1E-5& 3.4E-6& 1.5E-12& 2.5E-11& 5.9E-12& 1.0E-9& 2.7E-12& 4.9E-10& 6.7E-12& 3.4E-9&79.0\\
     14&  138.3&    0.6& 3.3E-5& 4.3E-10& 1.1E-5& 3.2E-6& 3.2E-12& 6.2E-11& 2.3E-11& 2.8E-9& 6.4E-12& 3.3E-10& 7.4E-12& 4.8E-9&80.3\\
     14&  103.8&    0.8& 1.5E-5& 7.6E-10& 1.2E-5& 3.0E-6& 7.0E-12& 1.6E-10& 1.1E-10& 6.1E-9& 1.2E-11& 2.0E-10& 7.4E-12& 7.9E-9&85.1\\
     14&   83.0&    1.0& 2.8E-6& 4.4E-9& 1.1E-5& 1.7E-6& 1.5E-11& 7.8E-10& 1.4E-9& 7.7E-9& 6.7E-12& 1.5E-10& 8.2E-12& 9.4E-9&77.9\\
     14&   69.2&    1.2& 1.1E-7& 2.0E-6& 6.2E-6& 2.7E-7& 2.0E-11& 3.2E-9& 2.4E-10& 4.4E-9& 2.4E-12& 1.5E-10& 1.1E-11& 6.2E-9&44.6\\
&&&&&&&&$\zeta_{17}=    3.0$ \\
   0.14&  207.5&    0.4& 3.8E-5& 1.1E-9& 1.3E-7& 3.4E-6& 1.1E-11& 1.8E-13& 1.8E-13& 3.2E-11& 6.6E-12& 1.9E-9& 2.8E-11& 3.5E-9&95.9\\
   0.14&  138.3&    0.6& 2.1E-5& 2.7E-9& 1.3E-7& 3.5E-6& 2.2E-11& 5.4E-13& 7.1E-13& 7.3E-11& 1.6E-11& 1.8E-9& 3.2E-11& 3.5E-9&95.8\\
   0.14&  103.8&    0.8& 1.3E-5& 1.1E-8& 1.4E-7& 4.5E-6& 3.4E-11& 9.8E-13& 2.0E-12& 1.9E-10& 4.9E-11& 1.7E-9& 3.3E-11& 3.7E-9&97.5\\
   0.14&   83.0&    1.0& 2.2E-6& 3.5E-7& 1.2E-7& 4.3E-6& 1.1E-10& 5.0E-12& 1.9E-11& 2.8E-10& 5.7E-11& 1.5E-9& 3.7E-11& 3.9E-9&87.9\\
   0.14&   69.2&    1.2& 3.9E-8& 1.3E-6& 4.5E-8& 1.6E-7& 2.6E-10& 4.2E-11& 6.8E-12& 4.2E-11& 8.8E-12& 1.5E-9& 4.5E-11& 3.6E-9&32.2\\
  0.443&  207.5&    0.4& 3.8E-5& 1.1E-9& 4.2E-7& 3.4E-6& 1.1E-11& 8.0E-13& 6.1E-13&10.0E-11& 6.6E-12& 1.8E-9& 2.7E-11& 3.6E-9&94.5\\
  0.443&  138.3&    0.6& 2.1E-5& 2.7E-9& 4.2E-7& 3.5E-6& 2.2E-11& 2.4E-12& 2.6E-12& 2.3E-10& 1.6E-11& 1.7E-9& 3.1E-11& 3.7E-9&94.5\\
  0.443&  103.8&    0.8& 1.3E-5& 1.1E-8& 4.3E-7& 4.3E-6& 3.5E-11& 4.6E-12& 7.9E-12& 5.7E-10& 4.9E-11& 1.6E-9& 3.3E-11& 4.0E-9&96.8\\
  0.443&   83.0&    1.0& 2.2E-6& 3.7E-7& 3.8E-7& 4.2E-6& 1.1E-10& 2.1E-11& 6.9E-11& 8.3E-10& 5.7E-11& 1.3E-9& 3.6E-11& 4.3E-9&86.9\\
  0.443&   69.2&    1.2& 4.1E-8& 1.4E-6& 1.4E-7& 1.6E-7& 2.4E-10& 1.4E-10& 2.0E-11& 1.5E-10& 8.8E-12& 1.4E-9& 4.5E-11& 3.6E-9&32.3\\
    1.4&  207.5&    0.4& 3.9E-5& 1.1E-9& 1.3E-6& 3.3E-6& 1.1E-11& 4.2E-12& 2.3E-12& 3.0E-10& 6.7E-12& 1.7E-9& 2.7E-11& 3.7E-9&91.2\\
    1.4&  138.3&    0.6& 2.2E-5& 2.7E-9& 1.3E-6& 3.4E-6& 2.1E-11& 1.2E-11& 9.9E-12& 6.8E-10& 1.6E-11& 1.6E-9& 3.0E-11& 4.0E-9&91.4\\
    1.4&  103.8&    0.8& 1.3E-5& 1.1E-8& 1.3E-6& 4.1E-6& 3.6E-11& 2.5E-11& 3.4E-11& 1.6E-9& 4.8E-11& 1.3E-9& 3.1E-11& 4.8E-9&94.9\\
    1.4&   83.0&    1.0& 2.2E-6& 4.2E-7& 1.2E-6& 4.3E-6& 1.0E-10& 9.3E-11& 2.5E-10& 2.4E-9& 5.5E-11& 1.0E-9& 3.4E-11& 5.5E-9&85.3\\
    1.4&   69.2&    1.2& 4.8E-8& 1.6E-6& 4.6E-7& 1.9E-7& 1.9E-10& 5.0E-10& 5.6E-11& 5.7E-10& 8.7E-12& 1.2E-9& 4.4E-11& 3.9E-9&32.9\\
   4.43&  207.5&    0.4& 4.3E-5& 1.0E-9& 3.8E-6& 3.3E-6& 9.2E-12& 2.0E-11& 7.8E-12& 8.5E-10& 6.9E-12& 1.4E-9& 2.4E-11& 4.2E-9&85.8\\
   4.43&  138.3&    0.6& 2.4E-5& 2.4E-9& 3.8E-6& 3.3E-6& 1.9E-11& 5.4E-11& 3.3E-11& 2.0E-9& 1.6E-11& 1.2E-9& 2.7E-11& 5.0E-9&86.7\\
   4.43&  103.8&    0.8& 1.4E-5& 8.3E-9& 4.1E-6& 3.9E-6& 3.5E-11& 1.1E-10& 1.2E-10& 4.5E-9& 4.5E-11& 8.2E-10& 2.7E-11& 7.2E-9&91.9\\
   4.43&   83.0&    1.0& 2.5E-6& 5.2E-7& 3.7E-6& 4.7E-6& 8.1E-11& 3.1E-10& 7.5E-10& 6.5E-9& 4.5E-11& 5.9E-10& 2.8E-11& 9.1E-9&84.3\\
   4.43&   69.2&    1.2& 8.1E-8& 2.3E-6& 1.6E-6& 2.8E-7& 1.2E-10& 1.4E-9& 1.3E-10& 2.5E-9& 8.7E-12& 7.5E-10& 4.0E-11& 5.3E-9&37.2\\
     14&  207.5&    0.4& 5.6E-5& 8.5E-10& 1.2E-5& 3.8E-6& 6.1E-12& 6.2E-11& 1.8E-11& 2.5E-9& 7.8E-12& 9.6E-10& 1.8E-11& 5.6E-9&83.0\\
     14&  138.3&    0.6& 3.2E-5& 1.7E-9& 1.2E-5& 3.9E-6& 1.3E-11& 1.4E-10& 6.9E-11& 6.4E-9& 1.9E-11& 6.0E-10& 1.9E-11& 8.9E-9&85.6\\
     14&  103.8&    0.8& 1.6E-5& 4.6E-9& 1.3E-5& 4.4E-6& 2.7E-11& 2.8E-10& 2.9E-10& 1.3E-8& 3.8E-11& 3.6E-10& 1.7E-11& 1.5E-8&90.4\\
     14&   83.0&    1.0& 3.0E-6& 3.4E-7& 1.2E-5& 4.8E-6& 5.0E-11& 8.4E-10& 2.1E-9& 1.7E-8& 2.8E-11& 2.7E-10& 1.7E-11& 2.0E-8&85.6\\
     14&   69.2&    1.2& 2.5E-7& 3.6E-6& 7.8E-6& 6.6E-7& 6.5E-11& 3.0E-9& 3.3E-10& 1.1E-8& 8.5E-12& 2.9E-10& 2.5E-11& 1.3E-8&55.5\\
&&&&&&&&$\zeta_{17}=   10.0$ \\
   0.14&  207.5&    0.4& 3.9E-5& 3.5E-9& 1.4E-7& 3.7E-6& 5.0E-11& 3.7E-13& 6.8E-13& 7.8E-11& 2.2E-11& 3.8E-9& 8.3E-11& 5.7E-9&97.2\\
   0.14&  138.3&    0.6& 2.1E-5& 8.6E-9& 1.4E-7& 4.1E-6& 9.8E-11& 1.0E-12& 2.5E-12& 1.7E-10& 5.2E-11& 3.7E-9& 9.4E-11& 5.9E-9&97.1\\
   0.14&  103.8&    0.8& 1.3E-5& 3.3E-8& 1.4E-7& 6.0E-6& 1.5E-10& 1.6E-12& 6.2E-12& 4.4E-10& 1.7E-10& 3.4E-9& 9.7E-11& 6.3E-9&97.9\\
   0.14&   83.0&    1.0& 2.7E-6& 7.0E-7& 1.2E-7& 6.0E-6& 4.0E-10& 5.9E-12& 2.7E-11& 6.8E-10& 1.8E-10& 2.9E-9& 1.1E-10& 6.9E-9&85.6\\
   0.14&   69.2&    1.2& 1.4E-7& 2.4E-6& 5.8E-8& 5.6E-7& 6.8E-10& 2.8E-11& 1.1E-11& 1.8E-10& 3.3E-11& 2.9E-9& 1.3E-10& 6.0E-9&41.3\\
  0.443&  207.5&    0.4& 3.9E-5& 3.5E-9& 4.3E-7& 3.7E-6& 5.0E-11& 1.7E-12& 2.3E-12& 2.4E-10& 2.2E-11& 3.7E-9& 8.2E-11& 5.9E-9&96.3\\
  0.443&  138.3&    0.6& 2.2E-5& 8.6E-9& 4.3E-7& 4.1E-6& 9.7E-11& 4.6E-12& 8.7E-12& 5.4E-10& 5.2E-11& 3.6E-9& 9.2E-11& 6.1E-9&96.3\\
  0.443&  103.8&    0.8& 1.3E-5& 3.4E-8& 4.3E-7& 6.0E-6& 1.5E-10& 7.4E-12& 2.3E-11& 1.3E-9& 1.7E-10& 3.1E-9& 9.4E-11& 6.9E-9&97.4\\
  0.443&   83.0&    1.0& 2.8E-6& 7.8E-7& 3.8E-7& 6.1E-6& 3.9E-10& 2.3E-11& 8.7E-11& 2.0E-9& 1.8E-10& 2.5E-9& 1.0E-10& 7.9E-9&85.4\\
  0.443&   69.2&    1.2& 1.5E-7& 2.5E-6& 1.8E-7& 5.9E-7& 6.3E-10& 9.7E-11& 3.4E-11& 6.0E-10& 3.3E-11& 2.7E-9& 1.3E-10& 6.3E-9&41.6\\
    1.4&  207.5&    0.4& 4.0E-5& 3.5E-9& 1.3E-6& 3.7E-6& 4.8E-11& 9.0E-12& 7.8E-12& 7.4E-10& 2.2E-11& 3.5E-9& 7.9E-11& 6.2E-9&94.1\\
    1.4&  138.3&    0.6& 2.2E-5& 8.6E-9& 1.3E-6& 4.1E-6& 9.5E-11& 2.4E-11& 3.1E-11& 1.6E-9& 5.3E-11& 3.1E-9& 8.8E-11& 6.9E-9&94.3\\
    1.4&  103.8&    0.8& 1.4E-5& 3.4E-8& 1.4E-6& 6.0E-6& 1.6E-10& 3.8E-11& 8.8E-11& 3.8E-9& 1.7E-10& 2.4E-9& 8.6E-11& 8.8E-9&96.5\\
    1.4&   83.0&    1.0& 2.9E-6& 9.8E-7& 1.2E-6& 6.7E-6& 3.5E-10& 8.9E-11& 2.7E-10& 5.5E-9& 1.8E-10& 1.8E-9& 8.9E-11& 1.1E-8&85.2\\
    1.4&   69.2&    1.2& 1.8E-7& 3.0E-6& 6.0E-7& 6.7E-7& 5.2E-10& 3.5E-10& 9.0E-11& 2.1E-9& 3.3E-11& 2.3E-9& 1.2E-10& 7.3E-9&43.0\\
   4.43&  207.5&    0.4& 4.3E-5& 3.4E-9& 4.0E-6& 3.9E-6& 4.3E-11& 4.6E-11& 2.6E-11& 2.2E-9& 2.3E-11& 2.8E-9& 7.0E-11& 7.3E-9&90.1\\
   4.43&  138.3&    0.6& 2.5E-5& 8.2E-9& 4.0E-6& 4.3E-6& 8.6E-11& 1.1E-10& 1.0E-10& 4.8E-9& 5.5E-11& 2.2E-9& 7.5E-11& 9.3E-9&91.3\\
   4.43&  103.8&    0.8& 1.5E-5& 3.5E-8& 4.2E-6& 6.4E-6& 1.5E-10& 1.6E-10& 3.1E-10& 1.1E-8& 1.8E-10& 1.5E-9& 6.6E-11& 1.5E-8&95.2\\
   4.43&   83.0&    1.0& 3.4E-6& 1.4E-6& 3.8E-6& 8.0E-6& 2.6E-10& 2.8E-10& 6.5E-10& 1.5E-8& 1.6E-10& 1.0E-9& 6.4E-11& 1.9E-8&86.3\\
   4.43&   69.2&    1.2& 2.9E-7& 4.5E-6& 2.2E-6& 1.0E-6& 3.5E-10& 9.9E-10& 1.9E-10& 7.8E-9& 3.7E-11& 1.3E-9& 9.9E-11& 1.2E-8&49.7\\
     14&  207.5&    0.4& 5.5E-5& 3.1E-9& 1.2E-5& 4.5E-6& 3.0E-11& 1.5E-10& 6.2E-11& 6.7E-9& 2.5E-11& 1.8E-9& 5.0E-11& 1.1E-8&87.7\\
     14&  138.3&    0.6& 3.2E-5& 7.2E-9& 1.3E-5& 5.4E-6& 6.3E-11& 2.9E-10& 2.3E-10& 1.5E-8& 6.2E-11& 1.1E-9& 4.6E-11& 1.9E-8&90.8\\
     14&  103.8&    0.8& 1.7E-5& 3.6E-8& 1.3E-5& 7.9E-6& 1.1E-10& 4.1E-10& 7.4E-10& 3.2E-8& 1.6E-10& 6.4E-10& 3.5E-11& 3.5E-8&94.6\\
     14&   83.0&    1.0& 4.3E-6& 2.1E-6& 1.3E-5&10.0E-6& 1.5E-10& 6.9E-10& 1.3E-9& 4.1E-8& 1.2E-10& 4.9E-10& 3.1E-11& 4.5E-8&89.5\\
     14&   69.2&    1.2& 7.2E-7& 7.1E-6& 9.7E-6& 2.2E-6& 1.9E-10& 2.0E-9& 4.1E-10& 3.1E-8& 4.1E-11& 5.2E-10& 4.7E-11& 3.4E-8&69.1\\

                \bottomrule
        \end{tabular}
\end{table*}

\end{document}